\documentclass[superscriptaddress,twocolumn,amsmath,amssymb,aps,prc]{revtex4-2}
\usepackage{amsmath}
\usepackage{braket}
\usepackage{booktabs}
\usepackage{multirow}
\usepackage{physics}
\usepackage{xcolor}
\usepackage{framed}
\usepackage[normalem]{ulem}
\usepackage[dvipsnames]{xcolor}
\usepackage{graphicx}
\usepackage{dcolumn}
\usepackage{multirow}
\usepackage{longtable}
\usepackage{bm}
\usepackage{float}
\usepackage{hyperref}
\hypersetup{
    colorlinks=true,
    linkcolor=blue,
    filecolor=blue,
    citecolor=blue,
    urlcolor=blue,
    pdftitle={Overleaf Example},
    pdfpagemode=FullScreen,
    pdfauthor=author
    }
\begin{document}

\preprint{APS / Physical Review C}

\title{One- and two-nucleon transfer in $^{\mathbf{116}}$Sn+$^{\mathbf{60}}$Ni:
A coupled reaction channel analysis}

\author{Chandra Kumar}
\altaffiliation{Present address: Govt. T.C.L. P.G. College, Khokhra Bhata,
Janjgir 495668, Chhattisgarh, India}
\author{S. Nath}
\email{subir@iuac.res.in} 
\affiliation{Nuclear Physics Group, Inter-University Accelerator Centre,
Aruna Asaf Ali Marg, New Delhi 110067, India}

\begin{abstract}
Recent studies of multi-nucleon transfer in heavy ion collisions have employed both macroscopic and microscopic models. Although macroscopic approaches offer useful insights, microscopic analyses of high-precision experimental data provide a more reliable framework for understanding the nucleon transfer mechanisms. The present study aims to carry out a comprehensive theoretical investigation of the $^{116}$Sn+$^{60}$Ni system using microscopic coupled reaction channel (CRC) calculations. The calculations employ microscopic double-folding S$\tilde{a}$o Paulo potentials, incorporating all relevant inelastic and transfer couplings guided by observed $\gamma$-ray transitions, wherever available. For the one-nucleon transfer channels, spectroscopic amplitudes are also obtained from large-scale shell-model calculations. In the case of two-nucleon transfer, sequential, microscopic cluster and extreme cluster mechanisms are considered to reproduce the data. Results for quasielastic scattering and one-neutron ($1n$) transfer show excellent agreement with experimental data. Measured one-proton ($1p$) transfer probabilities are best described by incorporating experimental spectroscopic amplitudes in the CRC calculations. For transfer of two-nucleons, the extreme cluster mechanism is found to best reproduce the data. This study highlights that microscopic description of one- and two-nucleon transfer between two heavy ions in the CRC framework, without taking recourse to arbitrary normalization of the cross sections, is quite feasible. Nonetheless, lack of experimental corroboration for all the transitions included in the calculations and practical limits of computational resources, affecting accuracy of shell-model results and causing a cap on the number of states, leave room for further refinement of the results.
\end{abstract}
\maketitle
\section{Introduction}
\label{sec:introduction}
The study of multi-nucleon transfer (MNT) plays a significant role in
exploring pairing correlations~\cite{Winther1981,Oertzen2001}
the nuclear Josephson effect~\cite{Potel2021,Broglia2022,Szilner2024},
neutrinoless double $\beta$-decay~\cite{Cappuzzello2023} and the
transition from quasielastic (QEL) to deep-inelastic collision (DIC)
regimes~\cite{Diklic2023,Corradi2022}, while also offering a new
pathway to produce heavy neutron-rich nuclei~\cite{Devaraja2020,Devaraja2025}.
In addition, MNT channels are known to have a noticeable impact on
heavy-ion fusion dynamics, where a significant enhancement in the
sub-barrier fusion cross sections has been observed due to the
presence of MNT channels~\cite{Chandra2025PRC}.

An extensive amount of theoretical and experimental work on MNT
in intermediate and heavy-mass systems has been reported, focusing
on neutron / proton transfer channels above the Coulomb barrier
~\cite{Corradi2009,Cappuzzello2016,Mijatovi2022}. However, below
the Coulomb barrier, only a few experimental investigations exist,
primarily due to measurement constraints. The study of proton
transfer channels in heavy systems is limited as a result of the
low cross section of these channels. Some examples are
$^{92}$Mo+$^{54}$Fe \cite{Mijatovic2026}, $^{144}$Sm+$^{88}$Sr
\cite{Kunkel1988,Kunkel1990}, $^{144}$Sm+$^{208}$Pb \cite{Speer1991}
and $^{88}$Kr+$^{54}$Fe \cite{Wilpert1991}. 

These experimental results have been predominantly interpreted
using theoretical tools such as the Complex-\texttt{WKB} (CWKB)
approach~\cite{Winther1994,Winther1995} and the \textsc{grazing}
model~\cite{grazing}. However, the predictive power of these models
remains limited, as they are not fully microscopic and rely on
several assumptions. With advances in nuclear reaction theory and
the increasing availability of computational resources, it has
become feasible to perform more realistic quantum mechanical
calculations for both one- and two-nucleon transfer processes.
These developments have allowed the incorporation of detailed
nuclear structure inputs and realistic coupling schemes within
the Coupled Reaction Channel (CRC) framework~\cite{Thompson1988,fresco}.
As a result, several medium-mass systems have recently been explored
using such approaches. Notable examples include $^{28}$Si+$^{90,94}$Zr
~\cite{Chandra2023}, $^{16}$O+$^{142}$Ce~\cite{Biswas2023},
$^{18}$O+$^{28}$Si~\cite{Cardozo2018} and $^{13}$C+$^{18}$O
~\cite{Carbone2017} among others, where realistic CRC calculations
have been employed to interpret experimental MNT data. Consequently,
many systems have now been successfully described using fully
microscopic approaches for MNT channels, eliminating the need for
empirical or arbitrary scaling factors (see, for example, Refs.
~\cite{Chandra2023,Biswas2023,Cardozo2018,Cavallaro2013,Carbone2017,Ferreira2021}).
 
Despite these progresses, significant uncertainties still remain
in the quantum-mechanical modeling of MNT observables. These
uncertainties stem primarily from the choice of interaction
potentials, the complexity of nuclear structure inputs, and the
selection of states to be included in channel couplings
~\cite{Chandra2023}. Among these, the ambiguity in choosing
appropriate coupling schemes and state configurations continues
to pose a major challenge, as such choices can strongly influence
the theoretical outcomes and their agreement with experimental data.
 
This problem can be partially addressed by analyzing the energy
spectra of ejectiles or recoils~\cite{Cappuzzello2016}. However,
a more effective method is to detect characteristic $\gamma$-ray
transitions using particle-$\gamma$ coincidence techniques
~\cite{Szilner2007,Corradi2000,Peter2003, Oertzen2003}.
These techniques help identify which nuclear states are populated
during the reaction and provide important constraints for theoretical
models. It should be noted that particle-$\gamma$ coincidence
techniques are more effective in reactions involving heavy ions,
where charged-particle spectroscopy alone often fails to resolve
the quantum states of reaction products.

However, there are only a few examples in the literature where
such experimental data have been combined with microscopic CRC
calculations. One such notable case is the $^{118}$Sn+$^{206}$Pb
system ~\cite{Peter2003,Oertzen2003}, where CRC calculations
have been performed using the \textsc{fresco} code
\cite{Thompson1988,fresco}, in conjunction with particle-$\gamma$
coincidence data, offering a reliable framework for understanding
complex MNT processes.

This approach has motivated the present investigation of existing
experimental data from a system characterized by a closed proton shell
and an open neutron shell, \textit{viz.} $^{116}$Sn+$^{60}$Ni
\cite{Montanari2014,Montanari2016,Corradi2022}, which is one of the
few systems for which extensive measurements are available. These
include quasielastic (elastic + inelastic) scattering, as well as
one- and two-nucleon transfer measured around the Coulomb barrier.
Notably, particle-$\gamma$ coincidence data were also reported
for the one-neutron ($1n$) and two-neutron ($2n$) transfer probabilities
($P_{\mathrm{tr}}$) in this system~\cite{Montanari2016}, enabling a
detailed investigation of the population of specific nuclear states
and providing stronger constraints on the theoretical model. The
$P_{\mathrm{tr}}$ for $1n$- and $2n$-transfer in this system was
previously investigated using coupled-channels approaches
~\cite{Scamps2015}. However, these calculations exhibited
noticeable discrepancies compared to semiclassical results
~\cite{Montanari2014}.

The experimental work by Corradi \textit{et al.}~\cite{Corradi2022}
provided evidence for strong proton–proton correlations in the
$^{116}$Sn+$^{60}$Ni system. The transfer probabilities of one-proton
($1p$) and two-proton ($2p$) channels measured at the sub-barrier
energies were analyzed using a semiclassical (\textsc{grazing}) model.
Although the theoretical model successfully reproduced the
$1p$-transfer data, it significantly underestimated the $2p$-transfer
probability and required a large scaling factor to match the
experimental results. This discrepancy was attributed to the
absence of proton–proton pairing correlations in the theoretical
treatment, which considered only successive transfer mechanisms.

Given the availability of diverse experimental observables
\textendash \textit{viz.}, quasielastic excitation function,
characteristic $\gamma$-spectra and transfer probabilities
\textendash and the limitations of previous theoretical approaches,
a more robust analysis is warranted. Therefore, in the present work,
we carry out CRC calculations using \textsc{fresco} to describe
the excitation functions for both one- and two-nucleon transfer
channels in the $^{116}$Sn+$^{60}$Ni system
\cite{Montanari2014,Montanari2016,Corradi2022}. The measurements
were carried out in both normal and inverse kinematics. For
consistency, the gain of nucleon(s) by the lighter ($^{60}$Ni)
and the heavier ($^{116}$Sn) collision partners has been termed
as the pickup and stripping channels, respectively. Thus, in this
work, we have studied one- and two-neutron pickup and one- and
two-proton stripping channels, besides the quasielastic scattering
excitation function, in the collision between $^{116}$Sn and $^{60}$Ni.

It may be noted that the transfer of two nucleons can occur
via either direct or sequential mechanism, or as a result of
interference between these two modes \cite{scott1975}. In the
direct mechanism, two nucleons are transferred through a single
step (initial to final state without any intermediate state(s)),
whereas in sequential mechanism the nucleons are transferred,
in two steps, through the intermediate state(s). It is not possible
experimentally to distinguish direct from sequential processes.
However, these two mechanisms can be treated distinctly within
the CRC framework.

The paper is organized as follows. Sections~\ref{sect:shellmodel}
and~\ref{sect:SA_cluster_model} focus on the use of large-scale
shell-model calculations to evaluate the overlap functions between
the projectile and the target nuclei, providing essential nuclear
structure inputs. Section~\ref{sect:methodology} presents the
theoretical framework, including a detailed description of the CRC
methodology used for modeling the entrance and exit channel
partitions, along with the definitions of transfer probabilities
for both neutron and proton transfer channels. In Section
~\ref{sec:CRC calculation}, the calculated quasielastic and
transfer probabilities are compared with experimental results.
Section~\ref{sec:discussion} provides a comprehensive discussion
of the theoretical results, and finally, Section~\ref{conclusion}
summarizes the findings and outlines future scope of work.

\section{Large-scale shell-model calculations}
\label{sect:shellmodel}
In CRC calculations, the spectroscopic amplitude ($\mathcal{S}$)
is a key ingredient that represents the overlap between the initial
and final nuclear states. These amplitudes were either taken from
the literature, if available, or obtained from large-scale
shell-model calculations. To this end, we used the
\textsc{kshell} code~\cite{Shimizu2019} to perform detailed
calculations of the shell-model for the collision partners and the
residual nuclei after transfer. The resulting overlaps were then
incorporated into the CRC framework to ensure a realistic
description of the reaction dynamics. 

In the shell-model calculations for the heavier nuclei
({\it i.e.}, $^{114,115,116}$Sn, $^{117}$Sb  and $^{118}$Te), we used
two different two-body matrix elements \textendash the \texttt{SN100PN}
interaction~\cite{Brown2004, Machleidt1996} and a monopole-optimized
effective interaction~\cite{Chong2012}. In both cases, the model
space was defined using $^{100}$Sn as the inert core, with the
$1g_{7/2}$, $2d_{5/2}$, $2d_{3/2}$, $3s_{1/2}$ and $1h_{11/2}$
orbitals included as valence space for both neutrons and protons.
No truncation was applied while extracting the wave functions for
$^{114,115,116}$Sn nuclei within this model space. The theoretical
level schemes are presented in Table I, along with comparisons with
experimental spectra~\cite{nndc}, in the \textsf{Appendix}.
For the nucleus $^{118}$Te, we imposed truncations in the
shell-model calculations due to its mid-shell character, which results
in an extremely large model space and significant computational
challenges. Specifically, the number of neutron excitations was
limited to a maximum of two and the occupation number of valence
protons in the $1h_{11/2}$ orbital was restricted to a minimum of
0 and a maximum of 2. The valence particle numbers in all other
orbitals remained unconstrained. The wave functions for $^{117}$Sb
were generated using a truncation scheme similar to that used
in Ref.~\cite{Banik2020} with the \texttt{SN100PN} interaction.
For the monopole-optimized interaction \texttt{(Monopole)},
a truncation similar to the one used for $^{118}$Te was adopted.

For the lighter nuclei, we employed four different effective interactions: 
\texttt{jj44bpn}~\cite{jj44bpn}, \texttt{JUN45}~\cite{Honma2009}, 
\texttt{kb3g}~\cite{Poves2001}, and \texttt{fpd6npn}~\cite{Richter1991}. 
Among these, only \texttt{kb3g} and \texttt{fpd6npn} were used to generate 
the wave functions for $^{60}$Ni, $^{59}$Co, and $^{58}$Fe due to 
limitations of the model space. The \texttt{jj44bpn} and \texttt{JUN45}
Hamiltonians are defined within  the $fpg$ model space, assuming
$^{56}$Ni as an inert core. The corresponding valence space for both
protons and neutrons consists of the $2p_{3/2}$, $1f_{5/2}$, $2p_{1/2}$
and $1g_{9/2}$ orbitals. Shell-model calculations for $^{59}$Co ($Z=27$)
and $^{58}$Fe ($Z=26$) could not be performed within this framework,
since these nuclei require proton configurations outside the adopted
model space relative to the $^{56}$Ni core. However, for $^{60,61,62}$Ni,
calculations were carried out using all four interactions within their
respective model spaces. In contrast, the \texttt{kb3g} and
\texttt{fpd6npn} interactions were employed within the $fp$ model space,
comprising the $1f_{7/2}$, $2p_{3/2}$, $1f_{5/2}$, and $2p_{1/2}$ orbitals, 
with $^{40}$Ca taken as the inert core. This configuration allowed us to 
generate the wave functions for $^{60,61,62}$Ni, $^{59}$Co, and $^{58}$Fe.
No truncation was applied in the $fpg$ model space. However, in the $fp$ 
model space, due to computational constraints, the number of valence 
neutrons occupying the $1f_{7/2}$ orbital was restricted to a maximum of 
eight, while no restrictions were imposed on the remaining orbitals.

A reasonably good agreement between the experimental and theoretical
energy spectra was obtained for all the nuclei considered, as shown
in Tables I–IV of the \textsf{Appendix}. The spectroscopic
factors for one-particle transfer are presented in Tables V–VIII. Here
Tables V and VI correspond to the lighter nuclei, while Tables VII and
VIII correspond to the heavier nuclei.

\begin{figure*}[ht!]
    \centering
    \includegraphics[width=0.85\textwidth]{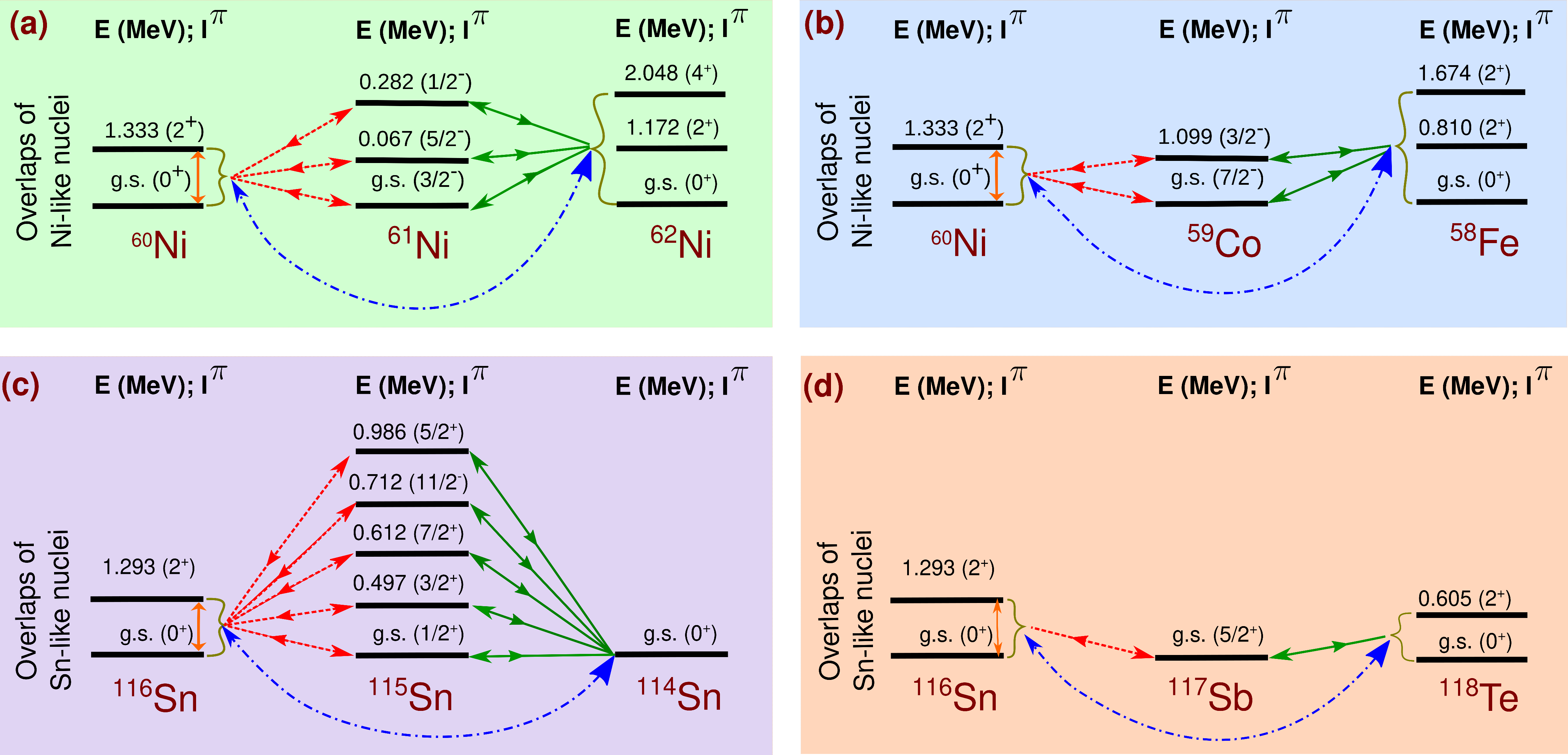}
    \caption{Overlaps of the Ni-like and the Sn-like
    nuclei considered for one- and two-nucleon transfer in CRC
    calculations. The (blue) dash-dotted lines denote direct
    transitions, while the (red) dotted lines and the (green)
    solid lines indicate sequential transfer of the first and the
    second nucleon, respectively. Panels (a) and (b) show the
    states of the lighter nuclei included in the calculations,
    respectively, for neutron pickup and proton stripping channels.
    The states of the corresponding heavier nuclei considered in
    the calculations for neutron pickup and proton stripping
    channels are depicted in panels (c) and (d), respectively.}
    \label{fig:fig01}
\end{figure*}

\section{Calculation of microscopic cluster amplitudes}
\label{sect:SA_cluster_model}
The simultaneous transfer of a pair of nucleons was studied using
the cluster model, in which motion of the nucleon pair was frozen
and segregated from the core. Further, two distinct approaches
were adopted, \textit{viz.}, (a) extreme cluster and (b) microscopic
cluster. In the extreme cluster approach, the total spin (S) of the
cluster was assumed to have 100$\%$ probability of being in the $S=0$
configuration with $1s$ internal states. The cluster $\mathcal{S}$s
were taken to be unity, as were widely accepted in the literature.
However, in the microscopic cluster approach there is no constraint
on the intrinsic spin of the cluster and the internal states. All
the configurations ($S=0$ and $S=1$) and both the internal states,
\textit{i.e.}, $1s$ and $1p$ are allowed. The definition of cluster
wave-function(s), the principal quantum number ($N$) and the orbital
angular momentum ($L$) with an internal state ($n$,$l$) were obtained
from the relation \cite{Moshinsky1959,Satchler1983}
\begin{equation}
\begin{split}
%\begin{dmath}
2(n_1-1)+l_1+2(n_2-1)+l_2 = \\ 2 (N-1)+L +2(n-1)+l.
%\end{dmath}
\end{split}
\end{equation}

\noindent
Here, $n_i$ and $l_i$ with $i=1,2$ are the principal quantum number
and the orbital angular momentum of individual nucleons, respectively. 

The $\mathcal{S}$ for the cluster wave function is computed using the
expression \cite{Carbone2017}
\begin{multline}
\label{eq:1}
\mathcal{S}_{\alpha \mathrm{J} \beta \mathrm{J}^{'}}[(nl)(NL)\Lambda S;J] = \\
\sum_{n_1l_1 n_2l_2}\sum_{j_1j_2} \hat{S}\hat{L}\hat{\textit{j}_1}\hat{\textit{j}_2}
\begin{Bmatrix} l_1 & 1/2 & j_1 \\ l_2 & 1/2 & j_2 \\ \Lambda & S & J \end{Bmatrix} \times \\
M^{\mathrm{L}}(n_1l_1n_2l_2;n l N L)
\ \ \mathcal{S}_{\alpha \mathrm{J} \beta \mathrm{J}^{'}}[n_1l_1j_1n_2l_2j_2;J]
\end{multline}

\noindent
Here, $J$ and $\Lambda$ refer to the total angular momentum and
the orbital angular momentum of the cluster with respect to the core.
The expression $M^{\mathrm{L}}(n_1l_1n_2l_2;n l N L)$ represents
the Moshinsky brackets, which were calculated by the code available
in \cite{Efros2021}. The values included in the braces represent
the $9j$ coefficients. The quantity 
$\mathcal{S}_{\alpha \mathrm{J} \beta \mathrm{J}^{'}}[n_1l_1j_1n_2l_2j_2;J]$
is the two nucleon spectroscopic amplitude in $j-j$ coupling which
was obtained from shell-model calculations. Thus the required cluster
$\mathcal{S}$s were obtained through the canonical transformation
of the spectroscopic amplitudes. $\mathcal{S}$s for the microscopic
cluster are listed in Tables IX \textendash XIII of the
\textsf{Appendix}.

\section{Theoretical Methodology}
\label{sect:methodology}
Exact finite-range CRC calculations were performed to study 
quasielastic (QEL) scattering (sum of elastic and inelastic scattering)
and the microscopic mechanisms of one- and two-nucleon transfer channels
in the $^{116}$Sn+$^{60}$Ni system using the code \textsc{fresco}.
The real and the imaginary parts of the optical potential for each
partition were obtained using the double-folding S$\tilde{a}$o Paulo
potential (SPP)~\cite{Chamon2002}, generated by the code \textsc{regina}
~\cite{Chamon2021}. Normalization factors of $N_\text{R} = 1.0$ and
$N_\text{I} = 0.6$ were applied to the real and the imaginary components
of the SPP, respectively. Similar normalization factors have been used
in previous works ~\cite{Pereira2009,Cardozo2018,Ferreira2022}. The
required reduced transition probabilities and deformation lengths were
obtained from the literature and are listed in Table~\ref{tab:table1}.
The Woods–Saxon form was assumed for the binding potential to generate
the bound-state wave functions for both the single nucleon and the nucleon
pair relative to their respective cores. In the case of one-step transfer,
the potential depth was adjusted to reproduce the experimental separation
energies for one- and two-nucleon. For two-step transfer, the depth was
varied at each step to match the corresponding experimental separation
energy. The reduced radius and diffuseness parameters were fixed at 1.30
fm and 0.7 fm, respectively~\cite{Paes2017,Biswas2023}, for all cases.
The transfer matrix elements for one-step transfer were calculated using
the prior representation of the potential, including complex remnant
terms. The two-step transfer was treated using the prior–post
representation of the transition potential to minimize the effects
of non-orthogonality~\cite{Thompson1988}. The coupling schemes adopted
for one- and two-nucleon transfer in CRC calculations are depicted
in Fig. \ref{fig:fig01}.

The transfer probabilities for neutron and proton channels were taken as

\begin{equation}
\label{equa:1nptr}
    P_{\mathrm{tr}}^{{\mathrm{1n,2n}}} = \frac{\left( \frac{d\sigma}{d\Omega} \right)^{\mathrm{1n,2n}}}{\left( \frac{d\sigma}{d\Omega} \right)^{\mathrm{1n,2n}} + \left( \frac{d\sigma}{d\Omega} \right)^{\mathrm{qel}}}
\end{equation}

and

\begin{equation}
\label{equa:1pptr}
    P_{\mathrm{tr}}^{{\mathrm{1p,2p}}} = \frac{\left( \frac{d\sigma}{d\Omega} \right)^{\mathrm{1p,2p}}}{\left( \frac{d\sigma}{d\Omega} \right)^{\mathrm{1p,2p}} + \left( \frac{d\sigma}{d\Omega} \right)^{\mathrm{el}}}
\end{equation}

\noindent
following the norms used for the experimental data
\cite{Montanari2014,Montanari2016,Corradi2022}.

The symbols $\left(\frac{d\sigma}{d\Omega} \right)^{\text{1n,2n,1p,2p}}$
and $\left(\frac{d\sigma}{d\Omega} \right)^{\text{el,qel}}$ represent
the differential cross sections for one- and two-nucleon (neutron and
proton) transfer channels and for elastic and quasielastic scattering,
respectively. The transfer probability for a Coulomb trajectory can be
expressed as a function of the distance of the closest approach, denoted
by $D$. This distance is related to the center-of-mass (c.m.) scattering
angle ($\theta_{\mathrm{c.m.}}$) and energy ($E_{\mathrm{c.m.}}$),
through the expression

\begin{equation}
    D=\frac{1.44 Z_{\textrm{p}}Z_{\textrm{t}}}{2{E_{\textrm{c.m.}}}}
    \left[1+\textrm{cosec}\left(\frac{\theta_{\textrm{c.m.}}}{2}\right)\right] \ .
\end{equation}

\noindent
Here, $Z_\mathrm{p}$ and $Z_\mathrm{t}$ are,respectively,
the atomic numbers of the projectile and the target.

\begin{table}[]
\centering
\caption{Reduced electromagnetic matrix elements
($\bra{I_{i}}ME(\lambda)\ket{I_f}$) and reduced deformation lengths
($\delta_{N}$) for the collision partners, used in CRC calculations.}
\label{tab:table1}
\begin{tabular}{llllll}
\hline
Nucleus & I$^\pi_{\mathrm{i}}\rightarrow$I$^\pi_{\mathrm{f}}$ & $\Delta \mathrm{E}$ & $\bra{\mathrm{I^\pi_{i}}}\mathrm{ME}(\lambda)\ket{\mathrm{I^\pi_f}}$ & $\delta_{N}$ & Ref. \\
 & & (MeV) & e$^{2}$fm$^\lambda$ & (fm) & \\
\hline
 & 0$^{+}\rightarrow 2^{+}$ & 1.293 & 45.49 &0.651 & \cite{Kundu2019} \\
$^{116}$Sn & 2$^{+}\rightarrow 4^{+}$ & 1.097  &60.798 & 0.870 & \cite{Petrache2019} \\
 & 0$^{+}\rightarrow 3^{-}$ & 2.266 & 308.220 & 0.542& \cite{Kundu2019} \\\hline 
 & 0$^{+}\rightarrow 2^{+}$ & 1.332 &30.1 &0.96 & \cite{Kenn2001} \\
 $^{60}$Ni & 2$^{+}\rightarrow 4^{+}$ & 1.173 &27.448 & 0.874 & \cite{Kenn2001} \\
 & 0$^{+}\rightarrow 2^{+}$ & 2.11 & 2.664 & 0.084 & \cite{Lesser1970} \\  
 \\ \hline
\end{tabular}
\end{table}

\section{Results of CRC calculations}
\label{sec:CRC calculation}
\subsection{Quasielastic scattering}
\label{sec:qes}
\begin{figure}[h]
%   \centering
    \includegraphics[width=0.9\linewidth]{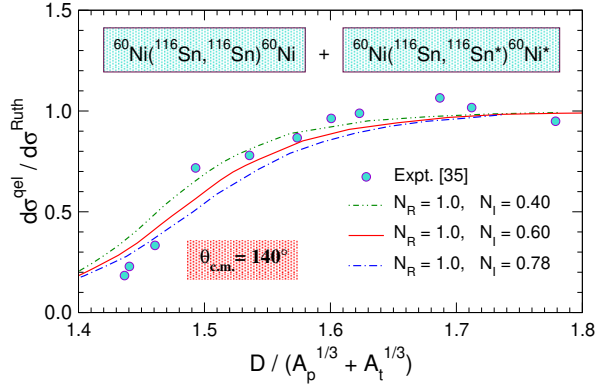}
    \caption{Quasielastic scattering excitation function for the
    system $^{116}$Sn+$^{60}$Ni. The experimental data points are
    denoted by filled circles whereas the lines stand for results of
    CRC calculations with different strengths for the imaginary
    component of the S$\tilde{a}$o Paulo potential.}
    \label{fig:fig02}
\end{figure}

We investigated the quasielastic (elastic + inelastic) scattering
excitation function of the $^{116}$Sn+$^{60}$Ni system
~\cite{Montanari2014,Montanari2016} by incorporating coupling
of low-lying states in the CRC calculation. The coupling scheme
was based on $\gamma$-ray transitions observed in the experiment,
performed in normal kinematics, using the AGATA array~\cite{Montanari2016}.
The $\gamma$-transitions, detected in the backward direction,
corresponded to $2^{+}$ (1293 keV), $4^{+}$ (2390 keV) and $3^{-}$
(2265 keV) states of $^{116}$Sn. For $^{60}$Ni, the identified
transitions included $2^{+}$ (1333 keV), $2^{+}$ (2159 keV) and
$4^{+}$ (2506 keV) states. The measured quasielastic excitation
function is compared with results of CRC calculations in
Fig.~\ref{fig:fig02}. Variations in calculated cross sections
as a function of $N_\text{I}$ is also shown in the figure. The
theoretical results reproduce the experimental cross sections
quite satisfactorily across the entire energy range. The ratio
of the $2^+$ state yield of $^{60}$Ni to other observed inelastic
channel yields at a laboratory energy ($E_{\mathrm{lab}}$) of 245
MeV (for $^{60}$Ni) and a laboratory angle ($\theta_\text{lab}$)
of $70^{\circ}$ was provided in Table 1 of Ref.~\cite{Montanari2016}.
We estimated these normalized yields from the present calculations.
This comparison (shown in Table~\ref{tab:table2}) serves as a test
of the reliability of the adopted optical model potential and the
transition matrix elements used in the CRC analysis. Our CRC results
show excellent agreement with the experimental data in most cases.
This provides evidence that the potential parameters and structural
inputs incorporated in the CRC calculations are reliable. 
\begin{table*}[ht]
\caption{Comparison between experimental yields (at
$E_\text{lab} = 245$ MeV (for $^{60}$Ni) and
$\theta_\text{lab} = 70^\circ$) and theoretical cross sections,
obtained using different interactions, for inelastic and transfer
channels. Results were normalized to the strength of the $2^+$
state in $^{60}$Ni.}
\label{tab:table2}
\begin{tabular}{|l|l|ccc|cc|}
\hline
\multirow{2}{*}{}                               & \multirow{6}{*}{$\mathcal{S}$ for KSHELL} & \multicolumn{3}{c|}{Quasielastic}                                                             & \multicolumn{2}{c|}{$1n$- transfer}                    \\ \cline{3-7} 
                                                &                                           & \multicolumn{2}{c|}{$^{116}$Sn}                                             & $^{60}$Ni       & \multicolumn{1}{c|}{$^{115}$Sn}      & $^{61}$Ni       \\ \cline{3-7} 
State I$^{\pi}$                                 &                                           & \multicolumn{1}{c|}{2$^+$}           & \multicolumn{1}{c|}{4$^+$}           & 4$^+$           & \multicolumn{1}{c|}{5/2$^{+}$}       & 1/2$^{-}$       \\
Experiment                                      &                                           & \multicolumn{1}{c|}{0.792$\pm0.160$} & \multicolumn{1}{c|}{0.042$\pm0.011$} & 0.060$\pm0.013$ & \multicolumn{1}{c|}{0.018$\pm0.003$} & 0.014$\pm0.003$ \\
Ref.~\cite{Montanari2016} &                                           & \multicolumn{1}{c|}{0.720}           & \multicolumn{1}{c|}{0.056}           & 0.11            & \multicolumn{1}{c|}{0.037}           & 0.033           \\
CRC$^a$                                         &                                           & \multicolumn{1}{c|}{0.802}           & \multicolumn{1}{c|}{0.053}           & 0.057           & \multicolumn{1}{c|}{}                &                 \\ \hline \hline
                                                & \texttt{SN100pn ;JUN45}      & \multicolumn{1}{c|}{0.809}           & \multicolumn{1}{c|}{0.041}           & 0.031           & \multicolumn{1}{c|}{0.034}           & 0.014           \\
                                                & \texttt{SN100pn ;fpd6npn}  & \multicolumn{1}{c|}{0.815}           & \multicolumn{1}{c|}{0.041}           & 0.031           & \multicolumn{1}{c|}{0.027}           & 0.012           \\
                                                & \texttt{SN100pn ;jj44bpn}    & \multicolumn{1}{c|}{0.811}           & \multicolumn{1}{c|}{0.041}           & 0.031           & \multicolumn{1}{c|}{0.032}           & 0.006           \\
CRC$^b$                                         & \texttt{SN100pn ;kb3g} & \multicolumn{1}{c|}{0.816}           & \multicolumn{1}{c|}{0.041}           & 0.031           & \multicolumn{1}{c|}{0.031}           & 0.016           \\
                                                & \texttt{Monopole ;JUN45}          & \multicolumn{1}{c|}{0.804}           & \multicolumn{1}{c|}{0.041}           & 0.031           & \multicolumn{1}{c|}{0.048}           & 0.02            \\
                                                & \texttt{Monopole ;fpd6npn}       & \multicolumn{1}{c|}{0.813}           & \multicolumn{1}{c|}{0.041}           & 0.031           & \multicolumn{1}{c|}{0.039}           & 0.017           \\
                                                & \texttt{Monopole ;jj44bpn}        & \multicolumn{1}{c|}{0.806}           & \multicolumn{1}{c|}{0.041}           & 0.031           & \multicolumn{1}{c|}{0.046}           & 0.008           \\
                                                & \texttt{Monopole ;kb3g}           & \multicolumn{1}{c|}{0.813}           & \multicolumn{1}{c|}{0.041}           & 0.031           & \multicolumn{1}{c|}{0.044}           & 0.022           \\ \hline
\end{tabular}
 \begin{footnotesize}
\\\textsuperscript{a} In the CRC calculations, coupling of only elastic and inelastic channels was considered.\\ \textsuperscript{b} In the CRC calculations, coupling of elastic, inelastic, and $1n$-pickup channels was considered.
\end{footnotesize}
\end{table*}

\subsection{One-neutron pickup}
\label{sect:1n}
In the calculations for $^{116}$Sn($^{60}$Ni,$^{61}$Ni)$^{115}$Sn,
the observed inelastic states of both projectile-like and target-like
nuclei, as reported in Ref.~\cite{Montanari2016}, were included in
the exit partition. In the entrance partition, only the first excited
states of projectile-like and target-like nuclei were considered
(see Fig.~\ref{fig:fig01}). Additionally, we tested the effect of
including all experimentally observed $\gamma$-transitions in the
CRC calculations for the $1n$-pickup channel.

\begin{figure*}[ht!]
%   \centering
    \includegraphics[width=0.75\linewidth]{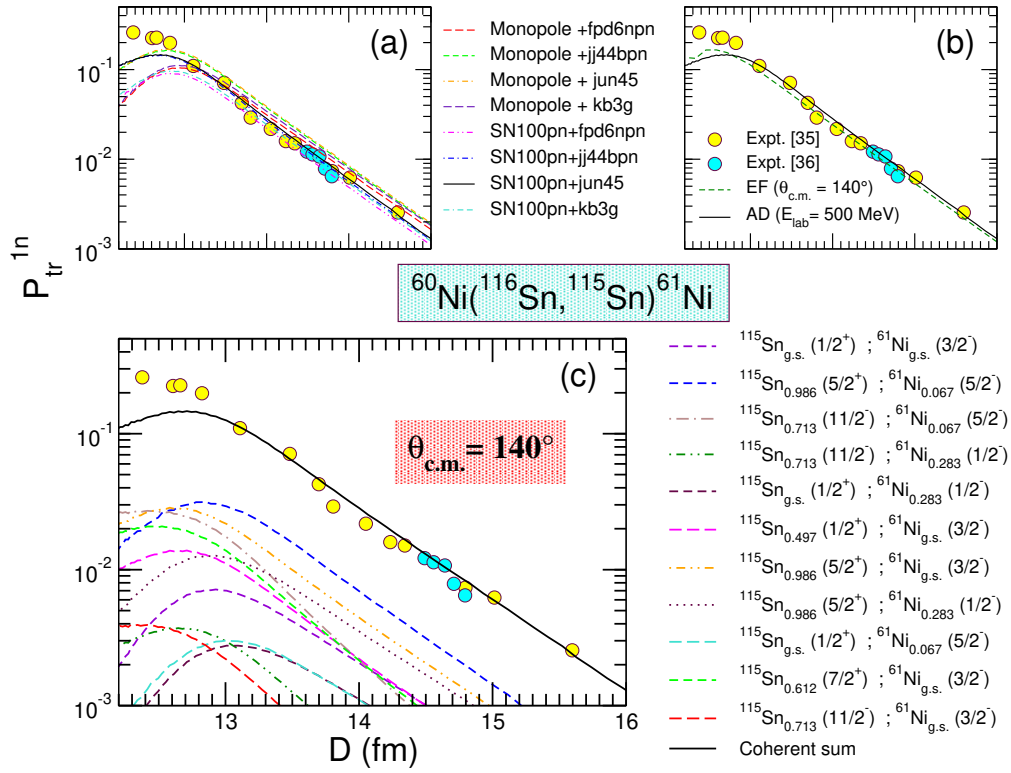}
    \caption{Comparison between the experimental (filled circles)
    and the theoretical transfer probabilities, obtained using
    different interactions in the shell model calculations, of
    the $1n$-pickup channel for the system $^{116}$Sn+$^{60}$Ni.
    (a) Results of CRC calculations using the angular distribution
    method at $E_\text{lab} = 500$ MeV (for $^{116}$Sn).
    (b) Comparison of results obtained from calculations of angular
    distributions and excitation functions
    (at $\theta_\text{c.m.} = 140^\circ$). (c) Contributions from
    specific states of the projectile and the target to $1n$-pickup
    cross sections. See text for details.}
    \label{fig:fig03}
\end{figure*}

We obtained the transfer probabilities using two methods: (i) analysis
of the angular distribution (AD) and (ii) extraction of the excitation
function (EF) at the mean angle $\theta_\text{c.m.} = 140^{\circ}$,
as measured using the heavy-ion magnetic spectrometer PRISMA
~\cite{Montanari2014}. The results from the angular distribution
calculations using different structural inputs, namely, the
$\mathcal{S}$s obtained from large-scale shell-model calculations
are shown in Figure~\ref{fig:fig03}(a). The theoretical curves
indicate that while various shell-model interactions lead to slight
differences in the magnitude of the transfer probabilities, the
overall shape of the distribution remained consistent. Among these
results, the $\mathcal{S}$s derived from \texttt{SN100PN},
\texttt{JUN45} and \texttt{jj44bpn} interactions provided reasonable
reproductions of the experimental data.

In the second approach, we extracted the $1n$-transfer excitation
function by varying $E_\text{c.m.}$ in the CRC calculations while
keeping the angle fixed at $\theta_\text{c.m.} = 140^{\circ}$.
Results from both methods are compared in Fig.~\ref{fig:fig03}(b).
The two theoretical curves exhibit similar shapes and magnitudes,
indicating consistency between the two methods.

It is to be noted that all the theoretical curves tend to underpredict
$P_\text{tr}$ at smaller values of $D$, which correspond to higher
values of $E_\text{c.m.}$. This behavior can be attributed to the
reaction dynamics: at larger distances, the interacting nuclei
follow a near-Rutherford trajectory, whereas at shorter distances,
the influence of the nuclear potential alters the trajectory.
This introduces uncertainties in defining the transfer probability
and may lead to the underestimation of the theoretical results
in this region, as also discussed in Ref.~\cite{Scamps2015}.

Furthermore, the individual contributions of each transition,
obtained from the microscopic CRC calculations, are presented
in Figure~\ref{fig:fig03}(c). It is evident from the figure that
the $5/2^+_{0.986}$ state of $^{115}$Sn and the $5/2^-_{0.067}$
state of $^{61}$Ni contribute most significantly to $P_\text{tr}$.
The second major contribution is from the $5/2^+_{0.986}$ state
of $^{115}$Sn and the g.s. of $^{61}$Ni. In Fig.~\ref{fig:fig03}(c),
the solid line, illustrates the coherent sum of contributions
from the different states of the collision partners involved
in the CRC calculations. Measured data and results of the CRC
calculations exhibit a significant level of agreement. Notably,
the use of a normalization constant \textendash often referred
to as the \textit{unhappiness factor} in conventional Distorted-Wave
Born Approximation (DWBA) finite-range calculations \textendash
is not required to reproduce the data. The theoretical results
for the normalized yield of $\gamma$-spectra in specific transfer
channels were compared with the reported experimental values
~\cite{Montanari2016} in Table~\ref{tab:table2}. These theoretical
predictions match well with the data, lending further credibility
to our microscopic approach.

In contrast, the semiclassical calculations presented in Ref.
~\cite{Montanari2014} employed a wide energy range for the exit
channels. Another study~\cite{Scamps2015} optimized calculations
by fitting the transfer probability of a single-neutron transfer,
focusing primarily on the low-lying states in the exit channels.
However, the theoretical curve shown in Ref.~\cite{Scamps2015}
did not provide sufficient insight into the specific role of
individual low-lying states, as it was derived from a best-fit
approach. Moreover, a discrepancy was evident in Ref.
~\cite{Montanari2014}, where the transfer probability was overestimated
by approximately 30\% \textendash a result attributed to the inclusion
of unobserved high-lying states in the exit channels, as pointed out
in Ref.~\cite{Scamps2015}. If this interpretation holds, it suggests
that the theoretical model used in those studies may not adequately
capture the impact of including such higher-lying states. On the other
hand, our CRC results indicate that the contribution of high-lying
states in $1n$-pickup channel is likely minimal, thereby allowing
for an accurate and consistent description of the experimental
data without overestimation.

\begin{figure*}[ht]
%   \centering
    \includegraphics[width=0.70\linewidth]{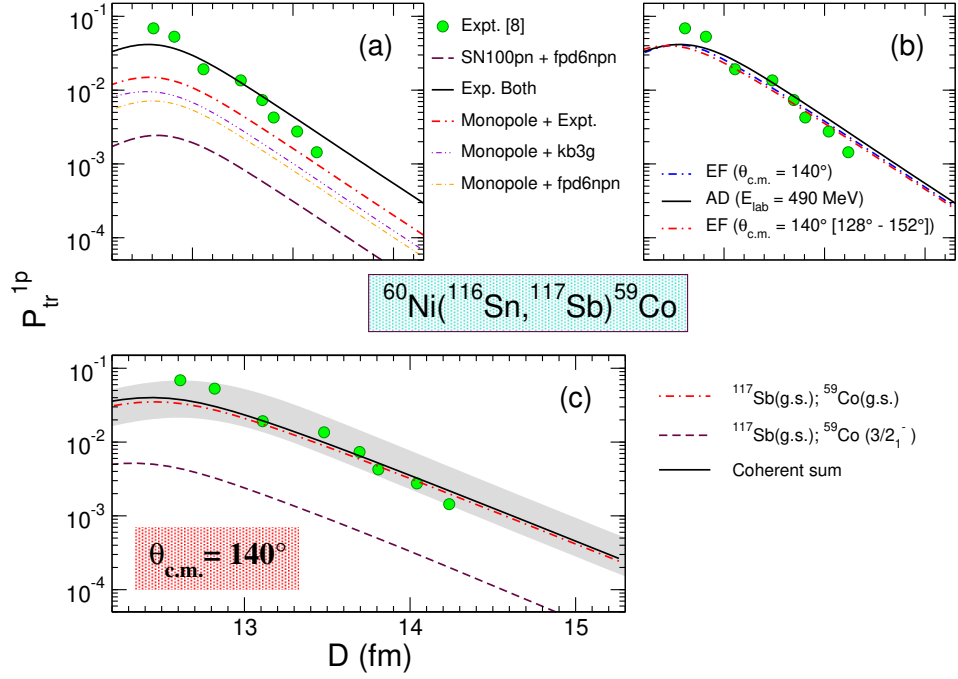}
    \caption{Comparison between the experimental (filled circles)
    and the theoretical transfer probabilities, obtained using
    different interactions in the shell model calculations, of
    the $1p$-stripping channel for the system $^{116}$Sn+$^{60}$Ni.
    (a) Results of CRC calculations using the angular distribution
    method at $E_\text{lab} = 490$ MeV (for $^{116}$Sn).
    (b) Comparison of results obtained from calculations of angular
    distributions and excitation functions
    (at $\theta_\text{c.m.} = 140^\circ$). (c) Contributions from
    specific states of the projectile and the target to $1p$-stripping
    cross sections. See text for details.}
    \label{fig:fig04}
\end{figure*}

\subsection{One-proton stripping}
\label{sect:1p}
Results of CRC calculations for the reaction
$^{116}$Sn($^{60}$Ni,$^{59}$Co)$^{117}$Sb are compared with the
experimental data in Fig.~\ref{fig:fig04}. Initially, spectroscopic
amplitudes derived from shell-model calculations were used in the
calculations. For the heavier nuclei, the \texttt{SN100PN} and
the monopole-optimized effective interactions were used, while
for the lighter nuclei, the \texttt{fpd6npn} and the \texttt{kb3g}
interactions were adopted. All combinations of these theoretical
inputs led to an underprediction of the experimental cross sections
(see Fig. \ref{fig:fig04}(a)). This discrepancy may be attributed
to significant deviations of the calculated $\mathcal{S}$ values
from the experimental ones (see Ref. \cite{Chandrafusion23}).
To address this, experimentally extracted $\mathcal{S}$s from the
literature~\cite{Blair1966,Ishimatsu1967} were incorporated into
the CRC calculations, which yielded a substantially improved
agreement with the data.

We employed three different approaches to deduce the theoretical
excitation function for the $1p$-stripping channel. The first two
methods are analogous to those used for the $1n$-pickup channel.
In the third approach, the excitation function was extracted by
integrating over the acceptance angle of the PRISMA, using a mean
c.m. angle of $\theta_\text{c.m.} = 140^{\circ}$. The results
obtained from all three approaches are compared in
Fig.~\ref{fig:fig04}(b), demonstrating consistency across the
methods. Similar to the $1n$-pickup channel, the CRC calculations
begin to underestimate the experimental excitation function at
smaller distances ({\it i.e.}, at higher energies). This behavior
can be understood in the same context as the $1n$-pickup case,
where deviations from Rutherford trajectories at short distances
reduce reliability of the definition of $P_\text{tr}$.

Furthermore, the individual contributions from various transitions
are depicted in Fig.~\ref{fig:fig04}(c). It is evident from the figure
that the g.s. to g.s. transition contributes the most to the measured
cross sections. The role of excitations in both collision partners
was also examined. Inclusion of the first ${\frac{3}{2}}^-$ excited
state of $^{59}$Co resulted in a marginal enhancement of the coherent
sum, whereas inclusion of the first excited state of $^{117}$Sb led
to an overestimation of the experimental cross sections
(not shown in the figure).

It is also worth noting that the previous study~\cite{Corradi2022}
on the $1p$-stripping reaction, based on Total Kinetic Energy Loss
(TKEL) spectra, reported that the peak position for the $1p$-stripping
channel had been observed near 5.13~MeV. As the beam energy increased,
the peak of the TKEL spectra did not shift toward higher energy,
indicating that higher excited states were not significantly populated.
Instead, the g.s. to g.s. transition remained dominant. This observation
supports our CRC results, as illustrated in Fig.~\ref{fig:fig04}(c),
where it is evident that primary contribution to the cross sections
arises from the g.s. to g.s. transition, while contributions from
higher excited states are comparatively small.

To examine the influence of uncertainties in the $\mathcal{S}$s
on calculated cross sections, the experimental $\mathcal{S}$s
used in the CRC calculations were varied by $\pm15\%$. This variation
successfully encompassed all experimental data points, as indicated
by the (grey) shaded band in Fig.~\ref{fig:fig04}(c).
Furthermore, it should be noted here that $1p$-stripping transfer
probabilities calculated using $\mathcal{S}$s obtained from the 
shell-model calculations do not lie within the shaded band.

\begin{figure}[ht]
     \centering
    \includegraphics[width=0.85\linewidth]{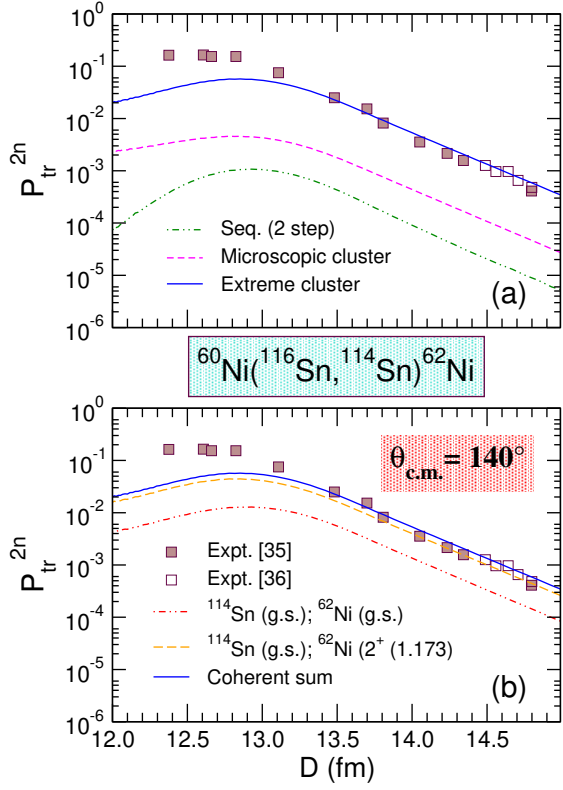}
    \caption{Comparison between the experimental (filled squares)
    and the theoretical transfer probabilities of the $2n$-pickup
    channel for the system $^{116}$Sn+$^{60}$Ni. (a) Results of
    CRC calculations using the angular distribution method at
    $E_\text{lab} = 500$ MeV (for $^{116}$Sn), showing variations
    among sequential, microscopic cluster and extreme cluster
    approaches. (b) Contributions from specific states of the
    projectile and the target in the extreme cluster model for
    $2n$-pickup cross sections. See text for details.}
    \label{fig:fig05}
\end{figure}

\subsection{Two-neutron pickup}
\label{sect:2n}
The reaction $^{116}$Sn($^{60}$Ni,$^{62}$Ni)$^{114}$Sn was studied
using three different approaches. For sequential transfer of two
neutrons the states in the intermediate partition were same as the
ones considered in the exit channel for the $1n$-pickup. In the final
partition, we included the states which were used for fitting the
TKEL spectra shown in Fig. 5 of Ref.~\cite{Montanari2016}. The
required spectroscopic amplitudes were taken from \textsc{kshell}
results. The results of sequential $2n$-pickup are shown in
Fig.~\ref{fig:fig05}(a) by the (green) dash-dotted line. Although
the theoretical curve underpredicts the experimental data by
approximately two orders of magnitude, the overall shape of
$P_\text{tr}$ is reasonably well reproduced. Despite the
underestimation, the analysis provides valuable insight into the
dominant states contributing to the $2n$-pickup process.

The underestimation of $P_\text{tr}$ by CRC calculations for
sequential transfer suggests that pairing between the two neutrons
might be playing a more significant role in enhancing the cross
sections. We next considered the microscopic cluster model for
transfer of a pair of neutrons. The required spectroscopic
amplitudes for projectile and target overlaps were calculated
by the methodology described in Section \ref{sect:SA_cluster_model}
and presented in Tables IX, XII and XIII of the
\textsf{Appendix}. Results of microscopic cluster
calculations are displayed in Fig.~\ref{fig:fig05}(a) which
clearly underpredict the data.

According to previous studies~\cite{Biswas2023,Chandra2023}, the
microscopic cluster model is highly sensitive to the value of
reduced radius ($r_0$) in generating bound-state wave functions.
To investigate this sensitivity, we varied the reduced radius
in our microscopic cluster model calculations from $r_0 = 1.3$~fm
to $r_0 = 1.4$~fm. Despite this adjustment, the calculated cross
sections continued to underpredict the data by nearly a factor
of two (not shown in the figure).

In Ref.~\cite{Montanari2014}, same experimental data were
reproduced using the sequential transfer mechanism using
the \texttt{CWKB} method. The discrepancy between this result and
the present one can be explained by the fact that we included only
those intermediate states which were observed in the normal kinematics
experiment~\cite{Montanari2016}. On the other hand, the previous
semiclassical calculations incorporated numerous intermediate
states with a wide distribution of excitation energies. This
explanation was also provided in Ref.~\cite{Scamps2015}, where the
authors parameterized the interaction potential between the nuclei
to fit the experimental $P_\text{tr}$. The theoretical curve presented
in Ref.~\cite{Montanari2016} overpredicted the $1n$-transfer probability
due to limitations in handling higher-lying states that served as
intermediate configurations in the theoretical model. This limitation
may have also affected the prediction of $2n$-pickup via the
sequential mechanism.

Finally, we applied the extreme cluster model, in which we
presupposed that the two neutrons were in their pure states and
that the intrinsic spin of the $2n$-cluster were in anti-parallel
($S=0$) mode. CRC results for the extreme cluster model are shown
in Fig. \ref{fig:fig05}(a). Here, we exclusively used the states
$^{62}$Ni(0$^{+}_{\mathrm{g.s.}}$), $^{62}$Ni(2$^{+}_{1.173}$)
and $^{114}$Sn(0$^{+}_{\mathrm{g.s.}}$) in the final partition.
We did not include the indirect path in the calculations. CRC
results for the extreme cluster model, depicted as a solid red
line in \ref{fig:fig05}(a), provide a better reproduction of
$P_\text{tr}$ for $2n$-pickup. It is evident from the results
that the first excited state of $^{62}$Ni(2$^{+}_{1.173}$) has
a larger contribution to the cross sections, compared to the
g.s. populations, as shown in Fig. \ref{fig:fig05}(b).

\begin{figure*}[ht!]
    \centering
    \includegraphics[width=0.7\linewidth]{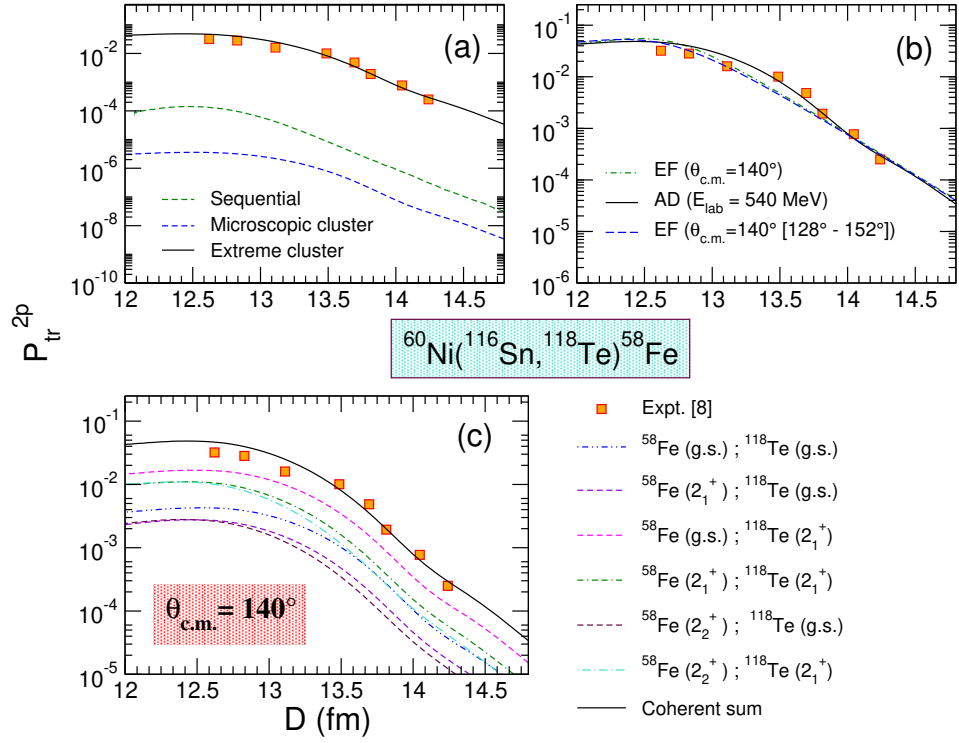}
    \caption{Comparison between the experimental (filled squares)
    and the theoretical transfer probabilities of the $2p$-stripping
    channel for the system $^{116}$Sn+$^{60}$Ni. (a) Results of CRC
    calculations using the angular distribution method at
    $E_\text{lab} = 540$ MeV (for $^{116}$Sn), showing variations
    among sequential, microscopic cluster and extreme cluster
    approaches. (b) Comparison of results obtained from calculations
    of angular distributions and excitation functions
    (at $\theta_\text{c.m.} = 140^\circ$) for the extreme cluster
    model. (c) Contributions from specific states of the projectile
    and the target in the extreme cluster model to $2p$-stripping
    cross sections. See text for details.}
    \label{fig:fig06}
\end{figure*}

\subsection{Two-proton stripping} 
\label{sec:2p}
Similar to the case of $2n$-pickup, we investigated the reaction
$^{60}$Ni($^{116}$Sn, $^{118}$Te)$^{58}$Fe by three different mechanisms.
Firstly, we performed sequential transfer calculations in which the
states in the intermediate partition were the same as those used for
the $1p$-stripping. The resulting $P_\text{tr}$ underpredicted the
data by nearly three orders of magnitude, as shown in
Fig.~\ref{fig:fig06}(a). This significant discrepancy indicates that
the two-step mechanism is not the dominant mode for the $2p$-stripping
process.

Subsequently, we carried out calculations using the microscopic
cluster model, with $\mathcal{S}$s which are presented in Tables
X and XI of the \textsf{Appendix}. The theoretical curve
obtained from the microscopic cluster model also underpredicted
the data (see Fig.~\ref{fig:fig06}(a)), falling by more than an
order of magnitude below the predictions of the sequential mechanism.
Here, we tested an alternative definition of the binding potential,
using $R = r_0 (A_\mathrm{c} + 2)^{1/3}$, where $A_\mathrm{c}$
is the mass number of the core nucleus with the $2p$-cluster,
following the approach in Refs.~\cite{Biswas2023,Chandra2023}.
This  modification led to a slight increase in the magnitude of
$P_\text{tr}$ (not shown in the figure). However, the overall
shape remained unsatisfactory.

Finally, we carried out the extreme cluster model calculations by
three different procedures, considering both excitation function and
angular distribution, similar to the case of $1p$-stripping. These
results, which are quite comparable, are displayed in
Fig.~\ref{fig:fig06}(b). However, the shape of $P_\text{tr}$ appear
to be best reproduced by the theoretical angular distribution.
We stress here that the indirect paths were not included in the
CRC calculations shown in the figure. Inclusion of these
transitions led to overprediction of the data at higher energies.

Individual contributions from various transitions are shown in
Fig.~\ref{fig:fig06}(c). The results indicate that population of
the $^{118}$Te(2$^+_{1}$) state in the exit channel contributes
significantly, as highlighted by the (magenta) dashed line in
Fig.~\ref{fig:fig06}(c). In Ref.~\cite{Corradi2022}, TKEL spectra
were obtained with a resolution of approximately 2~MeV. However,
$\gamma$-transitions were not measured in this experiment, unlike
the case of $2n$-pickup. A broadening of the TKEL spectra was
observed for higher TKEL, which may be attributed to transitions
involving higher-lying states. Additionally, contributions from 
deep-inelastic scattering could also play a role in this region
of the spectrum. Nevertheless, our theoretical CRC calculations
reproduce both the shape and magnitude of $P_\text{tr}$ for
$2p$-stripping quite well. This agreement suggests that the
observed broadening in the TKEL spectra might primarily arise
from inelastic transitions within the $2p$-stripping mechanism.
\begin{figure*}[ht!]
    \includegraphics[width=0.95\textwidth]{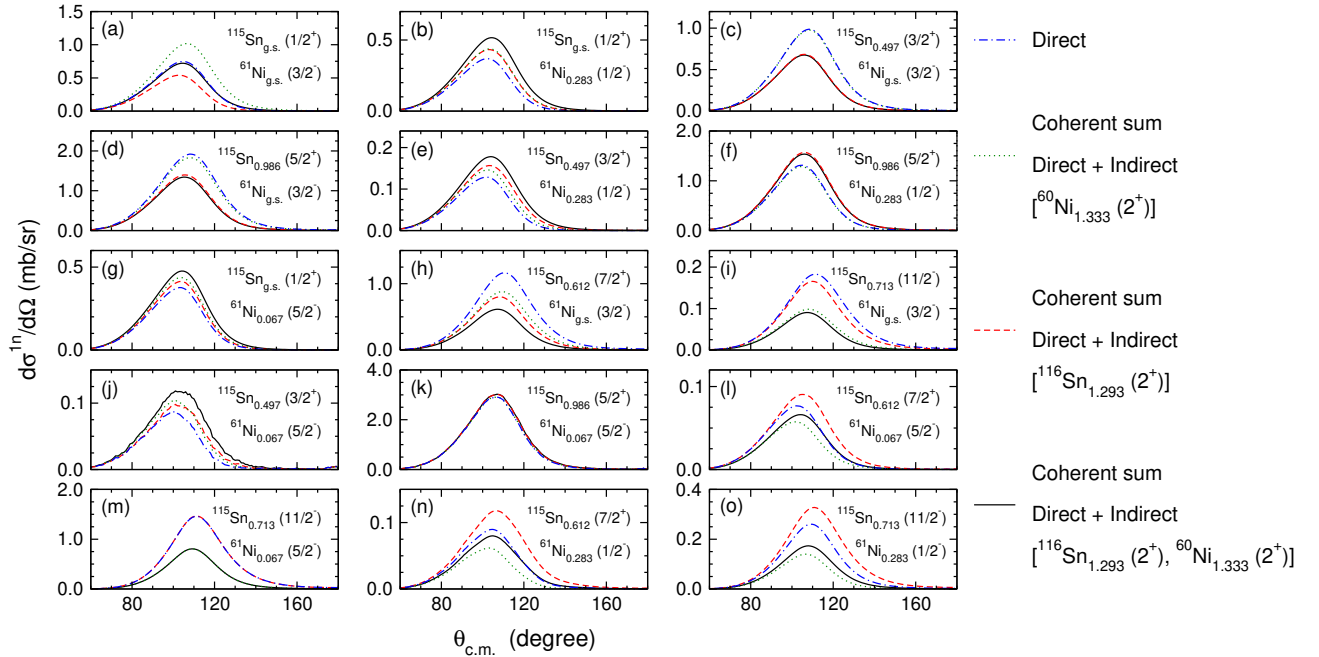}
    \caption{Theoretical $1n$-pickup angular distributions for
    the system $^{116}$Sn+$^{60}$Ni at $E_\text{lab} = 500$ MeV
    (for $^{116}$Sn). The $\mathcal{S}$s used in CRC calculations
    were genertaed using \texttt{SN100PN} and \texttt{jun45}
    interactions, respectively, for the heavier and the lighter
    nuclei. See text for details.}
    \label{fig:fig07}
\end{figure*}

\section{Discussion}
\label{sec:discussion}
The selection of states in the entrance channel, based on the observed
$\gamma$-ray transitions was not straightforward. This is because only
a subset of the possible channels might have actually participated
in the reaction via indirect pathways, such as inelastic excitations,
rather than all the channels observed experimentally. If only certain
states contribute to the reaction mechanism, it becomes possible to
observe interference patterns \textendash either constructive or
destructive \textendash depending on the inclusion of specific inelastic
transitions in the entrance channel. Analysis of these patterns can
provide valuable insights into the reaction dynamics. It was
demonstrated in Ref. ~\cite{Chandra2023} that presence of an
indirect route could significantly influence the direct route
in the reaction $^{94}$Zr({$^{28}$Si,$^{29}$Si})$^{93}$Zr at energies
above the Coulomb barrier. However, Kumar \textit{et al.}~\cite{Chandra2023}
did not explicitly  identify the entrance-channel states that most
significantly influence the interference pattern. In the present work,
we investigate in greater detail the role of inelastic transitions
in order to determine which indirect pathways \textendash arising
from either projectile or target excitations \textendash contribute
to the exit channels. These effects are examined by introducing the
corresponding couplings individually within the CRC calculations.

The CRC calculations reproduced the experimental $1n$-pickup excitation
function quite well, as shown in Fig.~\ref{fig:fig03}(c). The corresponding
differential angular distributions for specific states in the exit channel
are presented in Fig.~\ref{fig:fig07}. Here two routes for neutron transfer
are illustrated: (i) the direct route (denoted by (blue) dash-dotted line)
in which the constituents of the entrance channel is in the ground state
and (ii) the indirect route in which one or both constituents of the
entrance channel is/are in the first excited state (see Fig. \ref{fig:fig01})
prior to neutron transfer. The coherent sum of the angular distributions
were determined by the interference between the direct and the indirect paths.
Fig.~\ref{fig:fig07}(a) illustrates the angular distribution for the
population of both constituents of the exit channel in their ground states.
The results indicate that the inelastic excitation of $^{60}$Ni(2$^+$)
interferes constructively with the direct transition path. In contrast,
the contribution from the $^{116}$Sn(2$^+$) state exhibits destructive
interference of comparable magnitude. As a result, the opposing effects
largely cancel each other, yielding no significant impact from the
indirect transitions on the population of the ground states in the
exit channel.

Furthermore, when the ground state of the residual nucleus
$^{61}$Ni$_\text{g.s.}$(3/2$^-$) is populated, the indirect transitions
via $^{60}$Ni(2$^+$) and $^{116}$Sn(2$^+$) collectively exhibit a
destructive nature. However, for transitions populating excited states
of $^{61}$Ni, the interference between $^{60}$Ni(2$^+$) and the residual
states becomes constructive. This behavior highlights the interplay
of entrance channel couplings in shaping the transfer dynamics.

\begin{figure}[ht!]
%    \centering
    \includegraphics[width=1.0\linewidth]{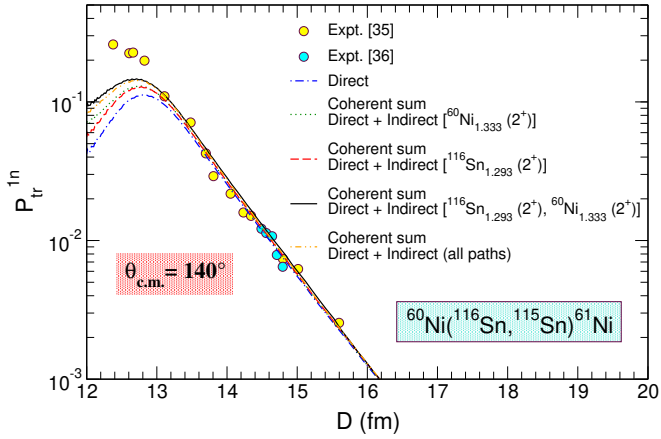}
    \caption{The effect of coupling to different indirect paths on the
    calculated $1n$-transfer probability for the system 
    $^{116}$Sn+$^{60}$Ni. See text for details.}
    \label{fig:fig08}
\end{figure}
The CRC results, illustrating the influence of indirect transition
routes are presented in Fig.~\ref{fig:fig08}, highlighting the
significant role of inelastic transitions in the $1n$-pickup channel,
particularly at higher energies. This behavior can be attributed to
the fact that, at shorter internuclear distances, the spatial
overlap between the colliding nuclei increases substantially.
Consequently, multi-step processes \textendash including inelastic
excitations preceding the transfer \textendash become more prominent.
This dominance of indirect pathways at lower distances is clearly
reflected in the results shown in Fig.~\ref{fig:fig08}.

One may also argue that the underprediction observed in the
theoretical results could stem from the omission of certain
inelastic transitions observed in the entrance channel, as not
all experimentally identified $\gamma$-transitions were included
in the initial CRC calculations. These missing inelastic couplings
might account for the deficit in the $1n$-pickup cross section at
smaller distances. To investigate this possibility, we expanded
the number of indirect transition pathways (which are not shown
in Fig. \ref{fig:fig01}) in the CRC calculations by including
additional states for which $\gamma$-transitions were identified
~\cite{Montanari2016}. However, the results obtained from this
extended CRC analysis indicate that the inclusion of these
additional multi-step processes does not significantly modify
the theoretical transfer probability, as illustrated by the
(orange) dash-double-dotted line in Fig.~\ref{fig:fig08}.

\begin{figure*}[ht!]
    \centering
    \includegraphics[width=0.75\textwidth]{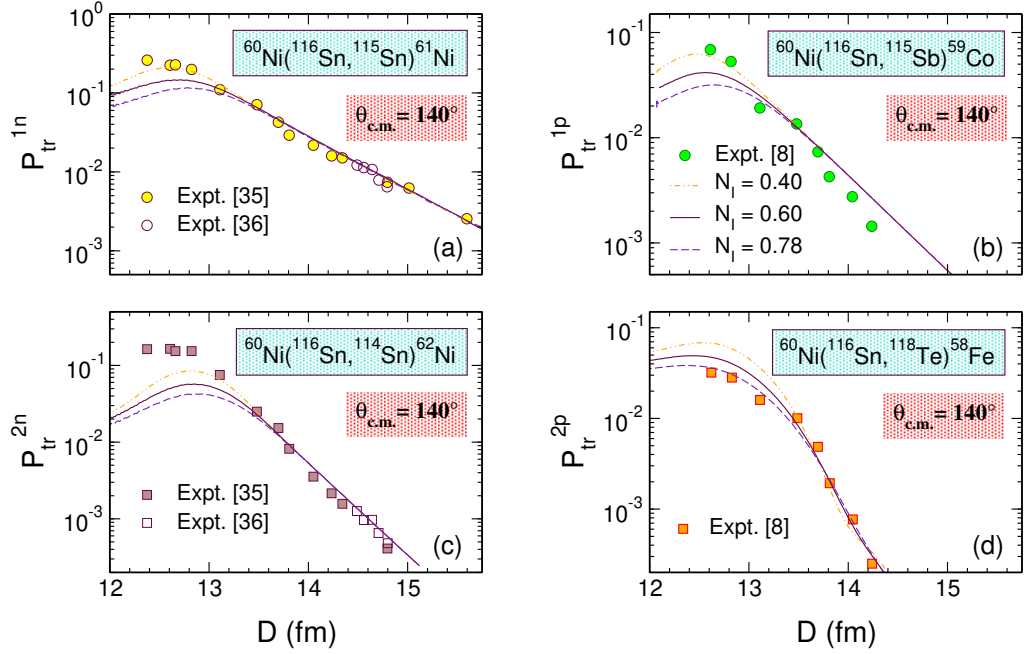}
    \caption{Absorption effects in (a) $1n$-pickup, (b) $1p$-stripping,
    (c) $2n$-pickup and (d) $2p$-stripping channels in the system
    $^{116}$Sn+$^{60}$Ni. See text for details.}
    \label{fig:fig09}
\end{figure*}

To further investigate the role of absorption, we examined its 
impact on the $1n$-pickup channel by varying the strength of the
imaginary component in the SPP across all partitions. The results,
presented in Fig.~\ref{fig:fig09}(a), show that reducing the
strength from $N_I = 0.6$ to $N_I = 0.4$ in the CRC calculations
significantly improves the reproduction of 
$P^\text{1n}_\text{tr}$ at smaller distances of closest approach.
This phenomenon, characterized by a decrease in $P_\text{tr}$ with
increasing absorption, has also been reported in previous studies
\cite{Scamps2015,Esbensen1998}, reinforcing the importance of
higher-order dynamics at higher energies.

We extended this analysis to the transfer probabilities of other
channels, and a similar sensitivity to the strength of the imaginary
potential was observed, as illustrated in Fig.~\ref{fig:fig09}(b-d).
Notably, absorption effects were found to be more pronounced in the
$2n$-pickup channel compared to the $1n$ case, in agreement with
earlier findings~\cite{Scamps2015}. In contrast, the $1p$- and
$2p$-stripping channels displayed different behavior, indicating
a nuanced dependence of transfer mechanisms on the type of nucleons
involved. This aspect of MNT requires further investigations.

As previously discussed in Section \ref{sect:shellmodel}, we
imposed some truncation on the model space to work within the
bounds of available computational resources. Consequently, certain
active orbits transitioned into inactive ones. It's worth noting
that some of these now-inactive orbits might still hold significance
in the transfer mechanism. While this truncation allowed us to achieve
a reasonable reproduction of energy spectra, it's important to
recognize that its impact was not limited to the spectrum of the
studied nuclei. Other observables, such as transition probabilities
and spectroscopic factors, could be affected as well, which in turn
might affect the outcomes of CRC calculations.

While performing the two-nucleon transfer calculations, only
those intermediate states that were included in the corresponding
one-nucleon transfer analysis, were considered. However, in the
case of two-nucleon transfer, additional states \textendash allowed
by angular momentum coupling \textendash could also be considered
in the intermediate partition for a more realistic description of
the underlying mechanism. The inclusion of such states would
require significantly more computational resources and was thus
beyond the scope of the present work.

\section{Summary and conclusions}
\label{conclusion}
We performed CRC calculations for QEL scattering and one- and
two-nucleon transfer in the $^{116}$Sn+$^{60}$Ni system. The
S$\tilde{a}$o Paulo potential was employed to generate the interaction
potentials between the participating nuclei. CRC calculations
incorporated couplings to the states that were experimentally
accessible through transfer channels and Coulomb excitations,
{\it i.e.}, inelastic transitions observed via $\gamma$-ray spectra.

The calculations for QEL scattering reproduced the inelastic
contributions quite well, in agreement with the experimentally
observed components \cite{Montanari2016}. The $1n$-pickup channel
was reasonably well described, particularly in the higher energy
region where indirect paths were found to be significant. For the
$2n$-pickup channel, the extreme cluster model provided a better
agreement with experimental $P_\text{tr}$ than the microscopic
cluster and sequential transfer models.

In the case of charged-particle transfer, {\it i.e.}, $1p$- and
$2p$-stripping, direct experimental guidance was limited, as $\gamma$-ray
spectra were not available for these channels ~\cite{Corradi2022}.
Nonetheless, our results indicate that in $1p$-stripping, the dominant
contribution comes from the g.s. to g.s. transition. For $2p$-stripping,
a good agreement with experimental data was achieved by treating the
two protons as a correlated cluster populating pure states in the final
nucleus, much like the case for $2n$-pickup.

To ensure robustness of the theoretical results, we extracted $P_\text{tr}$
using two approaches, {\it viz.}, via calculation of angular distributions
and excitation functions. Both methods yielded consistent trends in both
magnitude and shape of the transfer probabilities. Notably, a single set
of reduced radius and diffuseness parameter was used for both one- and
two-nucleon transfer channels. Despite this simplification, the CRC
results show an overall good agreement with experimental data across all
the channels. However, systematic underprediction of $P_\text{tr}$
was observed at higher energies, suggesting the possible role of
higher-order dynamical effects and uncertainties in the definition of
transfer probabilities in regions of strong nuclear overlap. Absorption
effects were found to play a more significant role in $2n$-pickup
compared to $1n$-pickup. Interestingly, an opposite trend was observed
in the case of proton transfer.

This study also underscores the ambiguities associated with the choice
of nuclear states in CRC calculations. These ambiguities can be
significantly reduced if the individual transitions and population
patterns of the transfer products are experimentally corroborated
\textendash especially in charged-particle transfer. In the current
analysis, we attempted to reproduce measured cross sections by
systematically coupling the low-lying states within the CRC
framework. The population patterns inferred from detection of
all relevant $\gamma$-rays would provide crucial insight into
the reaction mechanism and allow validation of the theoretical models.

Furthermore, constraints imposed by the limited model space in shell
model calculations and the resulting discrepancies between experimental
and theoretical excitation energies may also contribute to the
underestimation of $2n$- and $2p$-transfer probabilities in both
sequential and microscopic cluster approaches. These limitations
highlight the need for further investigations to improve nuclear
structure inputs and computational strategies for MNT studies
involving heavy ions.

Currently, only a limited number of systems have been studied across
all transfer channels with detailed $\gamma$-ray spectroscopy.
Comprehensive CRC analyses, combined with experimental verification
of final states, are essential for improving our understanding of the
reaction dynamics and for minimizing model-dependent uncertainties
in theoretical description of MNT channels. Nevertheless, the present
work demonstrates that a fully microscopic description of MNT in
heavy-ion collisions \textendash without the need for arbitrary
normalization factors \textendash is quite feasible within the CRC
formalism. It establishes a strong foundation for exploring more complex
systems and for improving the predictive power of reaction models.

\section{acknowledgements}
One of the authors (C.K.) acknowledges financial support from the
Council of Scientific and Industrial Research (CSIR), New Delhi
via grant no. CSIR/09/760(0038)/2019-EMR-I. The authors acknowledge
the National Supercomputing Mission (NSM) for providing computing
resources of `PARAM Ganga' installed at the Indian Institute of
Technology Roorkee, Roorkee 247667, Uttarakhand, which is implemented
by the Centre for Development of Advanced Computing (C-DAC) and
supported by the Ministry of  Electronics and Information Technology
(MeitY) and the Department of Science and Technology (DST), Government
of India, for carrying out large-scale shell-model and coupled reaction
channel (CRC) calculations. Technical support from Mr. Abhishek Kumar
and illuminating discussions with Dr. Rohan Biswas and Mrs. Gonika
on CRC calculations are thankfully acknowledged.

\clearpage
\appendix
\onecolumngrid
\begin{center}
{
{\bf{\Large{\textsf{Appendix}}}}\\~\\
}
\end{center}
\begin{framed}
\begin{center}
\large{\bf Shell model results for $^{\mathbf{58}}$Fe,
$^{\mathbf{59}}$Co, $^{\mathbf{60,61,62}}$Ni, $^{\mathbf{114,115,116}}$Sn,
$^{\mathbf{117}}$Sb and $^{\mathbf{118}}$Te to extract one- and
two-nucleon spectroscopic amplitudes}\\
\end{center}
\end{framed}
The spectroscopic amplitudes ($\mathcal{S}$s) for one and two nucleons were determined using the Large-Scale Shell Model (\texttt{LSSM}) calculation implemented in the \textsc{kshell} code \cite{Shimizu2019} for the $^{116}$Sn+$^{60}$Ni system. These $\mathcal{S}$s were then fed as input to the Coupled Reaction Channel (CRC) code, \textsc{fresco} \cite{fresco}.
For $^{60}$Ni-like nuclei, the \texttt{fp} or \texttt{fpg} model space was employed, incorporating interactions such as \texttt{fpd6npn} \cite{Richter1991}, \texttt{JUN45}\cite{Honma2009}, \texttt{KB3G}\cite{Poves2001}, and \texttt{JJ44BPN}\cite{jj44bpn}. For $^{116}$Sn-like nuclei, wave-functions were generated using the \texttt{SN100PN} \cite{Machleidt1996,Brown2004} and monopole-optimized effective interaction (\texttt{Monopole}) \cite{Chong2012}. %, alongside their corresponding model spaces, spanning nuclei from 50 to 82 nucleons. 
The results are presented in the tables below.
%\vspace{5mm}
%\section{Comparison of Spectra: Shell Model (\texttt{SM}) Calculations vs. Experimental Results}

\begin{table}[H]
\caption{A comparison between the calculated spectra of $^{60}$Ni and $^{61}$Ni, obtained using the \texttt{LSSM} framework with the \textsc{kshell} code, and the corresponding experimental data \cite{nndc}. Here, $I^\pi$ denotes the spin and parity of the states.}

\resizebox{\textwidth}{!}{%
\begin{tabular}{cccllllccclll}
\cmidrule(r){1-6} \cmidrule(lr){8-13}% \cmidrule(l){13-16}
%\cline{1-6} \cline{8-13}
\multirow{4}{*}{$I^\pi$} & \multicolumn{5}{c}{$^{60}$Ni} & \multicolumn{1}{c}{} & \multirow{4}{*}{$I^\pi$} & \multicolumn{5}{c}{$^{61}$Ni} \\ \cline{2-6} \cline{9-13} 
        & \multicolumn{5}{c}{Energy (MeV)}                        & \multirow{3}{*}{} &           & \multicolumn{5}{c}{Energy (MeV)}                        \\ \cline{2-6} \cline{9-13} 
        & \multirow{2}{*}{Exp.} & \multicolumn{4}{c}{\texttt{LSSM}}            &                   &           & \multirow{2}{*}{Exp.} & \multicolumn{4}{c}{\texttt{LSSM}}            \\ \cline{3-6} \cline{10-13} 
        &                       & \texttt{KB3G}  & \texttt{JUN45} & \texttt{fpd6npn} & \texttt{JJ44BPN} &                   &           &                       & \texttt{KB3G}  & \texttt{JUN45} & \texttt{fpd6npn} & \texttt{JJ44BPN} \\ \cline{1-6} \cline{8-13} 
0$^{+}$ & 0.000                 & 0.000     & 0.000     & 0.000       & 0.000       &                   & 3/2$^{-}$ & 0.000                 & 0.000   & 0.000  & 0.000  & 0.000   \\
2$^{+}$ & 1.332                 & 1.267 & 1.635 & 1.335   & 1.553   &                   & 5/2$^{-}$ & 0.67                  & 0.025 & -0.08 & 0.227   & 0.093   \\
2$^{+}$ & 2.158                 & 1.988 & 2.141 & 2.175   & 2.398   &                   & 1/2$^{-}$ & 0.282                 & 0.734 & 0.512 & 0.204   & 0.181   \\
0$^{+}$ & 2.284                 & 2.177 & 2.191 & 2.467   & 2.058   &                   & 1/2$^{-}$ & 0.656                 & 0.968 & 1.443 & 0.721   & 0.677   \\
4$^{+}$ & 2.505                 & 2.087 & 2.255 & 2.726   & 2.546   &                   & 5/2$^{-}$ & 0.908                 & 0.920 & 1.197 & 0.766   & 0.848   \\
\cline{1-6} \cline{8-13}
\end{tabular}%
}
\end{table}

\begin{table}[H]
\centering
\caption{A comparison between the calculated spectra of $^{62}$Ni, $^{58}$Fe, and $^{59}$Co, obtained using the \texttt{LSSM} framework with the \textsc{kshell} code, and the corresponding experimental data \cite{nndc}. $I^\pi$ stands for the spin and parity of the states.}
\label{tab:62Ni_59Ni_energy}
\fontsize{5mm}{5mm}\selectfont
\resizebox{\textwidth}{!}{%
\begin{tabular}{@{}cccllllcllllclll@{}}
\cmidrule(r){1-6} \cmidrule(lr){8-11} \cmidrule(l){13-16}
\multirow{4}{*}{$I^\pi$} & \multicolumn{5}{c}{$^{62}$Ni} &  & \multirow{4}{*}{$I^\pi$} & \multicolumn{3}{c}{$^{59}$Co} &  & \multirow{4}{*}{$I^\pi$} & \multicolumn{3}{c}{$^{58}$Fe} \\ \cmidrule(lr){2-6} \cmidrule(lr){9-11} \cmidrule(l){14-16} 
 & \multicolumn{5}{c}{Energy (MeV)} &  &  & \multicolumn{3}{c}{Energy (MeV)} &  &  & \multicolumn{3}{c}{Energy (MeV)} \\ \cmidrule(lr){2-6} \cmidrule(lr){9-11} \cmidrule(l){14-16} 
 & \multirow{2}{*}{Exp.} & \multicolumn{4}{c}{\texttt{LSSM}} &  &  & \multicolumn{1}{c}{\multirow{2}{*}{Exp.}} & \multicolumn{2}{c}{\texttt{LSSM}} &  &  & \multicolumn{1}{c}{\multirow{2}{*}{Exp.}} & \multicolumn{2}{c}{\texttt{LSSM}} \\ \cmidrule(lr){3-6} \cmidrule(lr){10-11} \cmidrule(l){15-16} 
 &  & \texttt{KB3G} & \texttt{JUN45} & \texttt{fpd6npn} & \texttt{JJ44BPN} &  &  & \multicolumn{1}{c}{} & \texttt{KB3G} &  \texttt{JUN45} &  &  & \multicolumn{1}{c}{} & \texttt{KB3G} & \texttt{fpd6npn} \\ \cmidrule(r){1-6} \cmidrule(lr){8-11} \cmidrule(l){13-16} 
0$^{+}$ & 0.000 & 0.000 & 0.000 & 0.000 & 0.000 &  & 7/2$^{-}$ & 0.000 & 0.000 & 0.000 &  & 0$^{+}$ & 0.000 & 0.000 & 0.000 \\
2$^{+}$ & 1.172 & 1.240 & 1.820 & 1.012 & 1.423 &  & 3/2$^{-}$ & 1.099 & 1.235 & 0.704 &  & 2$^{+}$ & 0.810 & 0.712 & 0.839 \\
0$^{+}$ & 2.048 & 1.864 & 2.193 & 1.736 & 1.997 &  & 9/2$^{-}$ & 1.190 & 1.186 & 1.250 &  & 2$^{+}$ & 1.674 & 1.485 & 1.652 \\
2$^{+}$ & 2.301 & 1.996 & 2.446 & 2.242 & 2.159 &  &  &  &  &  &  & 4$^{+}$ & 2.076 & 1.881 & 2.349 \\
4$^{+}$ & 2.336 & 2.018 & 3.088 & 2.315 & 2.689 &  &  &  &  &  &  & 3$^{+}$ & 2.133 & 1.855 & 2.211 \\ \cmidrule(r){1-6} \cmidrule(lr){8-11} \cmidrule(l){13-16} 
\end{tabular}%
}
\end{table}
%\vspace{10mm}

\begin{table}[H]
\centering
\caption{A comparison of the spectra obtained from \texttt{LSSM}
calculations using \textsc{kshell} and experimental data \cite{nndc}
for the nuclei $^{117}$Sb and $^{118}$Te. Here, $I^\pi$ denotes the
spin and parity of the states.}
\label{tab:117Sb_118Te_energy}
%\resizebox{\textwidth}{!}{%
\begin{tabular}{cllllclll}
\cmidrule(r){1-4} \cmidrule(r){6-9}
\multirow{4}{*}{$I^\pi$} & \multicolumn{3}{c}{$^{117}$Sb} & & \multirow{4}{*}{$I^\pi$} & \multicolumn{3}{c}{$^{118}$Te} \\ \cline{2-4} \cline{7-9} 
 & \multicolumn{3}{c}{Energy (MeV)} &  &  & \multicolumn{3}{c}{Energy (MeV)} \\ \cline{2-4} \cline{7-9} 
 & \multicolumn{1}{c}{\multirow{2}{*}{Exp.}} & \multicolumn{2}{c}{\texttt{LSSM}} &  &  & \multicolumn{1}{c}{\multirow{2}{*}{Exp.}} & \multicolumn{2}{c}{\texttt{LSSM}} \\ \cline{3-4} \cline{8-9} 
 & \multicolumn{1}{c}{} & \multicolumn{1}{c}{\texttt{SN100PN}} & \multicolumn{1}{c}{\texttt{Monopole}} &  &  & \multicolumn{1}{c}{} & \multicolumn{1}{c}{\texttt{SN100PN}} &\multicolumn{1}{c}{\texttt{Monopole}} \\ \cline{1-4} \cline{6-9} 
5/2$^{+}$ & 0.000 & 0.000 & 0.000 &  & 0$^{+}$ & 0.000 & \multicolumn{1}{c}{0.000} & 0.000 \\
7/2$^{+}$ & 0.527 & 0.532 & 0.599 &  & 2$^{+}$ & 0.605 & 0.275 & 0.415 \\
1/2$^{+}$ & 0.719 & 0.278 & 0.266 &  & 0$^{+}$ & 0.957 & 0.762 & 0.609 \\
3/2$^{+}$ & 0.923 & 0.601 & 1.154 &  & 2$^{+}$ & 1.150 & 0.934 & 0.787 \\ \cline{1-4} \cline{6-9} 
\end{tabular}%
%}
\end{table}

\begin{table}[H]
\centering
\caption{A comparison of the spectra obtained from \texttt{LSSM}
calculations using \textsc{kshell} and experimental data \cite{nndc}
for the nuclei $^{114,115,116}$Sn. $I^\pi$ stands for the spin and
parity of the states.}
\label{tab:116Sn_115Sn_114Sn}
\resizebox{\textwidth}{!}{%
\begin{tabular}{@{}cllllcllllclll@{}}
\cmidrule(r){1-4} \cmidrule(lr){6-9} \cmidrule(l){11-14}
\multirow{4}{*}{$I^\pi$} & \multicolumn{3}{c}{$^{116}$Sn} & & \multirow{4}{*}{$I^\pi$} & \multicolumn{3}{c}{$^{115}$Sn} &  & \multirow{4}{*}{$I^\pi$} & \multicolumn{3}{c}{$^{114}$Sn} \\ \cmidrule(lr){2-4} \cmidrule(lr){7-9} \cmidrule(l){12-14} 
 & \multicolumn{3}{c}{Energy (MeV)} & \multicolumn{1}{c}{} &  & \multicolumn{3}{c}{Energy (MeV)} &  &  & \multicolumn{3}{c}{Energy (MeV)} \\ \cmidrule(lr){2-4} \cmidrule(lr){7-9} \cmidrule(l){12-14} 
 & \multicolumn{1}{c}{\multirow{2}{*}{Exp.}} & \multicolumn{2}{c}{\texttt{LSSM}} & \multicolumn{1}{c}{} &  & \multicolumn{1}{c}{\multirow{2}{*}{Exp.}} & \multicolumn{2}{c}{\texttt{LSSM}} &  &  & \multicolumn{1}{c}{\multirow{2}{*}{Exp.}} & \multicolumn{2}{c}{\texttt{LSSM}} \\ \cmidrule(lr){3-4} \cmidrule(lr){8-9} \cmidrule(l){13-14} 
 & \multicolumn{1}{c}{} &\texttt{SN100PN} & \texttt{Monopole} &  &  &   & \texttt{SN100PN} & \texttt{Monopole} &  &  & \multicolumn{1}{c}{} & \texttt{SN100PN} &  \texttt{Monopole} \\ \cmidrule(r){1-4} \cmidrule(lr){6-9} \cmidrule(l){11-14} 
0$^{+}$ & 0.00 & 0.00 & 0.00 &  & 1/2$^{+}$ & 0.00 & 0.00 & 0.0 &  & 0$^{+}$ & 0.00 & 0.00 & 0.0 \\
2$^{+}$ & 1.293 & 1.160 & 1.202 &  & 3/2$^{+}$ & 0.497 & 0.163 & -0.045 &  & 2$^{+}$ & 1.299 & 1.228 & 1.392 \\
0$^{+}$ & 1.756 & 2.507 & 1.621 &  & 7/2$^{+}$ & 0.613 & 0.285 & 0.403 &  & 0$^{+}$ & 1.953 & 2.410 & 2.094 \\
0$^{+}$ & 2.027 & 2.546 & 2.588 &  & 11/2$^{-}$ & 0.713 & 0.659 & 0.237 &  & 0$^{+}$ & 2.156 & 2.710 & 2.209 \\
2$^{+}$ & 2.112 & 2.163 & 2.083 &  & 5/2$^{+}$ & 0.986 & 0.389 & 0.724 &  & 4$^{+}$ & 2.187 & 2.271 &  \\
4$^{+}$ & 2.529 & 2.204 & 2.229 &  & 3/2$^{+}$ & \multicolumn{1}{c}{1.280} & 0.773 & 1.015 &  & 2$^{+}$ & \multicolumn{1}{c}{2.238} & 2.175 & 2.287 \\ \cmidrule(r){1-4} \cmidrule(lr){6-9} \cmidrule(l){11-14} 
\end{tabular}%
}
\end{table}
%\newpage
\begin{table}[h]
\centering
\caption{One-neutron $\mathcal{S}$s relevant to the overlaps
with the lighter collision partner. The initial and the final
nuclei are characterized by spins and parities $I^\pi_{\mathrm{i}}$
and $I^\pi_{\mathrm{f}}$, respectively. Here, $n$, $\ell$ and $j$
denote the principal quantum number, orbital angular momentum and
total angular momentum of the valence particle state, respectively.}
\label{tab:my-table}
%\resizebox{\columnwidth}{!}{%
\begin{tabular}{cccccccccccc}
\midrule
$I^\pi_\text{i}$ &  & $n\ell_j$ &  & $I^\pi_\text{f}$ &  & &  & \texttt{KB3G} & \texttt{fpd6npn} & \texttt{JUN45} & \texttt{JJ44BPN} \\ \midrule
\multirow{3}{*}{$^{60}$Ni$_\text{g.s.}$(0$^+$)} &  & 2p$_{3/2}$ &  & $^{61}$Ni$_\text{g.s.}$(3/2$^-$) &  &  &  & -0.6701 & 0.6688 & -0.5952 & 0.7091 \\
 &  & 1f$_{5/2}$ &  & $^{61}$Ni$_{0.067}$(5/2$^-$) &  &  &  & 0.7856 & 0.7249 & -0.8470 & 0.8352 \\
 &  & 2p$_{1/2}$ &  & $^{61}$Ni$_{0.283}$(1/2$^-$) &  &  &  & -0.7577 & -0.6741 & 0.7611 & 0.4939 \\
 &  &  &  &  &  &  &  &  &  &  &  \\
\multirow{10}{*}{$^{60}$Ni$_{1.333}$(2$^+$)} &  & 1f$_{5/2}$ &  &  &  &  &  & -0.1707 & 0.1649 & 0.4066 & -0.3069 \\
 &  & 2p$_{3/2}$ &  & $^{61}$Ni$_\text{g.s.}$(3/2$^-$) &  &  &  & 0.8016 & -0.6750 & -0.7327 & 0.4198 \\
 &  & 2p$_{1/2}$ &  &  &  &  &  & 0.0553 & -0.0193 & -0.1358 & -0.0661 \\
 &  &  &  &  &  &  &  &  &  &  &  \\
 &  & 1f$_{5/2}$ &  &  &  &  &  & -0.4350 & -0.3456 & -0.2364 & 0.3797 \\
 &  & 2p$_{3/2}$ &  & $^{61}$Ni$_{0.067}$(5/2$^-$) &  &  &  & -0.2422 & -0.2877 & -0.6189 & 0.3169 \\
 &  & 2p$_{1/2}$ &  &  &  &  &  & -0.0666 & -0.0870 & -0.0621 & 0.1424 \\
 &  &  &  &  &  &  &  &  &  &  &  \\
 &  & 1f$_{5/2}$ &  & \multirow{2}{*}{$^{61}$Ni$_{0.283}$(1/2$^-$)} &  &  &  & 0.2853 & 0.4023 & -0.2301 & -0.3416 \\
 &  & 2p$_{3/2}$ &  &  &  &  &  & 0.1708 & 0.2002 & 0.1920 & 0.9765 \\ \hline
 &  &  &  &  &  &  &  &  &  &  &  \\
$^{61}$Ni$_\text{g.s.}$(3/2$^-$) &  & 2p$_{3/2}$ &  &  &  &  &  & 1.6062 & -1.3169 & -1.5695 & -1.5246 \\
$^{61}$Ni$_{0.067}$(5/2$^-$) &  & 1f$_{5/2}$ &  & $^{62}$Ni$_\text{g.s.}$(0$^+$) &  &  &  & -1.5548 & -1.5562 & -1.4384 & -1.5666 \\
$^{61}$Ni$_{0.283}$(1/2$^-$) &  & 2p$_{1/2}$ &  &  &  &  &  & 0.6308 & 0.8666 & 0.4738 & -0.3146 \\
 &  &  &  &  &  &  &  &  &  &  &  \\
 &  & 1f$_{5/2}$ &  & \multirow{10}{*}{$^{62}$Ni$_{1.172}$(2$^+$)} &  &  &  & 0.0864 & 0.0809 & -0.3182 & -0.2402 \\
$^{61}$Ni$_\text{g.s.}$(3/2$^-$) &  & 2p$_{3/2}$ &  &  &  &  &  & -0.0682 & 0.0251 & -0.0001 & -0.1031 \\
 &  & 2p$_{1/2}$ &  &  &  &  &  & 0.4606 & 0.5018 & -0.3003 & -0.3903 \\
 &  &  &  &  &  &  &  &  &  &  &  \\
 &  & 1f$_{5/2}$ &  &  &  &  &  & -0.8921 & 0.4468 & -0.7669 & -0.7816 \\
$^{61}$Ni$_{0.067}$(5/2$^-$) &  & 2p$_{3/2}$ &  &  &  &  &  & 0.0100 & -0.0523 & -0.0128 & 0.1329 \\
 & \multicolumn{1}{l}{} & 2p$_{1/2}$ & \multicolumn{1}{l}{} &  & \multicolumn{1}{l}{} & \multicolumn{1}{l}{} & \multicolumn{1}{l}{} & \multicolumn{1}{l}{-0.2660} & \multicolumn{1}{l}{0.3864} & \multicolumn{1}{l}{-0.4589} & \multicolumn{1}{l}{-0.4382} \\
 & \multicolumn{1}{l}{} & \multicolumn{1}{l}{} & \multicolumn{1}{l}{} &  & \multicolumn{1}{l}{} & \multicolumn{1}{l}{} & \multicolumn{1}{l}{} & \multicolumn{1}{l}{} & \multicolumn{1}{l}{} & \multicolumn{1}{l}{} & \multicolumn{1}{l}{} \\
\multirow{2}{*}{$^{61}$Ni$_{0.283}$(1/2$^-$)} & \multicolumn{1}{l}{} & 1f$_{5/2}$ & \multicolumn{1}{l}{} &  & \multicolumn{1}{l}{} & \multicolumn{1}{l}{} & \multicolumn{1}{l}{} & \multicolumn{1}{l}{0.2032} & \multicolumn{1}{l}{-0.3264} & \multicolumn{1}{l}{0.5201} & \multicolumn{1}{l}{-0.5471} \\
 & \multicolumn{1}{l}{} & 2p$_{3/2}$ & \multicolumn{1}{l}{} &  & \multicolumn{1}{l}{} & \multicolumn{1}{l}{} & \multicolumn{1}{l}{} & \multicolumn{1}{l}{-0.4340} & \multicolumn{1}{l}{0.3984} & \multicolumn{1}{l}{0.1674} & \multicolumn{1}{l}{-0.2918} \\ \hline
\end{tabular}%
%}
\end{table}

\begin{table}[ht!]
\centering
 \caption{One-proton $\mathcal{S}$s relevant to the overlaps
 with the lighter collision partner. The initial and the final
 nuclei are characterized by spins and parities $I^\pi_{\mathrm{i}}$
 and $I^\pi_{\mathrm{f}}$, respectively. Here, $n$, $\ell$ and $j$
 denote the principal quantum number, orbital angular momentum and
 total angular momentum of the valence particle state, respectively.}
\begin{tabular}{llllllllll}
\midrule
$I^\pi_\text{i}$ & & $n\ell_j$ &  & $I^\pi_\text{f}$ & &  & & \texttt{KB3G} & \texttt{fpd6npn} \\ \midrule
\multirow{2}{*}{$^{60}$Ni$_\text{g.s.}$(0$^+$)} & & 1f$_{7/2}$ && $^{59}$Co$_\text{g.s.}$(7/2$^-$) & &  &  & 2.0995 & -1.8841 \\
 &  & 2p$_{3/2}$ &  & $^{59}$Co$_{1.099}$(3/2$^-$) &  & &  & 0.1579  & 0.1779  \\  \\
\multirow{8}{*}{$^{60}$Ni$_\text{g.s.}$(2$^+$)} & & 1f$_{7/2}$ & & \multirow{3}{*}{$^{59}$Co$_\text{g.s.}$(7/2$^-$)} & & & & -0.7447 & 0.7650 \\
&  & 2p$_{3/2}$ &  &   &  &      &      & 0.1166  & -0.1620 \\
 &  & 1f$_{5/2}$ &  &  &  &      &      & 0.0553  & -0.0672 \\
  &  &  &  &  &  &      &      &         &         \\
&  & 1f$_{7/2}$ &  & \multirow{4}{*}{$^{59}$Co$_{1.099}$(3/2$^-$)} &  &      &      & 0.4238  & -0.0913 \\
 &  & 2p$_{3/2}$ &  &  &  &      &      & -0.0021 & 0.1218  \\
&  & 1f$_{5/2}$ &  &  &  &      &      & 0.0110  & 0.0567  \\
 &  & 2p$_{1/2}$ &  &  &  &      &      & 0.0316  & 0.1138  \\ \hline
&  &   &  &    &  &      &      &    &  \\
\multirow{9}{*}{$^{59}$Co$_\text{g.s.}$(7/2$^-$)}  &  & 1f$_{7/2}$ &  & $^{58}$Fe$_\text{g.s.}$(0$^+$)  &  &      &      & 0.4960  & -0.5038 \\
&  &&  &  &  &      &      &         &         \\
  &  & 1f$_{7/2}$ &  &   &  &      &      & -0.8960 & -0.8437 \\
  &  & 2p$_{3/2}$ &  & $^{58}$Fe$_{0.810}$(2$^+$)                    &  &      &      & 0.0289  & 0.0230  \\
 &  & 1f$_{5/2}$ &  &   &  &      &      & 0.0263  & 0.0450  \\
 &  &            &  &    &  &      &      &         &         \\
 &  & 1f$_{7/2}$ &  &     &  &      &      & -0.1322 & 0.1410 \\
&  & 2p$_{3/2}$ &  & $^{58}$Fe$_{1.674}$(2$^+$)     &  &      &      &  -0.0396 & 0.0720  \\
&  & 1f$_{5/2}$ &  &   &  &      &      &  0.0015 & 0.0063  \\ \\
\multirow{6}{*}{$^{59}$Co$_{1.099}$(3/2$^-$)} &  & 2p$_{3/2}$ &  & $^{58}$Fe$_\text{g.s.}$(0$^+$)   &  &      &      & 0.0692  & 0.4724  \\
  &  &       &  &   &  &      &      &         &         \\
  &  & 1f$_{7/2}$ &  & \multirow{4}{*}{$^{58}$Fe$_{0.810}$(2$^+$)}   &  &      &      & 0.0635  & 0.4102  \\
 &  & 2p$_{3/2}$ &  &   &  &      &      & -0.1518 & -0.3825 \\
 &  & 1f$_{5/2}$ &  &    &  &      &      & 0.0228  & 0.1898  \\
 &  & 2p$_{1/2}$ &  &   &  &      &      & 0.0146  & 0.4528  \\ 
&  &      &  &     &  &      &      &         &        \\
&  & 1f$_{7/2}$ &  & \multirow{4}{*}{$^{58}$Fe$_{1.674}$(2$^+$)}   &  &      &      & -0.7565  & 0.4187  \\
&  & 2p$_{3/2}$ &  &    &  &      &      & -0.0236 &-0.0122 \\
 &  & 1f$_{5/2}$ &  &    &  &      &      & 0.0009  & 0.1298  \\
  &  & 2p$_{1/2}$ &  &     &  &      &      & 0.0146  &  -0.0156 \\ \hline
&  &    &  &   &  &   &  &     &     \\
\end{tabular}%
\end{table}

\begin{table}[ht]
\centering
\caption{One-proton $\mathcal{S}$s relevant to the overlaps with
the heavier collision partner. The initial and the final nuclei
are characterized by spins and parities $I^\pi_{\mathrm{i}}$ and
$I^\pi_{\mathrm{f}}$, respectively. Here, $n$, $\ell$ and $j$
denote the principal quantum number, orbital angular momentum and
total angular momentum of the valence particle state, respectively.}

    \label{tab:one_partitcle}
    \fontsize{9pt}{9pt}\selectfont

    \begin{tabular}{cccccccccc}
    \toprule
     $I^\pi_\text{i}$ &  & {$n \ell_j$} &  & $I^\pi_\text{f}$ &  & 
       &  & \textbf{\texttt{SN100PN}} & \textbf{\texttt{Monopole}} \\
    \midrule
       &  & 2d$_{5/2}$ &  & $^{117}$Sb$_{\mathrm{g.s.}}$(5/2$^+$) &  &  &  & -0.0968 & -0.5762 \\
       $^{116}$Sn$_{\mathrm{g.s.}}$(0$^+$) &  & 1g$_{7/2}$ &  & $^{117}$Sb$_{0.527}$(7/2$^+$) &  &  &  & 0.0991 & 0.3883 \\
       &  & 3s$_{1/2}$ &  & $^{117}$Sb$_{0.719}$(1/2$^+$) &  &  &  & 0.0537 & -0.5035 \\
     \midrule
     $^{117}$Sb$_\text{g.s.}$(5/2$^+$) &  & 2d$_{5/2}$ &  &   &  & &  & 0.0955 & -0.9448 \\
$^{117}$Sb$_{0.527}$(7/2$^+$) &  & 1g$_{7/2}$ &  & $^{118}$Te$_\text{g.s.}$(0$^+$) &  & & & -0.0998 & 0.6488 \\
$^{117}$Sb$_{0.719}$(1/2$^+$) &  & 3s$_{1/2}$ &  &  &  &  &  & -0.0299 & -0.4790 \\
 &  &  &  &  &  &  &  &  \\
\multirow{4}{*}{$^{117}$Sb$_\text{g.s.}$(5/2$^+$)} &  & 3s$_{1/2}$ &  & \multirow{10}{*}{$^{118}$Te$_{0.605}$(2$^+$)} &  &  & & & -0.2021 \\
 &  & 2d$_{3/2}$ &  &  &   & &  & -0.0062  & -0.0407 \\
 &  & 2d$_{5/2}$ &  &  &  &  & & -0.0288 & -0.4719 \\
 &  & 1g$_{7/2}$ & \multicolumn{1}{l}{} &  & \multicolumn{1}{l}{} &  & &  -0.0055& -0.0915 \\
 & \multicolumn{1}{l}{} & \multicolumn{1}{l}{} & \multicolumn{1}{l}{} &  & \multicolumn{1}{l}{} & \multicolumn{1}{l}{} & \multicolumn{1}{l}{} &  \\
\multirow{3}{*}{$^{117}$Sb$_{0.527}$(7/2$^+$)} & \multicolumn{1}{l}{} & 2d$_{3/2}$ & \multicolumn{1}{l}{} &  & \multicolumn{1}{l}{} &  & & \multicolumn{1}{l}{0.0097}& 0.0480 \\
 &  & 2d$_{5/2}$ & \multicolumn{1}{l}{} &  & \multicolumn{1}{l}{} &  &&  \multicolumn{1}{l}{-0.0062}& -0.2302 \\
 &  & 1g$_{7/2}$ & \multicolumn{1}{l}{} &  & \multicolumn{1}{l}{} &  &  &\multicolumn{1}{l}{0.0465}& 0.5028 \\
 & \multicolumn{1}{l}{} &  & \multicolumn{1}{l}{} &  & \multicolumn{1}{l}{} & \multicolumn{1}{l}{} & \multicolumn{1}{l}{} &  \\
\multirow{2}{*}{$^{117}$Sb$_{0.719}$(1/2$^+$)} & \multicolumn{1}{l}{} & 2d$_{3/2}$ & \multicolumn{1}{l}{} &  & \multicolumn{1}{l}{} &  & &\multicolumn{1}{l}{0.0059} & -0.0978 \\
 & \multicolumn{1}{l}{} & 2d$_{5/2}$ &  &  & \multicolumn{1}{l}{} &  & & \multicolumn{1}{l}{0.0122} & -0.2378 \\ \hline
\end{tabular}
\end{table}

\begin{table}[ht!]
\centering
\caption{One-neutron $\mathcal{S}$s relevant to the overlaps with the
heavier collision partner. The initial and the final nuclei are
characterized by spins and parities $I^\pi_{\mathrm{i}}$ and
$I^\pi_{\mathrm{f}}$, respectively. Here, $n$, $\ell$ and $j$
denote the principal quantum number, orbital angular momentum and
total angular momentum of the valence particle state, respectively.}
\label{tab:my-table}
%\resizebox{\columnwidth}{!}{%
%{\small
\begin{tabular}{clccclll}
\hline
$I^\pi_\text{i}$  & \multicolumn{1}{c}{} & $n\ell_j$ &  & $I^\pi_\text{f}$& & \texttt{SN100PN} & \texttt{Mono-pole} \\ \hline
\multirow{5}{*}{$^{116}$Sn$_\text{g.s.}$(0$^+$)} & \multicolumn{1}{c}{} & 3s$_{1/2}$ &  & $^{115}$Sn$_\text{g.s.}$(1/2$^+$) & \multicolumn{1}{c}{} & \multicolumn{1}{c}{-0.5612} & \multicolumn{1}{c}{-0.7604} \\
 & \multicolumn{1}{c}{} & 2d$_{3/2}$ &  & $^{115}$Sn$_{0.497}$(3/2$^+$) & \multicolumn{1}{c}{} & \multicolumn{1}{c}{-0.8016} & \multicolumn{1}{c}{-0.9755} \\
 & \multicolumn{1}{c}{} & 1g$_{7/2}$ &  & $^{115}$Sn$_{0.613}$(7/2$^+$) & \multicolumn{1}{c}{} & \multicolumn{1}{c}{1.7841} & \multicolumn{1}{c}{2.2392} \\
 & \multicolumn{1}{c}{} & 1h$_{11/2}$ &  & $^{115}$Sn$_{0.713}$(11/2$^-$) & \multicolumn{1}{c}{} & \multicolumn{1}{c}{-1.4435} & \multicolumn{1}{c}{1.5022} \\
 & \multicolumn{1}{c}{} & 2d$_{5/2}$ &  & $^{115}$Sn$_{0.986}$(5/2$^+$) & \multicolumn{1}{c}{} & \multicolumn{1}{c}{-1.6551} & \multicolumn{1}{c}{2.0148} \\
\multirow{19}{*}{$^{116}$Sn$_{1.293}$(2$^+$)} & \multicolumn{1}{c}{} &  &  &  & \multicolumn{1}{c}{} & \multicolumn{1}{c}{} & \multicolumn{1}{c}{} \\
 & \multicolumn{1}{c}{} & 2d$_{3/2}$ &  & \multirow{2}{*}{$^{115}$Sn$_\text{g.s.}$(1/2$^+$)} & \multicolumn{1}{c}{} & \multicolumn{1}{c}{-0.2476} & \multicolumn{1}{c}{-0.3437} \\
 & \multicolumn{1}{c}{} & 2d$_{5/2}$ &  &  & \multicolumn{1}{c}{} & \multicolumn{1}{c}{0.0154} & \multicolumn{1}{c}{0.0701} \\
 &  & \multicolumn{1}{l}{} & \multicolumn{1}{l}{} & \multicolumn{1}{l}{} &  &  &  \\
 & \multicolumn{1}{c}{} & 3s$_{1/2}$ &  &  & \multicolumn{1}{c}{} & \multicolumn{1}{c}{0.2255} & \multicolumn{1}{c}{0.2848} \\
 & \multicolumn{1}{c}{} & 2d$_{3/2}$ &  &  & \multicolumn{1}{c}{} & \multicolumn{1}{c}{-0.2474} & \multicolumn{1}{c}{-0.3830} \\
 & \multicolumn{1}{c}{} & 2d$_{5/2}$ &  & $^{115}$Sn$_{0.497}$(3/2$^+$) & \multicolumn{1}{c}{} & \multicolumn{1}{c}{-0.0235} & \multicolumn{1}{c}{-0.0440} \\
 & \multicolumn{1}{c}{} & 1g$_{7/2}$ &  &  & \multicolumn{1}{c}{} & \multicolumn{1}{c}{-0.0643} & \multicolumn{1}{c}{-0.0431} \\
 & \multicolumn{1}{c}{} &  &  &  & \multicolumn{1}{c}{} & \multicolumn{1}{c}{} & \multicolumn{1}{c}{} \\
 & \multicolumn{1}{c}{} & 2d$_{3/2}$ &  &  & \multicolumn{1}{c}{} & \multicolumn{1}{c}{0.3477} & \multicolumn{1}{c}{0.4683} \\
 & \multicolumn{1}{c}{} & 2d$_{5/2}$ &  & \multirow{2}{*}{$^{115}$Sn$_{0.613}$(7/2$^+$)} & \multicolumn{1}{c}{} & \multicolumn{1}{c}{0.0345} & \multicolumn{1}{c}{0.0282} \\
 & \multicolumn{1}{c}{} & 1g$_{7/2}$ &  &  & \multicolumn{1}{c}{} & \multicolumn{1}{c}{-0.1603} & \multicolumn{1}{c}{-0.2392} \\
 & \multicolumn{1}{c}{} &  &  &  & \multicolumn{1}{c}{} & \multicolumn{1}{c}{} & \multicolumn{1}{c}{} \\
 & \multicolumn{1}{c}{} & 1h$_{11/2}$ &  & $^{115}$Sn$_{0.713}$(11/2$^-$) & \multicolumn{1}{c}{} & \multicolumn{1}{c}{-0.9209} & \multicolumn{1}{c}{0.9507} \\
 & \multicolumn{1}{c}{} &  &  &  & \multicolumn{1}{c}{} & \multicolumn{1}{c}{} & \multicolumn{1}{c}{} \\
 & \multicolumn{1}{c}{} & 3s$_{1/2}$ &  &  & \multicolumn{1}{c}{} & \multicolumn{1}{c}{-0.2841} & \multicolumn{1}{c}{0.1919} \\
 & \multicolumn{1}{c}{} & 2d$_{3/2}$ &  &  & \multicolumn{1}{c}{} & \multicolumn{1}{c}{-0.1720} & \multicolumn{1}{c}{0.2737} \\
 & \multicolumn{1}{c}{} & 2d$_{5/2}$ &  & $^{115}$Sn$_{0.986}$(5/2$^+$) & \multicolumn{1}{c}{} & \multicolumn{1}{c}{0.2136} & \multicolumn{1}{c}{-0.2008} \\
 & \multicolumn{1}{c}{} & 1g$_{7/2}$ & \multicolumn{1}{l}{} & \multicolumn{1}{l}{} &  & 0.0585 & \multicolumn{1}{c}{-0.0520} \\ \hline
\multicolumn{1}{l}{} &  & \multicolumn{1}{l}{} & \multicolumn{1}{l}{} & \multicolumn{1}{l}{} &  &  &  \\
$^{115}$Sn$_\text{g.s.}$(1/2$^+$) &  & 3s$_{1/2}$ & \multicolumn{1}{l}{} & \multirow{5}{*}{$^{114}$Sn$_\text{g.s.}$(0$^+$)} &  & -0.7246 & 0.7848 \\
$^{115}$Sn$_{0.497}$(3/2$^+$) &  & 2d$_{3/2}$ & \multicolumn{1}{l}{} &  &  & -0.7328 & 0.8264 \\
$^{115}$Sn$_{0.613}$(7/2$^+$) &  & 1g$_{7/2}$ & \multicolumn{1}{l}{} &  &  & 0.4764 & -0.4666 \\
$^{115}$Sn$_{0.713}$(11/2$^-$) &  & 1h$_{11/2}$ & \multicolumn{1}{l}{} &  &  & 0.5447 & 0.8426 \\
$^{115}$Sn$_{0.986}$(5/2$^+$) &  & 2d$_{5/2}$ & \multicolumn{1}{l}{} &  &  & -0.4295 & -0.3649 \\
\multicolumn{1}{l}{} &  &  &  &  &  &  &  \\
\multirow{2}{*}{$^{115}$Sn$_\text{g.s.}$(1/2$^+$)} &  & 2d$_{3/2}$ &  & \multirow{18}{*}{$^{114}$Sn$_{1.299}$(2$^+$)} &  & 0.1068 & 0.0556 \\
 &  & 2d$_{5/2}$ &  &  &  & 0.5801 & -0.6690 \\
\multicolumn{1}{l}{} &  & \multicolumn{1}{l}{} & \multicolumn{1}{l}{} &  &  &  &  \\
\multicolumn{1}{l}{} &  & 3s$_{1/2}$ &  &  &  & -0.0208 & -0.1364 \\
\multicolumn{1}{l}{} &  & 2d$_{3/2}$ &  &  &  & -0.0784 & 0.0789 \\
$^{115}$Sn$_{0.497}$(3/2$^+$) &  & 2d$_{5/2}$ &  &  &  & 0.2584 & -0.2515 \\
\multicolumn{1}{l}{} &  & 1g$_{7/2}$ &  &  &  & 0.3636 & -0.5478 \\
%\multicolumn{1}{l}{} &  &  &  &  &  &  &  \\
\multicolumn{1}{l}{} &  & 2d$_{3/2}$ &  &  &  & 0.0066 & 0.0450 \\
\multirow{2}{*}{$^{115}$Sn$_{0.613}$(7/2$^+$)} &  & 2d$_{5/2}$ &  &  &  & -0.0862 & 0.0612 \\
 &  & 1g$_{7/2}$ &  &  &  & -0.5134 & 0.5188 \\
%\multicolumn{1}{l}{} &  &  &  &  &  &  &  \\
$^{115}$Sn$_{0.713}$(11/2$^-$) &  & 1h$_{11/2}$ &  &  &  & 0.0603 & 0.1450 \\
\multicolumn{1}{l}{} &  &  &  &  &  &  &  \\
\multicolumn{1}{l}{} &  & 3s$_{1/2}$ &  &  &  & 0.1031 & 0.3045 \\
\multicolumn{1}{l}{} &  & 2d$_{3/2}$ &  &  &  & -0.0387 & 0.0851 \\
$^{115}$Sn$_{0.986}$(5/2$^+$) &  & 2d$_{5/2}$ &  &  &  & 0.4128 & 0.3041 \\
\multicolumn{1}{l}{} &  & 1g$_{7/2}$ & \multicolumn{1}{l}{} &  &  & -0.1148 & -0.0645 \\
\hline \\
\end{tabular}%
%}
%}
\end{table}

\begin{table}[ht!]
\centering
\setlength{\tabcolsep}{4pt}
\caption{Cluster spectroscopic amplitude ($\mathcal{S}$ (c.m.)) for
di-neutron transfer involving the heavier nucleus $^{116}$Sn are
presented for the relevant overlaps. The calculations were performed
using \textsc{kshell} with the \texttt{SN100PN} and the \texttt{Monopole}
effective interactions. Here, $j_1$ and $j_2$ stand for the total
angular momenta of nucleons 1 and 2, respectively. The quantum numbers
$n$, $\ell$ ($N$, $L$) represent the principal quantum numbers and
orbital angular momenta of the nucleons relative to each other
(and to the core), respectively. $J$ and $\Lambda$ denote the total
angular momentum and the orbital angular momentum of the cluster
with respect to the core, respectively, and $S$ is the intrinsic
spin of the two-neutron cluster. The spins and parities of the
initial and final nuclei are given by $I^{\pi}_{\mathrm{i}}$ and
$I^{\pi}_{\mathrm{f}}$, respectively.}
\label{tab:116Sn_117Sb_tna}
\resizebox{\textwidth}{!}{%
\begin{tabular}{lllllllllllllllllllllllll}
\cmidrule{1-12} \cmidrule{14-25}
$I^\pi_\text{i}$  & $j_1$$j_2$ & $J$ & $I^\pi_\text{f}$  & $\mathcal{S}$ ($j-j$) \texttt{(SN100PN)} & $n$ & $\ell$ & $N$ & $L$ & $\Lambda$ & $S$ & $\mathcal{S}$ (c.m.) \texttt{(SN100PN)} & & $I^\pi_\text{i}$ & $j_1$$j_2$ & $J$ & $I^\pi_\text{f}$ & $\mathcal{S}$ ($j-j$)\texttt{(Monopole)} & $n$ & $\ell$ & $N$ & $L$ & $\Lambda$ & $S$ & $\mathcal{S}$ (c.m.) \texttt{(Monopole)}\\ \cmidrule{1-12} \cmidrule{14-25}
 & (1g$_{7/2})^2$ & \multirow{7}{*}{0} &  & -0.91777 &  &  &  &  &  &  &  &  &  & (1g$_{7/2})^2$ & \multirow{7}{*}{0} & \multirow{7}{*}{$^{114}$Sn$_\text{g.s.}$(0$^{+}$)} & 0.84224 &  &  &  &  &  &  &  \\
\multirow{55}{*}{$^{116}$Sn$_\text{g.s.}$(0$^{+}$)} & (2d$_{5/2})^2$ &  & \multirow{6}{*}{$^{114}$Sn$_\text{g.s.}$(0$^{+}$)} & -0.79657 & 1 & 0 & 5 & 0 & 0 & 0 & -0.61427 &  & \multirow{55}{*}{$^{116}$Sn$_\text{g.s.}$(0$^{+}$)} & (2d$_{5/2})^2$ &  &  & 0.62585 & 1 & 0 & 5 & 0 & 0 & 0 & 0.58225 \\
 & (2d$_{3/2})^2$ &  &  & -0.61462 &  &  &  &  &  &  &  &  &  & (2d$_{3/2})^2$ &  &  & 0.66089 &  &  &  &  &  &  &  \\
 & (3s$_{1/2})^2$ &  &  & -0.44234 &  &  &  &  &  &  &  &  &  & (3s$_{1/2})^2$ &  &  & 0.49075 &  &  &  &  &  &  &  \\
 &  &  &  &  &  &  &  &  &  &  &  &  &  &  &  &  &  &  &  &  &  &  &  &  \\
 & \multirow{2}{*}{(1h$_{11/2})^2$} &  &  & \multirow{2}{*}{1.15529} & 1 & 0 & 6 & 0 & 0 & 0 & 0.05375 &  &  & \multirow{2}{*}{(1h$_{11/2})^2$} &  &  & \multirow{2}{*}{-0.98199} & 1 & 0 & 6 & 0 & 0 & 0 & -0.04569 \\
 &  &  &  &  & 1 & 1 & 5 & 1 & 1 & 1 & -0.20808 &  &  &  &  &  &  & 1 & 1 & 5 & 1 & 1 & 1 & 0.17687 \\
 &  &  &  &  &  &  &  &  &  &  &  &  &  &  &  &  &  &  &  &  &  &  &  &  \\
 & (1g$_{7/2})^2$ &  &  & -0.91777 &  &  &  &  &  &  &  &  &  & (1g$_{7/2})^2$ &  &  & 0.84224 &  &  &  &  &  &  &  \\
 & (2d$_{5/2})^2$ &  &  & -0.79657 & 1 & 1 & 4 & 1 & 1 & 1 & -0.17 &  &  & (2d$_{5/2})^2$ &  &  & 0.62585 & 1 & 1 & 4 & 1 & 1 & 1 & 0.226 \\
 & (2d$_{3/2})^2$ &  &  & -0.61462 &  &  &  &  &  &  &  &  &  & (2d$_{3/2})^2$ &  &  & 0.66089 &  &  &  &  &  &  &  \\ \cline{2-12} \cline{15-25} 
 &  &  &  &  &  &  &  &  &  &  &  &  &  &  &  &  &  &  &  &  &  &  &  &  \\
 & (1g$_{7/2})^2$ & 2 & \multirow{43}{*}{$^{114}$Sn$_{1.299}$(2$^{+}$)} & 0.56890 &  &  &  &  &  &  &  &  &  & (1g$_{7/2})^2$ &  & \multirow{43}{*}{$^{114}$Sn$_{1.299}$(2$^{+}$)} & -0.61047 &  &  &  &  &  &  &  \\
 & 1g$_{7/2}$ 2d$_{5/2}$ &  &  & -0.14981 &  &  &  &  &  &  &  &  &  & 1g$_{7/2}$ 2d$_{5/2}$ &  &  & 0.12496 &  &  &  &  &  &  &  \\
 & 1g$_{7/2}$ 2d$_{3/2}$ &  &  & 0.23111 &  &  &  &  &  &  &  &  &  & 1g$_{7/2}$ 2d$_{3/2}$ &  &  & -0.39477 &  &  &  &  &  &  &  \\
 & (2d$_{5/2})^2$ &  &  & 0.39403 & 1 & 0 & 4 & 2 & 2 & 0 & 0.20995 &  &  & (2d$_{5/2})^2$ &  &  & -0.30764 & 1 & 0 & 4 & 2 & 2 & 0 & -0.28395 \\
 & 2d$_{5/2}$ 2d$_{3/2}$ &  &  & 0.13848 &  &  &  &  &  &  &  &  &  & 2d$_{5/2}$ 2d$_{3/2}$ &  &  & -0.19000 &  &  &  &  &  &  &  \\
 & 2d$_{5/2}$ 3s$_{1/2}$ &  &  & 0.25323 &  &  &  &  &  &  &  &  &  & 2d$_{5/2}$ 3s$_{1/2}$ & 2 &  & -0.41964 &  &  &  &  &  &  &  \\
 & (2d$_{3/2})^2$ &  &  & 0.05130 &  &  &  &  &  &  &  &  &  & (2d$_{3/2})^2$ &  &  & -0.07984 &  &  &  &  &  &  &  \\
 & 2d$_{3/2}$ 3s$_{1/2}$ &  &  & -0.07823 &  &  &  &  &  &  &  &  &  & 2d$_{3/2}$ 3s$_{1/2}$ &  &  & 0.23399 &  &  &  &  &  &  &  \\
 &  &  &  &  &  &  &  &  &  &  &  &  &  &  &  &  &  &  &  &  &  &  &  &  \\
 &  &  &  &  & 1 & 0 & 5 & 2 & 2 & 0 & -0.00609 &  &  &  &  &  &  & 1 & 0 & 5 & 2 & 2 & 0 & 8E-05 \\
 & (1h$_{11/2})^2$ &  &  & -0.23991 & 1 & 1 & 5 & 1 & 1 & 1 & -0.02030 &  &  & (1h$_{11/2})^2$ &  &  & 0.00283 & 1 & 1 & 5 & 1 & 1 & 1 & 0.00023 \\
 &  &  &  &  & 1 & 1 & 4 & 3 & 3 & 1 & 0.01195 &  &  &  &  &  &  & 1 & 1 & 4 & 3 & 3 & 1 & -0.00014 \\
 &  &  &  &  &  &  &  &  &  &  &  &  &  &  &  &  &  &  &  &  &  &  &  &  \\
 & (1g$_{7/2})^2$ &  &  & 0.56890 &  &  &  &  &  &  &  &  &  & (1g$_{7/2})^2$ &  &  & -0.61047 &  &  &  &  &  &  &  \\
 & 1g$_{7/2}$ 2d$_{5/2}$ &  &  & -0.14981 &  &  &  &  &  &  &  &  &  & 1g$_{7/2}$ 2d$_{5/2}$ &  &  & 0.12496 &  &  &  &  &  &  &  \\
 & 1g$_{7/2}$ 2d$_{3/2}$ &  &  & 0.23111 &  &  &  &  &  &  &  &  &  & 1g$_{7/2}$ 2d$_{3/2}$ &  &  & -0.39477 &  &  &  &  &  &  &  \\
 & (2d$_{5/2})^2$ &  &  & 0.39403 & 1 & 1 & 4 & 1 & 2 & 1 & -0.02217 &  &  & (2d$_{5/2})^2$ &  &  & -0.30764 & 1 & 1 & 4 & 1 & 2 & 1 & 0.0355 \\
 & 2d$_{5/2}$ 2d$_{3/2}$ &  &  & 0.13848 &  &  &  &  &  &  &  &  &  & 2d$_{5/2}$ 2d$_{3/2}$ &  &  & -0.19000 &  &  &  &  &  &  &  \\
 & 2d$_{5/2}$ 3s$_{1/2}$ &  &  & 0.25323 &  &  &  &  &  &  &  &  &  & 2d$_{5/2}$ 3s$_{1/2}$ &  &  & -0.41964 &  &  &  &  &  &  &  \\
 & (2d$_{3/2})^2$ &  &  & 0.05130 &  &  &  &  &  &  &  &  &  & (2d$_{3/2})^2$ &  &  & -0.07984 &  &  &  &  &  &  &  \\
 & 2d$_{3/2}$ 3s$_{1/2}$ &  &  & -0.07823 &  &  &  &  &  &  &  &  &  & 2d$_{3/2}$ 3s$_{1/2}$ &  &  & 0.23399 &  &  &  &  &  &  &  \\
 &  &  &  &  &  &  &  &  &  &  &  &  &  &  &  &  &  &  &  &  &  &  &  &  \\
 & (2d$_{3/2})^2$ &  &  & 0.05130 &  &  &  &  &  &  &  &  &  & (2d$_{3/2})^2$ &  &  & -0.07984 &  &  &  &  &  &  &  \\
 & 2d$_{5/2}$ 2d$_{3/2}$ &  &  & 0.13848 & 1 & 1 & 4 & 1 & 1 & 1 & 0.04094 &  &  & 2d$_{5/2}$ 2d$_{3/2}$ &  &  & -0.19000 & 1 & 1 & 4 & 1 & 1 & 1 & 0.00678 \\
 & (2d$_{5/2})^2$ &  &  & 0.39403 &  &  &  &  &  &  &  &  &  & (2d$_{5/2})^2$ &  &  & -0.30764 &  &  &  &  &  &  &  \\
 & (1g$_{7/2})^2$ &  &  & 0.56890 &  &  &  &  &  &  &  &  &  & (1g$_{7/2})^2$ &  &  & -0.61047 &  &  &  &  &  &  &  \\
 &  &  &  &  &  &  &  &  &  &  &  &  &  &  &  &  &  &  &  &  &  &  &  &  \\
 & (1g$_{7/2})^2$ &  &  & 0.56890 &  &  &  &  &  &  &  &  &  & (1g$_{7/2})^2$ &  &  & -0.61047 &  &  &  &  &  &  &  \\
 & 1g$_{7/2}$ 2d$_{5/2}$ &  &  & -0.14981 &  &  &  &  &  &  &  &  &  & 1g$_{7/2}$ 2d$_{5/2}$ &  &  & 0.12496 &  &  &  &  &  &  &  \\
 & 1g$_{7/2}$ 2d$_{3/2}$ &  &  & 0.23111 &  &  &  &  &  &  &  &  &  & 1g$_{7/2}$ 2d$_{3/2}$ &  &  & -0.39477 &  &  &  &  &  &  &  \\
 & (2d$_{5/2})^2$ &  &  & 0.39403 & 1 & 1 & 3 & 3 & 2 & 1 & -0.03744 &  &  & (2d$_{5/2})^2$ &  &  & -0.30764 & 1 & 1 & 3 & 3 & 2 & 1 & 0.05987 \\
 & 2d$_{5/2}$ 2d$_{3/2}$ &  &  & 0.13848 &  &  &  &  &  &  &  &  &  & 2d$_{5/2}$ 2d$_{3/2}$ &  &  & -0.19000 &  &  &  &  &  &  &  \\
 & 2d$_{5/2}$ 3s$_{1/2}$ &  &  & 0.25323 &  &  &  &  &  &  &  &  &  & 2d$_{5/2}$ 3s$_{1/2}$ &  &  & -0.41964 &  &  &  &  &  &  &  \\
 & (2d$_{3/2})^2$ &  &  & 0.05130 &  &  &  &  &  &  &  &  &  & (2d$_{3/2})^2$ &  &  & -0.07984 &  &  &  &  &  &  &  \\
 & 2d$_{3/2}$ 3s$_{1/2}$ &  &  & -0.07823 &  &  &  &  &  &  &  &  &  & 2d$_{3/2}$ 3s$_{1/2}$ &  &  & 0.23399 &  &  &  &  &  &  &  \\
 &  &  &  &  &  &  &  &  &  &  &  &  &  &  &  &  &  &  &  &  &  &  &  &  \\
 & (1g$_{7/2})^2$ &  &  & 0.56890 &  &  &  &  &  &  &  &  &  & (1g$_{7/2})^2$ &  &  & -0.61047 &  &  &  &  &  &  &  \\
 & 1g$_{7/2}$ 2d$_{5/2}$ &  &  & -0.14981 &  &  &  &  &  &  &  &  &  & 1g$_{7/2}$ 2d$_{5/2}$ &  &  & 0.12496 &  &  &  &  &  &  &  \\
 & 1g$_{7/2}$ 2d$_{3/2}$ &  &  & 0.23111 &  &  &  &  &  &  &  &  &  & 1g$_{7/2}$ 2d$_{3/2}$ &  &  & -0.39477 &  &  &  &  &  &  &  \\
 & (2d$_{5/2})^2$ &  &  & 0.39403 & 1 & 0 & 4 & 2 & 2 & 1 & 0.01894 &  &  & (2d$_{5/2})^2$ &  &  & -0.30764 & 1 & 0 & 4 & 2 & 2 & 1 & -0.01275 \\
 & 2d$_{5/2}$ 2d$_{3/2}$ &  &  & 0.13848 &  &  &  &  &  &  &  &  &  & 2d$_{5/2}$ 2d$_{3/2}$ &  &  & -0.19000 &  &  &  &  &  &  &  \\
 & 2d$_{5/2}$ 3s$_{1/2}$ &  &  & 0.25323 &  &  &  &  &  &  &  &  &  & 2d$_{5/2}$ 3s$_{1/2}$ &  &  & -0.41964 &  &  &  &  &  &  &  \\
 & (2d$_{3/2})^2$ &  &  & 0.05130 &  &  &  &  &  &  &  &  &  & (2d$_{3/2})^2$ &  &  & -0.07984 &  &  &  &  &  &  &  \\
 & 2d$_{3/2}$ 3s$_{1/2}$ &  &  & -0.07823 &  &  &  &  &  &  &  &  &  & 2d$_{3/2}$ 3s$_{1/2}$ &  &  & 0.23399 &  &  &  &  &  &  &  \\ \cline{1-12} \cline{14-25} 
 &  &  &  &  &  &  &  &  &  &  &  &  &  &  &  &  &  &  &  &  &  &  &  &  \\
 &  &  &  &  &  &  &  &  &  &  &  &  &  &  &  &  &  &  &  &  &  &  &  &  \\
\multirow{51}{*}{$^{116}$Sn$_{1.290}$(2$^{+}$)} & (1g$_{7/2})^2$ & \multirow{51}{*}{2} &  & -0.02097 &  &  &  &  &  &  &  &  & \multirow{51}{*}{$^{116}$Sn$_{1.290}$(2$^{+}$)} & (1g$_{7/2})^2$ & \multirow{51}{*}{2} & \multirow{51}{*}{$^{114}$Sn$_{g.s,}$(0$^{+}$)} & -0.00123 &  &  &  &  &  &  &  \\
 & 1g$_{7/2}$ 2d$_{5/2}$ &  &  & 0.03446 &  &  &  &  &  &  &  &  &  & 1g$_{7/2}$ 2d$_{5/2}$ &  &  & -0.03308 &  &  &  &  &  &  &  \\
 & 1g$_{7/2}$ 2d$_{3/2}$ &  &  & -0.15039 &  &  &  &  &  &  &  &  &  & 1g$_{7/2}$ 2d$_{3/2}$ &  &  & 0.16567 &  &  &  &  &  &  &  \\
 & (2d$_{5/2})^2$ &  & $^{114}$Sn$_{g.s,}$(0$^{+}$) & -0.04972 & 1 & 0 & 4 & 2 & 2 & 0 & -0.11029 &  &  & (2d$_{5/2})^2$ &  &  & 0.04272 & 1 & 0 & 4 & 2 & 2 & 0 & 0.11357 \\
 & 2d$_{5/2}$ 2d$_{3/2}$ &  &  & -0.06708 &  &  &  &  &  &  &  &  &  & 2d$_{5/2}$ 2d$_{3/2}$ &  &  & 0.04066 &  &  &  &  &  &  &  \\
 & 2d$_{5/2}$ 3s$_{1/2}$ &  &  & -0.13163 &  &  &  &  &  &  &  &  &  & 2d$_{5/2}$ 3s$_{1/2}$ &  &  & 0.06394 &  &  &  &  &  &  &  \\
 & (2d$_{3/2})^2$ &  &  & -0.12988 &  &  &  &  &  &  &  &  &  & (2d$_{3/2})^2$ &  &  & 0.18526 &  &  &  &  &  &  &  \\
 & 2d$_{3/2}$ 3s$_{1/2}$ &  &  & 0.18734 &  &  &  &  &  &  &  &  &  & 2d$_{3/2}$ 3s$_{1/2}$ &  &  & -0.26432 &  &  &  &  &  &  &  \\
 &  &  &  &  &  &  &  &  &  &  &  &  &  &  &  &  &  &  &  &  &  &  &  &  \\
 &  &  &  &  & 1 & 0 & 5 & 2 & 2 & 0 & 0.01078 &  &  &  &  &  &  & 1 & 0 & 5 & 2 & 2 & 0 & -0.01134 \\
 & (1h$_{11/2})^2$ &  &  & 0.42487 & 1 & 1 & 5 & 1 & 1 & 1 & 0.03595 &  &  & (1h$_{11/2})^2$ &  &  & -0.44675 & 1 & 1 & 5 & 1 & 1 & 1 & -0.0378 \\
 &  &  &  &  & 1 & 1 & 4 & 3 & 3 & 1 & -0.02116 &  &  &  &  &  &  & 1 & 1 & 4 & 3 & 3 & 1 & 0.02225 \\
 &  &  &  &  &  &  &  &  &  &  &  &  &  &  &  &  &  &  &  &  &  &  &  &  \\
 & (1g$_{7/2})^2$ &  &  & -0.02097 &  &  &  &  &  &  &  &  &  & (1g$_{7/2})^2$ &  &  & -0.00123 &  &  &  &  &  &  &  \\
 & 1g$_{7/2}$ 2d$_{5/2}$ &  &  & 0.03446 &  &  &  &  &  &  &  &  &  & 1g$_{7/2}$ 2d$_{5/2}$ &  &  & -0.03308 &  &  &  &  &  &  &  \\
 & 1g$_{7/2}$ 2d$_{3/2}$ &  &  & -0.15039 &  &  &  &  &  &  &  &  &  & 1g$_{7/2}$ 2d$_{3/2}$ &  &  & 0.16567 &  &  &  &  &  &  &  \\
 & (2d$_{5/2})^2$ &  &  & -0.04972 & 1 & 1 & 4 & 1 & 2 & 1 & 0.0238 &  &  & (2d$_{5/2})^2$ &  &  & 0.04272 & 1 & 1 & 4 & 1 & 2 & 1 & -0.03758 \\
 & 2d$_{5/2}$ 2d$_{3/2}$ &  &  & -0.06708 &  &  &  &  &  &  &  &  &  & 2d$_{5/2}$ 2d$_{3/2}$ &  &  & 0.04066 &  &  &  &  &  &  &  \\
 & 2d$_{5/2}$ 3s$_{1/2}$ &  &  & -0.13163 &  &  &  &  &  &  &  &  &  & 2d$_{5/2}$ 3s$_{1/2}$ &  &  & 0.06394 &  &  &  &  &  &  &  \\
 & (2d$_{3/2})^2$ &  &  & -0.12988 &  &  &  &  &  &  &  &  &  & (2d$_{3/2})^2$ &  &  & 0.18526 &  &  &  &  &  &  &  \\
 & 2d$_{3/2}$ 3s$_{1/2}$ &  &  & 0.18734 &  &  &  &  &  &  &  &  &  & 2d$_{3/2}$ 3s$_{1/2}$ &  &  & -0.26432 &  &  &  &  &  &  &  \\
 &  &  &  &  &  &  &  &  &  &  &  &  &  &  &  &  &  &  &  &  &  &  &  &  \\
 & (2d$_{3/2})^2$ &  &  & -0.12988 &  &  &  &  &  &  &  &  &  & (2d$_{3/2})^2$ &  &  & 0.18526 &  &  &  &  &  &  &  \\
 & 2d$_{5/2}$ 2d$_{3/2}$ &  &  & -0.13163 & 1 & 1 & 4 & 1 & 1 & 1 & 0.01594 &  &  & 2d$_{5/2}$ 2d$_{3/2}$ &  &  & 0.04066 & 1 & 1 & 4 & 1 & 1 & 1 & -0.01579 \\
 & (2d$_{5/2})^2$ &  &  & -0.04972 &  &  &  &  &  &  &  &  &  & (2d$_{5/2})^2$ &  &  & 0.04272 &  &  &  &  &  &  &  \\
 & (1g$_{7/2})^2$ &  &  & -0.02097 &  &  &  &  &  &  &  &  &  & (1g$_{7/2})^2$ &  &  & -0.00123 &  &  &  &  &  &  &  \\
 &  &  &  &  &  &  &  &  &  &  &  &  &  &  &  &  &  &  &  &  &  &  &  &  \\
 & (1g$_{7/2})^2$ &  &  & -0.02097 &  &  &  &  &  &  &  &  &  & (1g$_{7/2})^2$ &  &  & -0.00123 &  &  &  &  &  &  &  \\
 & 1g$_{7/2}$ 2d$_{5/2}$ &  &  & 0.03446 &  &  &  &  &  &  &  &  &  & 1g$_{7/2}$ 2d$_{5/2}$ &  &  & -0.03308 &  &  &  &  &  &  &  \\
 & 1g$_{7/2}$ 2d$_{3/2}$ &  &  & -0.15039 &  &  &  &  &  &  &  &  &  & 1g$_{7/2}$ 2d$_{3/2}$ &  &  & 0.16567 &  &  &  &  &  &  &  \\
 & (2d$_{5/2})^2$ &  &  & -0.04972 & 1 & 1 & 3 & 3 & 2 & 1 & 0.04009 &  &  & (2d$_{5/2})^2$ &  &  & 0.04272 & 1 & 1 & 3 & 3 & 2 & 1 & -0.06331 \\
 & 2d$_{5/2}$ 2d$_{3/2}$ &  &  & -0.06708 &  &  &  &  &  &  &  &  &  & 2d$_{5/2}$ 2d$_{3/2}$ &  &  & 0.04066 &  &  &  &  &  &  &  \\
 & 2d$_{5/2}$ 3s$_{1/2}$ &  &  & -0.13163 &  &  &  &  &  &  &  &  &  & 2d$_{5/2}$ 3s$_{1/2}$ &  &  & 0.06394 &  &  &  &  &  &  &  \\
 & (2d$_{3/2})^2$ &  &  & -0.12988 &  &  &  &  &  &  &  &  &  & (2d$_{3/2})^2$ &  &  & 0.18526 &  &  &  &  &  &  &  \\
 & 2d$_{3/2}$ 3s$_{1/2}$ &  &  & 0.18734 &  &  &  &  &  &  &  &  &  & 2d$_{3/2}$ 3s$_{1/2}$ &  &  & -0.26432 &  &  &  &  &  &  &  \\
 &  &  &  &  &  &  &  &  &  &  &  &  &  &  &  &  &  &  &  &  &  &  &  &  \\
 & (1g$_{7/2})^2$ &  &  & -0.02097 &  &  &  &  &  &  &  &  &  & (1g$_{7/2})^2$ &  &  & -0.00123 &  &  &  &  &  &  &  \\
 & 1g$_{7/2}$ 2d$_{5/2}$ &  &  & 0.03446 &  &  &  &  &  &  &  &  &  & 1g$_{7/2}$ 2d$_{5/2}$ &  &  & -0.03308 &  &  &  &  &  &  &  \\
 & 1g$_{7/2}$ 2d$_{3/2}$ &  &  & -0.15039 &  &  &  &  &  &  &  &  &  & 1g$_{7/2}$ 2d$_{3/2}$ &  &  & 0.16567 &  &  &  &  &  &  &  \\
 & (2d$_{5/2})^2$ &  &  & -0.04972 & 1 & 0 & 4 & 2 & 2 & 1 & 0.02095 &  &  & (2d$_{5/2})^2$ &  &  & 0.04272 & 1 & 0 & 4 & 2 & 2 & 1 & -0.05525 \\
 & 2d$_{5/2}$ 2d$_{3/2}$ &  &  & -0.06708 &  &  &  &  &  &  &  &  &  & 2d$_{5/2}$ 2d$_{3/2}$ &  &  & 0.04066 &  &  &  &  &  &  &  \\
 & 2d$_{5/2}$ 3s$_{1/2}$ &  &  & -0.13163 &  &  &  &  &  &  &  &  &  & 2d$_{5/2}$ 3s$_{1/2}$ &  &  & 0.06394 &  &  &  &  &  &  &  \\
 & (2d$_{3/2})^2$ &  &  & -0.12988 &  &  &  &  &  &  &  &  &  & (2d$_{3/2})^2$ &  &  & 0.18526 &  &  &  &  &  &  &  \\
 & 2d$_{3/2}$ 3s$_{1/2}$ &  &  & 0.18734 &  &  &  &  &  &  &  &  &  & 2d$_{3/2}$ 3s$_{1/2}$ &  &  & -0.26432 &  &  &  &  &  &  &  \\
 \cline{2-12} \cline{15-25} 
\end{tabular}%
}
\end{table}
\pagebreak

\begin{table}[ht!]
\centering
\setlength{\tabcolsep}{4pt}
\renewcommand{\arraystretch}{0.9}
\label{tab:116Sn_114Sn_2nsa}
\resizebox{\textwidth}{!}{%
\begin{tabular}{llllllllllllllllllllllllll}
\cline{1-12} \cline{14-25}
 & (1g$_{7/2})^2$ &  &  & -0.81250 &  &  &  &  &  &  &  &  &  & (1g$_{7/2})^2$ &  &  & 0.76175 &  &  &  &  &  &  &  &  \\
 & (2d$_{5/2})^2$ &  &  & -0.71206 & 1 & 0 & 5 & 0 & 0 & 0 & -0.52559 &  &  & (2d$_{5/2})^2$ &  &  & 0.53488 & 1 & 0 & 5 & 0 & 0 & 0 & 0.44946 &  \\
 & (2d$_{3/2})^2$ &  &  & -0.51411 &  &  &  &  &  &  &  &  &  & (2d$_{3/2})^2$ &  &  & 0.46881 &  &  &  &  &  &  &  &  \\
 & (3s$_{1/2})^2$ &  &  & -0.35094 &  &  &  &  &  &  &  &  &  & (3s$_{1/2})^2$ &  &  & 0.33271 &  &  &  &  &  &  &  &  \\
 &  &  &  &  &  &  &  &  &  &  &  &  &  &  &  &  &  &  &  &  &  &  &  &  &  \\
 &  &  &  &  & 1 & 0 & 6 & 0 & 0 & 0 & 0.04471 &  &  &  &  &  &  & 1 & 0 & 6 & 0 & 0 & 0 & -0.03728 &  \\
 & \multirow{-2}{*}{(1h$_{11/2})^2$} &  &  & \multirow{-2}{*}{0.96097} & 1 & 1 & 5 & 1 & 1 & 1 & -0.17309 &  &  & \multirow{-2}{*}{(1h$_{11/2})^2$} &  &  & \multirow{-2}{*}{-0.80128} & 1 & 1 & 5 & 1 & 1 & 1 & 0.14432 &  \\
 &  &  &  &  &  &  &  &  &  &  &  &  &  &  &  &  &  &  &  &  &  &  &  &  &  \\
 & (1g$_{7/2})^2$ &  &  & -0.81250 &  &  &  &  &  &  &  &  &  & (1g$_{7/2})^2$ &  &  & 0.76175 &  &  &  &  &  &  &  &  \\
 & (2d$_{5/2})^2$ &  &  & -0.71206 & 1 & 1 & 4 & 1 & 1 & 1 & -0.137 &  &  & (2d$_{5/2})^2$ & \multirow{-10}{*}{0} &  & 0.53488 & 1 & 1 & 4 & 1 & 1 & 1 & 0.165 &  \\
 & (2d$_{3/2})^2$ & \multirow{-11}{*}{0} &  & -0.51411 &  &  &  &  &  &  &  &  &  & (2d$_{3/2})^2$ &  &  & 0.46881 &  &  &  &  &  &  &  &  \\ \cline{5-12} \cline{18-25}
 &  &  &  &  &  &  &  &  &  &  &  &  &  &  &  &  &  &  &  &  &  &  &  &  &  \\
 & (1g$_{7/2})^2$ &  &  & -0.02950 &  &  &  &  &  &  &  &  &  & (1g$_{7/2})^2$ &  &  & 0.01267 &  &  &  &  &  &  &  &  \\
 & 1g$_{7/2}$ 2d$_{5/2}$ &  &  & -0.03394 &  &  &  &  &  &  &  &  &  & 1g$_{7/2}$ 2d$_{5/2}$ &  &  & 0.02786 &  &  &  &  &  &  &  &  \\
 & 1g$_{7/2}$ 2d$_{3/2}$ &  &  & 0.18270 &  &  &  &  &  &  &  &  &  & 1g$_{7/2}$ 2d$_{3/2}$ &  &  & -0.21811 &  &  &  &  &  &  &  &  \\
 & (2d$_{5/2})^2$ &  &  & 0.05061 & 1 & 0 & 4 & 2 & 2 & 0 & 0.08488 &  &  & (2d$_{5/2})^2$ &  &  & -0.01838 & 1 & 0 & 4 & 2 & 2 & 0 & -0.0879 &  \\
 & 2d$_{5/2}$ 2d$_{3/2}$ &  &  & 0.05245 &  &  &  &  &  &  &  &  &  & 2d$_{5/2}$ 2d$_{3/2}$ &  &  & -0.03900 &  &  &  &  &  &  &  &  \\
 & 2d$_{5/2}$ 3s$_{1/2}$ &  &  & 0.14257 &  &  &  &  &  &  &  &  &  & 2d$_{5/2}$ 3s$_{1/2}$ &  &  & -0.06659 &  &  &  &  &  &  &  &  \\
 & (2d$_{3/2})^2$ &  &  & 0.04303 &  &  &  &  &  &  &  &  &  & (2d$_{3/2})^2$ &  &  & -0.10720 &  &  &  &  &  &  &  &  \\
 & 2d$_{3/2}$ 3s$_{1/2}$ &  &  & -0.09615 &  &  &  &  &  &  &  &  &  & 2d$_{3/2}$ 3s$_{1/2}$ &  &  & 0.17300 &  &  &  &  &  &  &  &  \\
 &  &  &  &  &  &  &  &  &  &  &  &  &  &  &  &  &  &  &  &  &  &  &  &  &  \\
 &  &  &  &  & 1 & 0 & 5 & 2 & 2 & 0 & 0.00254 &  &  &  &  &  &  & 1 & 0 & 5 & 2 & 2 & 0 & 0.00241 &  \\
 & (1h$_{11/2})^2$ &  &  & 0.09976 & 1 & 1 & 5 & 1 & 1 & 1 & 0.00844 &  &  & (1h$_{11/2})^2$ &  &  & 0.09466 & 1 & 1 & 5 & 1 & 1 & 1 & 0.00801 &  \\
 &  &  &  &  & 1 & 1 & 4 & 3 & 3 & 1 & -0.00497 &  &  &  &  &  &  & 1 & 1 & 4 & 3 & 3 & 1 & -0.00472 &  \\
 &  &  &  &  &  &  &  &  &  &  &  &  &  &  &  &  &  &  &  &  &  &  &  &  &  \\
 & (1g$_{7/2})^2$ &  &  & -0.02950 &  &  &  &  &  &  &  &  &  & (1g$_{7/2})^2$ &  &  & 0.01267 &  &  &  &  &  &  &  &  \\
 & 1g$_{7/2}$ 2d$_{5/2}$ &  &  & -0.03394 &  &  &  &  &  &  &  &  &  & 1g$_{7/2}$ 2d$_{5/2}$ &  &  & 0.02786 &  &  &  &  &  &  &  &  \\
 & 1g$_{7/2}$ 2d$_{3/2}$ &  &  & 0.18270 &  &  &  &  &  &  &  &  &  & 1g$_{7/2}$ 2d$_{3/2}$ &  &  & -0.21811 &  &  &  &  &  &  &  &  \\
 & (2d$_{5/2})^2$ &  &  & 0.05061 & 1 & 1 & 4 & 1 & 2 & 1 & -0.01693 &  &  & (2d$_{5/2})^2$ &  &  & -0.01838 & 1 & 1 & 4 & 1 & 2 & 1 & 0.03259 &  \\
 & 2d$_{5/2}$ 2d$_{3/2}$ &  &  & 0.05245 &  &  &  &  &  &  &  &  &  & 2d$_{5/2}$ 2d$_{3/2}$ &  &  & -0.03900 &  &  &  &  &  &  &  &  \\
 & 2d$_{5/2}$ 3s$_{1/2}$ &  &  & 0.14257 &  &  &  &  &  &  &  &  &  & 2d$_{5/2}$ 3s$_{1/2}$ &  &  & -0.06659 &  &  &  &  &  &  &  &  \\
 & (2d$_{3/2})^2$ &  &  & 0.04303 &  &  &  &  &  &  &  &  &  & (2d$_{3/2})^2$ &  &  & -0.10720 &  &  &  &  &  &  &  &  \\
 & 2d$_{3/2}$ 3s$_{1/2}$ &  &  & -0.09615 &  &  &  &  &  &  &  &  &  & 2d$_{3/2}$ 3s$_{1/2}$ &  &  & 0.17300 &  &  &  &  &  &  &  &  \\
 &  &  &  &  &  &  &  &  &  &  &  &  &  &  &  &  &  &  &  &  &  &  &  &  &  \\
 & (2d$_{3/2})^2$ &  &  & 0.04303 &  &  &  &  &  &  &  &  &  & (2d$_{3/2})^2$ &  &  & -0.10720 &  &  &  &  &  &  &  &  \\
 & 2d$_{5/2}$ 2d$_{3/2}$ &  &  & 0.05245 & 1 & 1 & 4 & 1 & 1 & 1 & 0.00164 &  &  & 2d$_{5/2}$ 2d$_{3/2}$ &  &  & -0.03900 & 1 & 1 & 4 & 1 & 1 & 1 & 0.014 &  \\
 & (2d$_{5/2})^2$ &  &  & 0.05061 &  &  &  &  &  &  &  &  &  & (2d$_{5/2})^2$ &  &  & -0.01838 &  &  &  &  &  &  &  &  \\
 & (1g$_{7/2})^2$ &  &  & -0.02950 &  &  &  &  &  &  &  &  &  & (1g$_{7/2})^2$ &  &  & 0.01267 &  &  &  &  &  &  &  &  \\
 &  &  &  &  &  &  &  &  &  &  &  &  &  &  &  &  &  &  &  &  &  &  &  &  &  \\
 & (1g$_{7/2})^2$ &  &  & -0.02950 &  &  &  &  &  &  &  &  &  & (1g$_{7/2})^2$ &  &  & 0.01267 &  &  &  &  &  &  &  &  \\
 & 1g$_{7/2}$ 2d$_{5/2}$ &  &  & -0.03394 &  &  &  &  &  &  &  &  &  & 1g$_{7/2}$ 2d$_{5/2}$ &  &  & 0.02786 &  &  &  &  &  &  &  &  \\
 & 1g$_{7/2}$ 2d$_{3/2}$ &  &  & 0.18270 &  &  &  &  &  &  &  &  &  & 1g$_{7/2}$ 2d$_{3/2}$ &  &  & -0.21811 &  &  &  &  &  &  &  &  \\
 & (2d$_{5/2})^2$ &  &  & 0.05061 & 1 & 1 & 3 & 3 & 2 & 1 & -0.02856 &  &  & (2d$_{5/2})^2$ &  &  & -0.01838 & 1 & 1 & 3 & 3 & 2 & 1 & 0.05494 &  \\
 & 2d$_{5/2}$ 2d$_{3/2}$ &  &  & 0.05245 &  &  &  &  &  &  &  &  &  & 2d$_{5/2}$ 2d$_{3/2}$ &  &  & -0.03900 &  &  &  &  &  &  &  &  \\
 & 2d$_{5/2}$ 3s$_{1/2}$ &  &  & 0.14257 &  &  &  &  &  &  &  &  &  & 2d$_{5/2}$ 3s$_{1/2}$ &  &  & -0.06659 &  &  &  &  &  &  &  &  \\
 & (2d$_{3/2})^2$ &  &  & 0.04303 &  &  &  &  &  &  &  &  &  & (2d$_{3/2})^2$ &  &  & -0.10720 &  &  &  &  &  &  &  &  \\
 & 2d$_{3/2}$ 3s$_{1/2}$ &  &  & -0.09615 &  &  &  &  &  &  &  &  &  & 2d$_{3/2}$ 3s$_{1/2}$ &  &  & 0.17300 &  &  &  &  &  &  &  &  \\
 &  &  &  &  &  &  &  &  &  &  &  &  &  &  &  &  &  &  &  &  &  &  &  &  &  \\
 & (1g$_{7/2})^2$ &  &  & -0.02950 &  &  &  &  &  &  &  &  &  & (1g$_{7/2})^2$ &  &  & 0.01267 &  &  &  &  &  &  &  &  \\
 & 1g$_{7/2}$ 2d$_{5/2}$ &  &  & -0.03394 &  &  &  &  &  &  &  &  &  & 1g$_{7/2}$ 2d$_{5/2}$ &  &  & 0.02786 &  &  &  &  &  &  &  &  \\
 & 1g$_{7/2}$ 2d$_{3/2}$ &  &  & 0.18270 &  &  &  &  &  &  &  &  &  & 1g$_{7/2}$ 2d$_{3/2}$ &  &  & -0.21811 &  &  &  &  &  &  &  &  \\
 & (2d$_{5/2})^2$ &  &  & 0.05061 & 1 & 0 & 4 & 2 & 2 & 1 & -0.00458 &  &  & (2d$_{5/2})^2$ &  &  & -0.01838 & 1 & 0 & 4 & 2 & 2 & 1 & 0.03921 &  \\
 & 2d$_{5/2}$ 2d$_{3/2}$ &  &  & 0.05245 &  &  &  &  &  &  &  &  &  & 2d$_{5/2}$ 2d$_{3/2}$ &  &  & -0.03900 &  &  &  &  &  &  &  &  \\
 & 2d$_{5/2}$ 3s$_{1/2}$ &  &  & 0.14257 &  &  &  &  &  &  &  &  &  & 2d$_{5/2}$ 3s$_{1/2}$ &  &  & -0.06659 &  &  &  &  &  &  &  &  \\
 & (2d$_{3/2})^2$ &  &  & 0.04303 &  &  &  &  &  &  &  &  &  & (2d$_{3/2})^2$ & \multirow{-44}{*}{2} &  & -0.10720 &  &  &  &  &  &  &  &  \\
 & 2d$_{3/2}$ 3s$_{1/2}$ & \multirow{-44}{*}{2} &  & -0.09615 &  &  &  &  &  &  &  & \multirow{-44}{*}{} &  & 2d$_{3/2}$ 3s$_{1/2}$ &  &  & 0.17300 &  &  &  &  &  &  &  &  \\ \cline{5-12} \cline{18-25} \\
 & (1g$_{7/2})^2$ &  &  & 0.09096 &  &  &  &  &  &  &  &  &  & (1g$_{7/2})^2$ &  &  & -0.04641 &  &  &  &  &  &  &  &  \\
 & 1g$_{7/2}$ 2d$_{5/2}$ &  &  & -0.06563 &  &  &  &  &  &  &  &  &  & 1g$_{7/2}$ 2d$_{5/2}$ &  &  & 0.05748 &  &  &  &  &  &  &  &  \\
 & 1g$_{7/2}$ 2d$_{3/2}$ &  &  & 0.09921 & 1 & 0 & 3 & 4 & 4 & 0 & 0.07903 &  &  & 1g$_{7/2}$ 2d$_{3/2}$ &  &  & -0.18193 & 1 & 0 & 3 & 4 & 4 & 0 & -0.10559 &  \\
 & 1g$_{7/2}$ 3s$_{1/2}$ &  &  & -0.10074 &  &  &  &  &  &  &  &  &  & 1g$_{7/2}$ 3s$_{1/2}$ &  &  & 0.18206 &  &  &  &  &  &  &  &  \\
 & (2d$_{5/2})^2$ &  &  & 0.05027 &  &  &  &  &  &  &  &  &  & (2d$_{5/2})^2$ &  &  & -0.03287 &  &  &  &  &  &  &  &  \\
 & 2d$_{5/2}$ 2d$_{3/2}$ &  &  & 0.17287 &  &  &  &  &  &  &  &  &  & 2d$_{5/2}$ 2d$_{3/2}$ &  &  & -0.23647 &  &  &  &  &  &  &  &  \\
 &  &  &  &  &  &  &  &  &  &  &  &  &  &  &  &  &  &  &  &  &  &  &  &  &  \\
 & (1g$_{7/2})^2$ &  &  & 0.09096 &  &  &  &  &  &  &  &  &  & (1g$_{7/2})^2$ &  &  & -0.04641 &  &  &  &  &  &  &  &  \\
 & 1g$_{7/2}$ 2d$_{5/2}$ &  &  & -0.06563 &  &  &  &  &  &  &  &  &  & 1g$_{7/2}$ 2d$_{5/2}$ &  &  & 0.05748 &  &  &  &  &  &  &  &  \\
 & 1g$_{7/2}$ 2d$_{3/2}$ &  &  & 0.09921 & 1 & 0 & 3 & 4 & 4 & 1 & 0.01944 &  &  & 1g$_{7/2}$ 2d$_{3/2}$ &  &  & -0.18193 & 1 & 0 & 3 & 4 & 4 & 1 & -0.02229 &  \\
 & 1g$_{7/2}$ 3s$_{1/2}$ &  &  & -0.10074 &  &  &  &  &  &  &  &  &  & 1g$_{7/2}$ 3s$_{1/2}$ &  &  & 0.18206 &  &  &  &  &  &  &  &  \\
 & (2d$_{5/2})^2$ &  &  & 0.05027 &  &  &  &  &  &  &  &  &  & (2d$_{5/2})^2$ &  &  & -0.03287 &  &  &  &  &  &  &  &  \\
 & 2d$_{5/2}$ 2d$_{3/2}$ &  &  & 0.17287 &  &  &  &  &  &  &  &  &  & 2d$_{5/2}$ 2d$_{3/2}$ &  &  & -0.23647 &  &  &  &  &  &  &  &  \\
 &  &  &  &  &  &  &  &  &  &  &  &  &  &  &  &  &  &  &  &  &  &  &  &  &  \\
 & (1g$_{7/2})^2$ &  &  & 0.09096 &  &  &  &  &  &  &  &  &  & (1g$_{7/2})^2$ &  &  & -0.04641 &  &  &  &  &  &  &  &  \\
 & 1g$_{7/2}$ 2d$_{5/2}$ &  &  & -0.06563 &  &  &  &  &  &  &  &  &  & 1g$_{7/2}$ 2d$_{5/2}$ &  &  & 0.05748 &  &  &  &  &  &  &  &  \\
 & 1g$_{7/2}$ 2d$_{3/2}$ &  &  & 0.09921 & 1 & 1 & 3 & 3 & 4 & 1 & -0.01825 &  &  & 1g$_{7/2}$ 2d$_{3/2}$ &  &  & -0.18193 & 1 & 1 & 3 & 3 & 4 & 1 & 0.02898 &  \\
 & 1g$_{7/2}$ 3s$_{1/2}$ &  &  & -0.10074 &  &  &  &  &  &  &  &  &  & 1g$_{7/2}$ 3s$_{1/2}$ &  &  & 0.18206 &  &  &  &  &  &  &  &  \\
 & (2d$_{5/2})^2$ &  &  & 0.05027 &  &  &  &  &  &  &  &  &  & (2d$_{5/2})^2$ &  &  & -0.03287 &  &  &  &  &  &  &  &  \\
 & 2d$_{5/2}$ 2d$_{3/2}$ &  &  & 0.17287 &  &  &  &  &  &  &  &  &  & 2d$_{5/2}$ 2d$_{3/2}$ &  &  & -0.23647 &  &  &  &  &  &  &  &  \\
 &  &  &  &  &  &  &  &  &  &  &  &  &  &  &  &  &  &  &  &  &  &  &  &  &  \\
 & (1g$_{7/2})^2$ &  &  & 0.09096 &  &  &  &  &  &  &  &  &  & (1g$_{7/2})^2$ &  &  & -0.04641 &  &  &  &  &  &  &  &  \\
 & 1g$_{7/2}$ 2d$_{5/2}$ &  &  & -0.06563 &  &  &  &  &  &  &  &  &  & 1g$_{7/2}$ 2d$_{5/2}$ &  &  & 0.05748 &  &  &  &  &  &  &  &  \\
 & 1g$_{7/2}$ 2d$_{3/2}$ &  &  & 0.09921 & 1 & 1 & 2 & 5 & 4 & 1 & -0.0369 &  &  & 1g$_{7/2}$ 2d$_{3/2}$ &  &  & -0.18193 & 1 & 1 & 2 & 5 & 4 & 1 & 0.05859 &  \\
 & 1g$_{7/2}$ 3s$_{1/2}$ &  &  & -0.10074 &  &  &  &  &  &  &  &  &  & 1g$_{7/2}$ 3s$_{1/2}$ &  &  & 0.18206 &  &  &  &  &  &  &  &  \\
 & (2d$_{5/2})^2$ &  &  & 0.05027 &  &  &  &  &  &  &  &  &  & (2d$_{5/2})^2$ &  &  & -0.03287 &  &  &  &  &  &  &  &  \\
 & 2d$_{5/2}$ 2d$_{3/2}$ &  &  & 0.17287 &  &  &  &  &  &  &  &  &  & 2d$_{5/2}$ 2d$_{3/2}$ &  &  & -0.23647 &  &  &  &  &  &  &  &  \\
 &  &  &  &  &  &  &  &  &  &  &  &  &  &  &  &  &  &  &  &  &  &  &  &  &  \\
 & (1g$_{7/2})^2$ &  &  & 0.09096 &  &  &  &  &  &  &  &  &  & (1g$_{7/2})^2$ &  &  & -0.04641 &  &  &  &  &  &  &  &  \\
 & 1g$_{7/2}$ 2d$_{5/2}$ &  &  & -0.06563 &  &  &  &  &  &  &  &  &  & 1g$_{7/2}$ 2d$_{5/2}$ &  &  & 0.05748 &  &  &  &  &  &  &  &  \\
 & 1g$_{7/2}$ 2d$_{3/2}$ &  &  & 0.09921 & 1 & 1 & 3 & 3 & 3 & 1 & -0.01623 &  &  & 1g$_{7/2}$ 2d$_{3/2}$ &  &  & -0.18193 & 1 & 1 & 3 & 3 & 3 & 1 & 0.0294 &  \\
 & (2d$_{5/2})^2$ &  &  & 0.05027 &  &  &  &  &  &  &  &  &  & (2d$_{5/2})^2$ &  &  & -0.03287 &  &  &  &  &  &  &  &  \\
 & 2d$_{5/2}$ 2d$_{3/2}$ &  &  & 0.17287 &  &  &  &  &  &  &  &  &  & 2d$_{5/2}$ 2d$_{3/2}$ &  &  & -0.23647 &  &  &  &  &  &  &  &  \\
 &  &  &  &  &  &  &  &  &  &  &  &  &  &  &  &  &  &  &  &  &  &  &  &  &  \\
 & (1g$_{7/2})^2$ &  &  & 0.09096 &  &  &  &  &  &  &  &  &  & (1g$_{7/2})^2$ &  &  & -0.04641 &  &  &  &  &  &  &  &  \\
 & 1g$_{7/2}$ 2d$_{5/2}$ &  &  & -0.06563 & 1 & 1 & 2 & 5 & 5 & 1 & 0.03712 &  &  & 1g$_{7/2}$ 2d$_{5/2}$ &  &  & 0.05748 & 1 & 1 & 2 & 5 & 5 & 1 & -0.05179 &  \\
 & 1g$_{7/2}$ 2d$_{3/2}$ &  &  & 0.09921 &  &  &  &  &  &  &  &  &  & 1g$_{7/2}$ 2d$_{3/2}$ & \multirow{-38}{*}{4} &  & -0.18193 &  &  &  &  &  &  &  &  \\
 &  & \multirow{-38}{*}{4} &  &  &  &  &  &  &  &  &  &  &  &  &  &  &  &  &  &  &  &  &  &  &  \\
 &  &  &  &  & 1 & 0 & 4 & 4 & 4 & 0 & -0.00571 &  &  &  &  &  &  & 1 & 0 & 4 & 4 & 4 & 0 & 0.00124 &  \\
 &  &  &  &  & 1 & 0 & 4 & 4 & 5 & 1 & 0 &  &  &  &  &  &  & 1 & 0 & 4 & 4 & 5 & 1 & 0 &  \\
\multirow{-103}{*}{$^{116}$Sn$_{1.290}$(2$^{+}$)} & \multirow{-3}{*}{(1h$_{11/2})^2$} &  & \multirow{-103}{*}{$^{114}$Sn$_{1.299}$(2$^{+}$)} &  & 1 & 1 & 3 & 5 & 5 & 1 & 0.00953 &  & \multirow{-103}{*}{$^{116}$Sn$_{1.290}$(2$^{+}$)} & \multirow{-3}{*}{(1h$_{11/2})^2$} &  & \multirow{-103}{*}{$^{114}$Sn$_{1.299}$(2$^{+}$)} &  & 1 & 1 & 3 & 5 & 5 & 1 & -0.00206 &  \\
 &  &  &  & \multirow{-4}{*}{-0.21766} & 1 & 1 & 4 & 3 & 3 & 1 & -0.01703 &  &  &  &  &  & \multirow{-4}{*}{0.04694} & 1 & 1 & 4 & 3 & 3 & 1 & 0.00368 &  \\ \cline{1-25}
\end{tabular}%
}
\end{table}

\pagebreak
\begin{table}[ht!]
\centering
\setlength{\tabcolsep}{6pt}
\renewcommand{\arraystretch}{0.9}
\caption{Cluster spectroscopic amplitude ($\mathcal{S}$ (c.m.)) for
di-proton transfer involving the heavier nucleus $^{116}$Sn are presented
for the relevant overlaps. The calculations were performed using
\textsc{kshell} with the \texttt{SN100PN} and the \texttt{Monopole}
effective interactions. Here, $j_1$ and $j_2$ stand for the total
angular momenta of nucleons 1 and 2, respectively. The quantum numbers
$n$, $\ell$ ($N$, $L$) represent the principal quantum numbers and
orbital angular momenta of the nucleons relative to each other
(and to the core), respectively. $J$ and $\Lambda$ denote the total
angular momentum and the orbital angular momentum of the cluster
with respect to the core, respectively, and $S$ is the intrinsic
spin of the two-proton cluster. The spins and parities of the
initial and final nuclei are given by $I^{\pi}_{\mathrm{i}}$
and $I^{\pi}_{\mathrm{f}}$, respectively.}

\label{tab:116Sn_118Te_2p}
\resizebox{\textwidth}{!}{%
\begin{tabular}{lllllllllllllllllllllllll}
\cmidrule{1-12} \cmidrule{14-25}
$I^\pi_\text{i}$ & $j_1$$j_2$ & $J$ & $I^\pi_\text{f}$  & $\mathcal{S}$ ($j-j$) \texttt{(SN100PN)} & $n$ & $\ell$ & $N$ & $L$ & $\Lambda$ & $S$ & $\mathcal{S}$ (c.m.) \texttt{(SN100PN)} & & $I^\pi_\text{i}$ & $j_1$$j_2$ & $J$ & $I^\pi_\text{f}$ & $\mathcal{S}$ ($j-j$)\texttt{(Monopole)} & $n$ & $\ell$ & $N$ & $L$ & $\Lambda$ & $S$ & $\mathcal{S}$ (c.m.) \texttt{(Monopole)}\\ \cmidrule{1-12} \cmidrule{14-25}

 & (1g$_{7/2})^2$ & \multirow{7}{*}{0} &  & -0.22802 &  &  &  &  &  &  &  &  &  & (1g$_{7/2})^2$ & \multirow{7}{*}{0} & \multirow{6}{*}{$^{118}$Te$_\text{g.s.}$(0$^{+}$)} & -0.22357 &  &  &  &  &  &  &  \\
\multirow{55}{*}{$^{116}$Sn$_\text{g.s.}$(0$^{+}$)} & (2d$_{5/2})^2$ &  & \multirow{6}{*}{$^{118}$Te$_\text{g.s.}$(0$^{+}$)} & -0.15238 & 1 & 0 & 5 & 0 & 0 & 0 & -0.10008 &  & \multirow{55}{*}{$^{116}$Sn$_\text{g.s.}$(0$^{+}$)} & (2d$_{5/2})^2$ &  &  & -0.39871 & 1 & 0 & 5 & 0 & 0 & 0 & -0.22735 \\
 & (2d$_{3/2})^2$ &  &  & -0.07377 &  &  &  &  &  &  &  &  &  & (2d$_{3/2})^2$ &  &  & -0.09972 &  &  &  &  &  &  &  \\
 & (3s$_{1/2})^2$ &  &  & -0.05105 &  &  &  &  &  &  &  &  &  & (3s$_{1/2})^2$ &  &  & -0.19363 &  &  &  &  &  &  &  \\
 &  &  &  &  &  &  &  &  &  &  &  &  &  &  &  &  &  &  &  &  &  &  &  &  \\
 & \multirow{2}{*}{(1h$_{11/2})^2$} &  &  & \multirow{2}{*}{0} & 1 & 0 & 6 & 0 & 0 & 0 & 0 &  &  & \multirow{2}{*}{(1h$_{11/2})^2$} &  &  & \multirow{2}{*}{0} & 1 & 0 & 6 & 0 & 0 & 0 & 0 \\
 &  &  &  &  & 1 & 1 & 5 & 1 & 1 & 1 & 0 &  &  &  &  &  &  & 1 & 1 & 5 & 1 & 1 & 1 & 0 \\
 &  &  &  &  &  &  &  &  &  &  &  &  &  &  &  &  &  &  &  &  &  &  &  &  \\
 & (1g$_{7/2})^2$ &  &  & -0.22802 &  &  &  &  &  &  &  &  &  & (1g$_{7/2})^2$ &  &  & -0.22357 &  &  &  &  &  &  &  \\
 & (2d$_{5/2})^2$ &  &  & -0.15238 & 1 & 1 & 4 & 1 & 1 & 1 & -0.027 &  &  & (2d$_{5/2})^2$ &  &  & -0.39871 & 1 & 1 & 4 & 1 & 1 & 1 & 0.042 \\
 & (2d$_{3/2})^2$ &  &  & -0.07377 &  &  &  &  &  &  &  &  &  & (2d$_{3/2})^2$ &  &  & -0.09972 &  &  &  &  &  &  &  \\ \cline{2-12} \cline{15-25} 
 &  &  &  &  &  &  &  &  &  &  &  &  &  &  &  &  &  &  &  &  &  &  &  &  \\
 & (1g$_{7/2})^2$ & 2 & \multirow{43}{*}{$^{118}$Te$_{0.605}$(2$^{+}$)} & 0.12687 &  &  &  &  &  &  &  &  &  & (1g$_{7/2})^2$ &  & \multirow{43}{*}{$^{118}$Te$_{0.605}$(2$^{+}$)} & -0.11182 &  &  &  &  &  &  &  \\
 & 1g$_{7/2}$ 2d$_{5/2}$ &  &  & -0.01516 &  &  &  &  &  &  &  &  &  & 1g$_{7/2}$ 2d$_{5/2}$ &  &  & 0.05060 &  &  &  &  &  &  &  \\
 & 1g$_{7/2}$ 2d$_{3/2}$ &  &  & 0.03077 &  &  &  &  &  &  &  &  &  & 1g$_{7/2}$ 2d$_{3/2}$ &  &  & -0.04848 &  &  &  &  &  &  &  \\
 & (2d$_{5/2})^2$ &  &  & 0.04994 & 1 & 0 & 4 & 2 & 2 & 0 & 0.03203 &  &  & (2d$_{5/2})^2$ &  &  & -0.17506 & 1 & 0 & 4 & 2 & 2 & 0 & -0.08715 \\
 & 2d$_{5/2}$ 2d$_{3/2}$ &  &  & 0.01953 &  &  &  &  &  &  &  &  &  & 2d$_{5/2}$ 2d$_{3/2}$ &  &  & -0.03419 &  &  &  &  &  &  &  \\
 & 2d$_{5/2}$ 3s$_{1/2}$ &  &  & 0.03039 &  &  &  &  &  &  &  &  &  & 2d$_{5/2}$ 3s$_{1/2}$ & 2 &  & -0.12998 &  &  &  &  &  &  &  \\
 & (2d$_{3/2})^2$ &  &  & 0.01755 &  &  &  &  &  &  &  &  &  & (2d$_{3/2})^2$ &  &  & -0.02407 &  &  &  &  &  &  &  \\
 & 2d$_{3/2}$ 3s$_{1/2}$ &  &  & -0.01915 &  &  &  &  &  &  &  &  &  & 2d$_{3/2}$ 3s$_{1/2}$ &  &  & 0.06163 &  &  &  &  &  &  &  \\
 &  &  &  &  &  &  &  &  &  &  &  &  &  &  &  &  &  &  &  &  &  &  &  &  \\
 &  &  &  &  & 1 & 0 & 5 & 2 & 2 & 0 & 0 &  &  &  &  &  &  & 1 & 0 & 5 & 2 & 2 & 0 & 0 \\
 & (1h$_{11/2})^2$ &  &  & 0 & 1 & 1 & 5 & 1 & 1 & 1 & 0 &  &  & (1h$_{11/2})^2$ &  &  & 0 & 1 & 1 & 5 & 1 & 1 & 1 & 0 \\
 &  &  &  &  & 1 & 1 & 4 & 3 & 3 & 1 & 0 &  &  &  &  &  &  & 1 & 1 & 4 & 3 & 3 & 1 & 0 \\
 &  &  &  &  &  &  &  &  &  &  &  &  &  &  &  &  &  &  &  &  &  &  &  &  \\
 & (1g$_{7/2})^2$ &  &  & 0.12687 &  &  &  &  &  &  &  &  &  & (1g$_{7/2})^2$ &  &  & -0.11182 &  &  &  &  &  &  &  \\
 & 1g$_{7/2}$ 2d$_{5/2}$ &  &  & -0.01516 &  &  &  &  &  &  &  &  &  & 1g$_{7/2}$ 2d$_{5/2}$ &  &  & 0.05060 &  &  &  &  &  &  &  \\
 & 1g$_{7/2}$ 2d$_{3/2}$ &  &  & 0.03077 &  &  &  &  &  &  &  &  &  & 1g$_{7/2}$ 2d$_{3/2}$ &  &  & -0.04848 &  &  &  &  &  &  &  \\
 & (2d$_{5/2})^2$ &  &  & 0.04994 & 1 & 1 & 4 & 1 & 2 & 1 & -0.00358 &  &  & (2d$_{5/2})^2$ &  &  & -0.17506 & 1 & 1 & 4 & 1 & 2 & 1 & 0.00488 \\
 & 2d$_{5/2}$ 2d$_{3/2}$ &  &  & 0.01953 &  &  &  &  &  &  &  &  &  & 2d$_{5/2}$ 2d$_{3/2}$ &  &  & -0.03419 &  &  &  &  &  &  &  \\
 & 2d$_{5/2}$ 3s$_{1/2}$ &  &  & 0.03039 &  &  &  &  &  &  &  &  &  & 2d$_{5/2}$ 3s$_{1/2}$ &  &  & -0.12998 &  &  &  &  &  &  &  \\
 & (2d$_{3/2})^2$ &  &  & 0.01755 &  &  &  &  &  &  &  &  &  & (2d$_{3/2})^2$ &  &  & -0.02407 &  &  &  &  &  &  &  \\
 & 2d$_{3/2}$ 3s$_{1/2}$ &  &  & -0.01915 &  &  &  &  &  &  &  &  &  & 2d$_{3/2}$ 3s$_{1/2}$ &  &  & 0.06163 &  &  &  &  &  &  &  \\
 &  &  &  &  &  &  &  &  &  &  &  &  &  &  &  &  &  &  &  &  &  &  &  &  \\
 & (2d$_{3/2})^2$ &  &  & 0.01755 &  &  &  &  &  &  &  &  &  & (2d$_{3/2})^2$ &  &  & -0.02407 &  &  &  &  &  &  &  \\
 & 2d$_{5/2}$ 2d$_{3/2}$ &  &  & 0.01953 & 1 & 1 & 4 & 1 & 1 & 1 & -0.0012 &  &  & 2d$_{5/2}$ 2d$_{3/2}$ &  &  & -0.03419 & 1 & 1 & 4 & 1 & 1 & 1 & -0.03691 \\
 & (2d$_{5/2})^2$ &  &  & 0.04994 &  &  &  &  &  &  &  &  &  & (2d$_{5/2})^2$ &  &  & -0.17506 &  &  &  &  &  &  &  \\
 & (1g$_{7/2})^2$ &  &  & 0.12687 &  &  &  &  &  &  &  &  &  & (1g$_{7/2})^2$ &  &  & -0.11182 &  &  &  &  &  &  &  \\
 &  &  &  &  &  &  &  &  &  &  &  &  &  &  &  &  &  &  &  &  &  &  &  &  \\
 & (1g$_{7/2})^2$ &  &  & 0.12687 &  &  &  &  &  &  &  &  &  & (1g$_{7/2})^2$ &  &  & -0.11182 &  &  &  &  &  &  &  \\
 & 1g$_{7/2}$ 2d$_{5/2}$ &  &  & -0.01516 &  &  &  &  &  &  &  &  &  & 1g$_{7/2}$ 2d$_{5/2}$ &  &  & 0.05060 &  &  &  &  &  &  &  \\
 & 1g$_{7/2}$ 2d$_{3/2}$ &  &  & 0.03077 &  &  &  &  &  &  &  &  &  & 1g$_{7/2}$ 2d$_{3/2}$ &  &  & -0.04848 &  &  &  &  &  &  &  \\
 & (2d$_{5/2})^2$ &  &  & 0.04994 & 1 & 1 & 3 & 3 & 2 & 1 & -0.00603 &  &  & (2d$_{5/2})^2$ &  &  & -0.17506 & 1 & 1 & 3 & 3 & 2 & 1 & 0.00822 \\
 & 2d$_{5/2}$ 2d$_{3/2}$ &  &  & 0.01953 &  &  &  &  &  &  &  &  &  & 2d$_{5/2}$ 2d$_{3/2}$ &  &  & -0.03419 &  &  &  &  &  &  &  \\
 & 2d$_{5/2}$ 3s$_{1/2}$ &  &  & 0.03039 &  &  &  &  &  &  &  &  &  & 2d$_{5/2}$ 3s$_{1/2}$ &  &  & -0.12998 &  &  &  &  &  &  &  \\
 & (2d$_{3/2})^2$ &  &  & 0.01755 &  &  &  &  &  &  &  &  &  & (2d$_{3/2})^2$ &  &  & -0.02407 &  &  &  &  &  &  &  \\
 & 2d$_{3/2}$ 3s$_{1/2}$ &  &  & -0.01915 &  &  &  &  &  &  &  &  &  & 2d$_{3/2}$ 3s$_{1/2}$ &  &  & 0.06163 &  &  &  &  &  &  &  \\
 &  &  &  &  &  &  &  &  &  &  &  &  &  &  &  &  &  &  &  &  &  &  &  &  \\
 & (1g$_{7/2})^2$ &  &  & 0.12687 &  &  &  &  &  &  &  &  &  & (1g$_{7/2})^2$ &  &  & -0.11182 &  &  &  &  &  &  &  \\
 & 1g$_{7/2}$ 2d$_{5/2}$ &  &  & -0.01516 &  &  &  &  &  &  &  &  &  & 1g$_{7/2}$ 2d$_{5/2}$ &  &  & 0.05060 &  &  &  &  &  &  &  \\
 & 1g$_{7/2}$ 2d$_{3/2}$ &  &  & 0.03077 &  &  &  &  &  &  &  &  &  & 1g$_{7/2}$ 2d$_{3/2}$ &  &  & -0.04848 &  &  &  &  &  &  &  \\
 & (2d$_{5/2})^2$ &  &  & 0.04994 & 1 & 0 & 4 & 2 & 2 & 1 & 0.00071 &  &  & (2d$_{5/2})^2$ &  &  & -0.17506 & 1 & 0 & 4 & 2 & 2 & 1 & -0.00683 \\
 & 2d$_{5/2}$ 2d$_{3/2}$ &  &  & 0.01953 &  &  &  &  &  &  &  &  &  & 2d$_{5/2}$ 2d$_{3/2}$ &  &  & -0.03419 &  &  &  &  &  &  &  \\
 & 2d$_{5/2}$ 3s$_{1/2}$ &  &  & 0.03039 &  &  &  &  &  &  &  &  &  & 2d$_{5/2}$ 3s$_{1/2}$ &  &  & -0.12998 &  &  &  &  &  &  &  \\
 & (2d$_{3/2})^2$ &  &  & 0.01755 &  &  &  &  &  &  &  &  &  & (2d$_{3/2})^2$ &  &  & -0.02407 &  &  &  &  &  &  &  \\
 & 2d$_{3/2}$ 3s$_{1/2}$ &  &  & -0.01915 &  &  &  &  &  &  &  &  &  & 2d$_{3/2}$ 3s$_{1/2}$ &  &  & 0.06163 &  &  &  &  &  &  &  \\ \cline{1-12} \cline{14-25} 
 &  &  &  &  &  &  &  &  &  &  &  &  &  &  &  &  &  &  &  &  &  &  &  &  \\
 &  &  &  &  &  &  &  &  &  &  &  &  &  &  &  &  &  &  &  &  &  &  &  &  \\
\multirow{51}{*}{$^{116}$Sn$_{1.290}$(2$^{+}$)} & (1g$_{7/2})^2$ & \multirow{51}{*}{2} & \multirow{44}{*}{$^{118}$Te$_{g.s,}$(0$^{+}$)} & -0.23110 &  &  &  &  &  &  &  &  & \multirow{51}{*}{$^{116}$Sn$_{1.290}$(2$^{+}$)} & (1g$_{7/2})^2$ & \multirow{51}{*}{2} & \multirow{44}{*}{$^{118}$Te$_{g.s,}$(0$^{+}$)} & -0.20674 &  &  &  &  &  &  &  \\
 & 1g$_{7/2}$ 2d$_{5/2}$ &  &  & 0.03605 &  &  &  &  &  &  &  &  &  & 1g$_{7/2}$ 2d$_{5/2}$ &  &  & 0.08747 &  &  &  &  &  &  &  \\
 & 1g$_{7/2}$ 2d$_{3/2}$ &  &  & -0.05922 &  &  &  &  &  &  &  &  &  & 1g$_{7/2}$ 2d$_{3/2}$ &  &  & -0.07654 &  &  &  &  &  &  &  \\
 & (2d$_{5/2})^2$ &  &  & -0.09710 & 1 & 0 & 4 & 2 & 2 & 0 & -0.06215 &  &  & (2d$_{5/2})^2$ &  &  & -0.28340 & 1 & 0 & 4 & 2 & 2 & 0 & -0.14267 \\
 & 2d$_{5/2}$ 2d$_{3/2}$ &  &  & -0.04011 &  &  &  &  &  &  &  &  &  & 2d$_{5/2}$ 2d$_{3/2}$ &  &  & -0.06372 &  &  &  &  &  &  &  \\
 & 2d$_{5/2}$ 3s$_{1/2}$ &  &  & -0.06100 &  &  &  &  &  &  &  &  &  & 2d$_{5/2}$ 3s$_{1/2}$ &  &  & -0.21201 &  &  &  &  &  &  &  \\
 & (2d$_{3/2})^2$ &  &  & -0.03348 &  &  &  &  &  &  &  &  &  & (2d$_{3/2})^2$ &  &  & -0.03848 &  &  &  &  &  &  &  \\
 & 2d$_{3/2}$ 3s$_{1/2}$ &  &  & 0.03678 &  &  &  &  &  &  &  &  &  & 2d$_{3/2}$ 3s$_{1/2}$ &  &  & 0.09641 &  &  &  &  &  &  &  \\
 &  &  &  &  &  &  &  &  &  &  &  &  &  &  &  &  &  &  &  &  &  &  &  &  \\
 &  &  &  &  & 1 & 0 & 5 & 2 & 2 & 0 & 0 &  &  &  &  &  &  & 1 & 0 & 5 & 2 & 2 & 0 & 0 \\
 & (1h$_{11/2})^2$ &  &  & 0 & 1 & 1 & 5 & 1 & 1 & 1 & 0 &  &  & (1h$_{11/2})^2$ &  &  & 0 & 1 & 1 & 5 & 1 & 1 & 1 & 0 \\
 &  &  &  &  & 1 & 1 & 4 & 3 & 3 & 1 & 0 &  &  &  &  &  &  & 1 & 1 & 4 & 3 & 3 & 1 & 0 \\
 &  &  &  &  &  &  &  &  &  &  &  &  &  &  &  &  &  &  &  &  &  &  &  &  \\
 & (1g$_{7/2})^2$ &  &  & -0.23110 &  &  &  &  &  &  &  &  &  & (1g$_{7/2})^2$ &  &  & -0.20674 &  &  &  &  &  &  &  \\
 & 1g$_{7/2}$ 2d$_{5/2}$ &  &  & 0.03605 &  &  &  &  &  &  &  &  &  & 1g$_{7/2}$ 2d$_{5/2}$ &  &  & 0.08747 &  &  &  &  &  &  &  \\
 & 1g$_{7/2}$ 2d$_{3/2}$ &  &  & -0.05922 &  &  &  &  &  &  &  &  &  & 1g$_{7/2}$ 2d$_{3/2}$ &  &  & -0.07654 &  &  &  &  &  &  &  \\
 & (2d$_{5/2})^2$ &  &  & -0.09710 & 1 & 1 & 4 & 1 & 2 & 1 & 0.00735 &  &  & (2d$_{5/2})^2$ &  &  & -0.28340 & 1 & 1 & 4 & 1 & 2 & 1 & 0.00784 \\
 & 2d$_{5/2}$ 2d$_{3/2}$ &  &  & -0.04011 &  &  &  &  &  &  &  &  &  & 2d$_{5/2}$ 2d$_{3/2}$ &  &  & -0.06372 &  &  &  &  &  &  &  \\
 & 2d$_{5/2}$ 3s$_{1/2}$ &  &  & -0.06100 &  &  &  &  &  &  &  &  &  & 2d$_{5/2}$ 3s$_{1/2}$ &  &  & -0.21201 &  &  &  &  &  &  &  \\
 & (2d$_{3/2})^2$ &  &  & -0.03348 &  &  &  &  &  &  &  &  &  & (2d$_{3/2})^2$ &  &  & -0.03848 &  &  &  &  &  &  &  \\
 & 2d$_{3/2}$ 3s$_{1/2}$ &  &  & 0.03678 &  &  &  &  &  &  &  &  &  & 2d$_{3/2}$ 3s$_{1/2}$ &  &  & 0.09641 &  &  &  &  &  &  &  \\
 &  &  &  &  &  &  &  &  &  &  &  &  &  &  &  &  &  &  &  &  &  &  &  &  \\
 & (2d$_{3/2})^2$ &  &  & -0.03348 &  &  &  &  &  &  &  &  &  & (2d$_{3/2})^2$ &  &  & -0.03848 &  &  &  &  &  &  &  \\
 & 2d$_{5/2}$ 2d$_{3/2}$ &  &  & -0.04011 & 1 & 1 & 4 & 1 & 1 & 1 & 0.00145 &  &  & 2d$_{5/2}$ 2d$_{3/2}$ &  &  & -0.06372 & 1 & 1 & 4 & 1 & 1 & 1 & -0.05553 \\
 & (2d$_{5/2})^2$ &  &  & -0.09710 &  &  &  &  &  &  &  &  &  & (2d$_{5/2})^2$ &  &  & -0.28340 &  &  &  &  &  &  &  \\
 & (1g$_{7/2})^2$ &  &  & -0.23110 &  &  &  &  &  &  &  &  &  & (1g$_{7/2})^2$ &  &  & -0.20674 &  &  &  &  &  &  &  \\
 &  &  &  &  &  &  &  &  &  &  &  &  &  &  &  &  &  &  &  &  &  &  &  &  \\
 & (1g$_{7/2})^2$ &  &  & -0.23110 &  &  &  &  &  &  &  &  &  & (1g$_{7/2})^2$ &  &  & -0.20674 &  &  &  &  &  &  &  \\
 & 1g$_{7/2}$ 2d$_{5/2}$ &  &  & 0.03605 &  &  &  &  &  &  &  &  &  & 1g$_{7/2}$ 2d$_{5/2}$ &  &  & 0.08747 &  &  &  &  &  &  &  \\
 & 1g$_{7/2}$ 2d$_{3/2}$ &  &  & -0.05922 &  &  &  &  &  &  &  &  &  & 1g$_{7/2}$ 2d$_{3/2}$ &  &  & -0.07654 &  &  &  &  &  &  &  \\
 & (2d$_{5/2})^2$ &  &  & -0.09710 & 1 & 1 & 3 & 3 & 2 & 1 & 0.0124 &  &  & (2d$_{5/2})^2$ &  &  & -0.28340 & 1 & 1 & 3 & 3 & 2 & 1 & 0.01321 \\
 & 2d$_{5/2}$ 2d$_{3/2}$ &  &  & -0.04011 &  &  &  &  &  &  &  &  &  & 2d$_{5/2}$ 2d$_{3/2}$ &  &  & -0.06372 &  &  &  &  &  &  &  \\
 & 2d$_{5/2}$ 3s$_{1/2}$ &  &  & -0.06100 &  &  &  &  &  &  &  &  &  & 2d$_{5/2}$ 3s$_{1/2}$ &  &  & -0.21201 &  &  &  &  &  &  &  \\
 & (2d$_{3/2})^2$ &  &  & -0.03348 &  &  &  &  &  &  &  &  &  & (2d$_{3/2})^2$ &  &  & -0.03848 &  &  &  &  &  &  &  \\
 & 2d$_{3/2}$ 3s$_{1/2}$ &  &  & 0.03678 &  &  &  &  &  &  &  &  &  & 2d$_{3/2}$ 3s$_{1/2}$ &  &  & 0.09641 &  &  &  &  &  &  &  \\
 &  &  &  &  &  &  &  &  &  &  &  &  &  &  &  &  &  &  &  &  &  &  &  &  \\
 & (1g$_{7/2})^2$ &  &  & -0.23110 &  &  &  &  &  &  &  &  &  & (1g$_{7/2})^2$ &  &  & -0.20674 &  &  &  &  &  &  &  \\
 & 1g$_{7/2}$ 2d$_{5/2}$ &  &  & 0.03605 &  &  &  &  &  &  &  &  &  & 1g$_{7/2}$ 2d$_{5/2}$ &  &  & 0.08747 &  &  &  &  &  &  &  \\
 & 1g$_{7/2}$ 2d$_{3/2}$ &  &  & -0.05922 &  &  &  &  &  &  &  &  &  & 1g$_{7/2}$ 2d$_{3/2}$ &  &  & -0.07654 &  &  &  &  &  &  &  \\
 & (2d$_{5/2})^2$ &  &  & -0.09710 & 1 & 0 & 4 & 2 & 2 & 1 & -0.00161 &  &  & (2d$_{5/2})^2$ &  &  & -0.28340 & 1 & 0 & 4 & 2 & 2 & 1 & -0.01317 \\
 & 2d$_{5/2}$ 2d$_{3/2}$ &  &  & -0.04011 &  &  &  &  &  &  &  &  &  & 2d$_{5/2}$ 2d$_{3/2}$ &  &  & -0.06372 &  &  &  &  &  &  &  \\
 & 2d$_{5/2}$ 3s$_{1/2}$ &  &  & -0.06100 &  &  &  &  &  &  &  &  &  & 2d$_{5/2}$ 3s$_{1/2}$ &  &  & -0.21201 &  &  &  &  &  &  &  \\
 & (2d$_{3/2})^2$ &  &  & -0.03348 &  &  &  &  &  &  &  &  &  & (2d$_{3/2})^2$ &  &  & -0.03848 &  &  &  &  &  &  &  \\
 & 2d$_{3/2}$ 3s$_{1/2}$ &  &  & 0.03678 &  &  &  &  &  &  &  &  &  & 2d$_{3/2}$ 3s$_{1/2}$ &  &  & 0.09641 &  &  &  &  &  &  &  \\
\\ \cline{2-12} \cline{15-25} 
\end{tabular}%
}
\end{table}

\begin{table}[]
\centering
\setlength{\tabcolsep}{4pt}
\renewcommand{\arraystretch}{0.8}
\resizebox{\textwidth}{!}{%
\begin{tabular}{llllllllllllllllllllllllll}
 & (1g$_{7/2})^2$ &  &  & 0.21767 &  &  &  &  &  &  &  &  &  & (1g$_{7/2})^2$ &  &  & -0.25596 &  &  &  &  &  &  &  &  \\
 & (2d$_{5/2})^2$ &  &  & 0.12488 & 1 & 0 & 5 & 0 & 0 & 0 & 0.08487 &  &  & (2d$_{5/2})^2$ &  &  & -0.35354 & 1 & 0 & 5 & 0 & 0 & 0 & -0.20572 &  \\
 & (2d$_{3/2})^2$ &  &  & 0.06200 &  &  &  &  &  &  &  &  &  & (2d$_{3/2})^2$ &  &  & -0.09343 &  &  &  &  &  &  &  &  \\
 & (3s$_{1/2})^2$ &  &  & 0.04186 &  &  &  &  &  &  &  &  &  & (3s$_{1/2})^2$ &  &  & -0.16613 &  &  &  &  &  &  &  &  \\
 &  &  &  &  &  &  &  &  &  &  &  &  &  &  &  &  &  &  &  &  &  &  &  &  &  \\
 &  &  &  &  & 1 & 0 & 6 & 0 & 0 & 0 & 0 &  &  &  &  &  &  & 1 & 0 & 6 & 0 & 0 & 0 & 0 &  \\
 & \multirow{-2}{*}{(1h$_{11/2})^2$} &  &  & \multirow{-2}{*}{0} & 1 & 1 & 5 & 1 & 1 & 1 & 0 &  &  & \multirow{-2}{*}{(1h$_{11/2})^2$} &  &  & \multirow{-2}{*}{0} & 1 & 1 & 5 & 1 & 1 & 1 & 0 &  \\
 &  &  &  &  &  &  &  &  &  &  &  &  &  &  &  &  &  &  &  &  &  &  &  &  &  \\
 & (1g$_{7/2})^2$ &  &  & 0.21767 &  &  &  &  &  &  &  &  &  & (1g$_{7/2})^2$ &  &  & -0.25596 &  &  &  &  &  &  &  &  \\
 & (2d$_{5/2})^2$ & \multirow{-10}{*}{0} &  & 0.12488 & 1 & 1 & 4 & 1 & 1 & 1 & 0.029 &  &  & (2d$_{5/2})^2$ & \multirow{-10}{*}{0} &  & -0.35354 & 1 & 1 & 4 & 1 & 1 & 1 & 0.024 &  \\
 & (2d$_{3/2})^2$ &  &  & 0.06200 &  &  &  &  &  &  &  &  &  & (2d$_{3/2})^2$ &  &  & -0.09343 &  &  &  &  &  &  &  &  \\ \cline{5-12} \cline{18-25}
 &  &  &  &  &  &  &  &  &  &  &  &  &  &  &  &  &  &  &  &  &  &  &  &  &  \\
 & (1g$_{7/2})^2$ &  &  & -0.12351 &  &  &  &  &  &  &  &  &  & (1g$_{7/2})^2$ &  &  & 0.11239 &  &  &  &  &  &  &  &  \\
 & 1g$_{7/2}$ 2d$_{5/2}$ &  &  & 0.00989 &  &  &  &  &  &  &  &  &  & 1g$_{7/2}$ 2d$_{5/2}$ &  &  & -0.03892 &  &  &  &  &  &  &  &  \\
 & 1g$_{7/2}$ 2d$_{3/2}$ &  &  & -0.00567 &  &  &  &  &  &  &  &  &  & 1g$_{7/2}$ 2d$_{3/2}$ &  &  & 0.00513 &  &  &  &  &  &  &  &  \\
 & (2d$_{5/2})^2$ &  &  & -0.03546 & 1 & 0 & 4 & 2 & 2 & 0 & -0.018 &  &  & (2d$_{5/2})^2$ &  &  & 0.08944 & 1 & 0 & 4 & 2 & 2 & 0 & 0.02342 &  \\
 & 2d$_{5/2}$ 2d$_{3/2}$ &  &  & -0.01050 &  &  &  &  &  &  &  &  &  & 2d$_{5/2}$ 2d$_{3/2}$ &  &  & -0.00044 &  &  &  &  &  &  &  &  \\
 & 2d$_{5/2}$ 3s$_{1/2}$ &  &  & -0.00767 &  &  &  &  &  &  &  &  &  & 2d$_{5/2}$ 3s$_{1/2}$ &  &  & -0.00849 &  &  &  &  &  &  &  &  \\
 & (2d$_{3/2})^2$ &  &  & -0.01086 &  &  &  &  &  &  &  &  &  & (2d$_{3/2})^2$ &  &  & 0.00573 &  &  &  &  &  &  &  &  \\
 & 2d$_{3/2}$ 3s$_{1/2}$ &  &  & 0.00728 &  &  &  &  &  &  &  &  &  & 2d$_{3/2}$ 3s$_{1/2}$ &  &  & -0.01141 &  &  &  &  &  &  &  &  \\
 &  &  &  &  &  &  &  &  &  &  &  &  &  &  &  &  &  &  &  &  &  &  &  &  &  \\
 &  &  &  &  & 1 & 0 & 5 & 2 & 2 & 0 & 0 &  &  &  &  &  &  & 1 & 0 & 5 & 2 & 2 & 0 & 0 &  \\
 & (1h$_{11/2})^2$ &  &  & 0 & 1 & 1 & 5 & 1 & 1 & 1 & 0 &  &  & (1h$_{11/2})^2$ &  &  & 0 & 1 & 1 & 5 & 1 & 1 & 1 & 0 &  \\
 &  &  &  &  & 1 & 1 & 4 & 3 & 3 & 1 & 0 &  &  &  &  &  &  & 1 & 1 & 4 & 3 & 3 & 1 & 0 &  \\
 &  &  &  &  &  &  &  &  &  &  &  &  &  &  &  &  &  &  &  &  &  &  &  &  &  \\
 & (1g$_{7/2})^2$ &  &  & -0.12351 &  &  &  &  &  &  &  &  &  & (1g$_{7/2})^2$ &  &  & 0.11239 &  &  &  &  &  &  &  &  \\
 & 1g$_{7/2}$ 2d$_{5/2}$ &  &  & 0.00989 &  &  &  &  &  &  &  &  &  & 1g$_{7/2}$ 2d$_{5/2}$ &  &  & -0.03892 &  &  &  &  &  &  &  &  \\
 & 1g$_{7/2}$ 2d$_{3/2}$ &  &  & -0.00567 &  &  &  &  &  &  &  &  &  & 1g$_{7/2}$ 2d$_{3/2}$ &  &  & 0.00513 &  &  &  &  &  &  &  &  \\
 & (2d$_{5/2})^2$ &  &  & -0.03546 & 1 & 1 & 4 & 1 & 2 & 1 & {\color[HTML]{000000} 0.00156} &  &  & (2d$_{5/2})^2$ &  &  & 0.08944 & 1 & 1 & 4 & 1 & 2 & 1 & -0.00604 &  \\
 & 2d$_{5/2}$ 2d$_{3/2}$ &  &  & -0.01050 &  &  &  &  &  &  &  &  &  & 2d$_{5/2}$ 2d$_{3/2}$ &  &  & -0.00044 &  &  &  &  &  &  &  &  \\
 & 2d$_{5/2}$ 3s$_{1/2}$ &  &  & -0.00767 &  &  &  &  &  &  &  &  &  & 2d$_{5/2}$ 3s$_{1/2}$ &  &  & -0.00849 &  &  &  &  &  &  &  &  \\
 & (2d$_{3/2})^2$ &  &  & -0.01086 &  &  &  &  &  &  &  &  &  & (2d$_{3/2})^2$ &  &  & 0.00573 &  &  &  &  &  &  &  &  \\
 & 2d$_{3/2}$ 3s$_{1/2}$ &  &  & 0.00728 &  &  &  &  &  &  &  &  &  & 2d$_{3/2}$ 3s$_{1/2}$ &  &  & -0.01141 &  &  &  &  &  &  &  &  \\
 &  &  &  &  &  &  &  &  &  &  &  &  &  &  &  &  &  &  &  &  &  &  &  &  &  \\
 & (2d$_{3/2})^2$ &  &  & 0.04303 &  &  &  &  &  &  &  &  &  & (2d$_{3/2})^2$ &  &  & 0.00573 &  &  &  &  &  &  &  &  \\
 & 2d$_{5/2}$ 2d$_{3/2}$ &  &  & 0.05245 & 1 & 1 & 4 & 1 & 1 & 1 & 0.00273 &  &  & 2d$_{5/2}$ 2d$_{3/2}$ &  &  & -0.00044 & 1 & 1 & 4 & 1 & 1 & 1 & 0.01922 &  \\
 & (2d$_{5/2})^2$ &  &  & 0.05061 &  &  &  &  &  &  &  &  &  & (2d$_{5/2})^2$ &  &  & 0.08944 &  &  &  &  &  &  &  &  \\
 & (1g$_{7/2})^2$ &  &  & -0.02950 &  &  &  &  &  &  &  &  &  & (1g$_{7/2})^2$ &  &  & 0.11239 &  &  &  &  &  &  &  &  \\
 &  &  &  &  &  &  &  &  &  &  &  &  &  &  &  &  &  &  &  &  &  &  &  &  &  \\
 & (1g$_{7/2})^2$ &  &  & -0.12351 &  &  &  &  &  &  &  &  &  & (1g$_{7/2})^2$ &  &  & 0.11239 &  &  &  &  &  &  &  &  \\
 & 1g$_{7/2}$ 2d$_{5/2}$ &  &  & 0.00989 &  &  &  &  &  &  &  &  &  & 1g$_{7/2}$ 2d$_{5/2}$ &  &  & -0.03892 &  &  &  &  &  &  &  &  \\
 & 1g$_{7/2}$ 2d$_{3/2}$ &  &  & -0.00567 &  &  &  &  &  &  &  &  &  & 1g$_{7/2}$ 2d$_{3/2}$ &  &  & 0.00513 &  &  &  &  &  &  &  &  \\
 & (2d$_{5/2})^2$ &  &  & -0.03546 & 1 & 1 & 3 & 3 & 2 & 1 & 0.00263 &  &  & (2d$_{5/2})^2$ &  &  & 0.08944 & 1 & 1 & 3 & 3 & 2 & 1 & -0.01019 &  \\
 & 2d$_{5/2}$ 2d$_{3/2}$ &  &  & -0.01050 &  &  &  &  &  &  &  &  &  & 2d$_{5/2}$ 2d$_{3/2}$ &  &  & -0.00044 &  &  &  &  &  &  &  &  \\
 & 2d$_{5/2}$ 3s$_{1/2}$ &  &  & -0.00767 &  &  &  &  &  &  &  &  &  & 2d$_{5/2}$ 3s$_{1/2}$ &  &  & -0.00849 &  &  &  &  &  &  &  &  \\
 & (2d$_{3/2})^2$ &  &  & -0.01086 &  &  &  &  &  &  &  &  &  & (2d$_{3/2})^2$ &  &  & 0.00573 &  &  &  &  &  &  &  &  \\
 & 2d$_{3/2}$ 3s$_{1/2}$ &  &  & 0.00728 &  &  &  &  &  &  &  &  &  & 2d$_{3/2}$ 3s$_{1/2}$ &  &  & -0.01141 &  &  &  &  &  &  &  &  \\
 &  &  &  &  &  &  &  &  &  &  &  &  &  &  &  &  &  &  &  &  &  &  &  &  &  \\
 & (1g$_{7/2})^2$ &  &  & -0.12351 &  &  &  &  &  &  &  &  &  & (1g$_{7/2})^2$ &  &  & 0.11239 &  &  &  &  &  &  &  &  \\
 & 1g$_{7/2}$ 2d$_{5/2}$ &  &  & 0.00989 &  &  &  &  &  &  &  &  &  & 1g$_{7/2}$ 2d$_{5/2}$ &  &  & -0.03892 &  &  &  &  &  &  &  &  \\
 & 1g$_{7/2}$ 2d$_{3/2}$ &  &  & -0.00567 &  &  &  &  &  &  &  &  &  & 1g$_{7/2}$ 2d$_{3/2}$ &  &  & 0.00513 &  &  &  &  &  &  &  &  \\
 & (2d$_{5/2})^2$ &  &  & -0.03546 & 1 & 0 & 4 & 2 & 2 & 1 & -0.00024 &  &  & (2d$_{5/2})^2$ &  &  & 0.08944 & 1 & 0 & 4 & 2 & 2 & 1 & -0.00818 &  \\
 & 2d$_{5/2}$ 2d$_{3/2}$ &  &  & -0.01050 &  &  &  &  &  &  &  &  &  & 2d$_{5/2}$ 2d$_{3/2}$ &  &  & -0.00044 &  &  &  &  &  &  &  &  \\
 & 2d$_{5/2}$ 3s$_{1/2}$ &  &  & -0.00767 &  &  &  &  &  &  &  &  &  & 2d$_{5/2}$ 3s$_{1/2}$ &  &  & -0.00849 &  &  &  &  &  &  &  &  \\
 & (2d$_{3/2})^2$ &  &  & -0.01086 &  &  &  &  &  &  &  &  &  & (2d$_{3/2})^2$ &  &  & 0.00573 &  &  &  &  &  &  &  &  \\
 & 2d$_{3/2}$ 3s$_{1/2}$ &  &  & 0.00728 &  &  &  &  &  &  &  &  &  & 2d$_{3/2}$ 3s$_{1/2}$ &  &  & -0.01141 &  &  &  &  &  &  &  &  \\
 &  & \multirow{-50}{*}{2} &  &  &  &  &  &  &  &  &  &  &  &  & \multirow{-50}{*}{2} &  &  &  &  &  &  &  &  &  &  \\
 &  &  &  &  &  &  &  &  &  &  &  &  &  &  &  &  &  &  &  &  &  &  &  &  &  \\ \cline{5-12} \cline{18-25}
 &  &  &  &  &  &  &  &  &  &  &  &  &  &  &  &  &  &  &  &  &  &  &  &  &  \\
 & (1g$_{7/2})^2$ &  &  & 0.07401 &  &  &  &  &  &  &  &  &  & (1g$_{7/2})^2$ &  &  & -0.06640 &  &  &  &  &  &  &  &  \\
 & 1g$_{7/2}$ 2d$_{5/2}$ &  &  & -0.02345 &  &  &  &  &  &  &  &  &  & 1g$_{7/2}$ 2d$_{5/2}$ &  &  & 0.05887 &  &  &  &  &  &  &  &  \\
 & 1g$_{7/2}$ 2d$_{3/2}$ &  &  & 0.01588 & 1 & 0 & 3 & 4 & 4 & 0 & 0.01704 &  &  & 1g$_{7/2}$ 2d$_{3/2}$ &  &  & -0.01905 & 1 & 0 & 3 & 4 & 4 & 0 & -0.04267 &  \\
 & 1g$_{7/2}$ 3s$_{1/2}$ &  &  & -0.01313 &  &  &  &  &  &  &  &  &  & 1g$_{7/2}$ 3s$_{1/2}$ &  &  & 0.03725 &  &  &  &  &  &  &  &  \\
 & (2d$_{5/2})^2$ &  &  & 0.02165 &  &  &  &  &  &  &  &  &  & (2d$_{5/2})^2$ &  &  & -0.09353 &  &  &  &  &  &  &  &  \\
 & 2d$_{5/2}$ 2d$_{3/2}$ &  &  & 0.02237 &  &  &  &  &  &  &  &  &  & 2d$_{5/2}$ 2d$_{3/2}$ &  &  & -0.05702 &  &  &  &  &  &  &  &  \\
 &  &  &  &  &  &  &  &  &  &  &  &  &  &  &  &  &  &  &  &  &  &  &  &  &  \\
 & (1g$_{7/2})^2$ &  &  & 0.07401 &  &  &  &  &  &  &  &  &  & (1g$_{7/2})^2$ &  &  & -0.06640 &  &  &  &  &  &  &  &  \\
 & 1g$_{7/2}$ 2d$_{5/2}$ &  &  & -0.02345 &  &  &  &  &  &  &  &  &  & 1g$_{7/2}$ 2d$_{5/2}$ &  &  & 0.05887 &  &  &  &  &  &  &  &  \\
 & 1g$_{7/2}$ 2d$_{3/2}$ &  &  & 0.01588 & 1 & 0 & 3 & 4 & 4 & 1 & 0.00065 &  &  & 1g$_{7/2}$ 2d$_{3/2}$ &  &  & -0.01905 & 1 & 0 & 3 & 4 & 4 & 1 & -0.00201 &  \\
 & 1g$_{7/2}$ 3s$_{1/2}$ &  &  & -0.01313 &  &  &  &  &  &  &  &  &  & 1g$_{7/2}$ 3s$_{1/2}$ &  &  & 0.03725 &  &  &  &  &  &  &  &  \\
 & (2d$_{5/2})^2$ &  &  & 0.02165 &  &  &  &  &  &  &  &  &  & (2d$_{5/2})^2$ &  &  & -0.09353 &  &  &  &  &  &  &  &  \\
 & 2d$_{5/2}$ 2d$_{3/2}$ &  &  & 0.02237 &  &  &  &  &  &  &  &  &  & 2d$_{5/2}$ 2d$_{3/2}$ &  &  & -0.05702 &  &  &  &  &  &  &  &  \\
 &  &  &  &  &  &  &  &  &  &  &  &  &  &  &  &  &  &  &  &  &  &  &  &  &  \\
 & (1g$_{7/2})^2$ &  &  & 0.07401 &  &  &  &  &  &  &  &  &  & (1g$_{7/2})^2$ &  &  & -0.06640 &  &  &  &  &  &  &  &  \\
 & 1g$_{7/2}$ 2d$_{5/2}$ &  &  & -0.02345 &  &  &  &  &  &  &  &  &  & 1g$_{7/2}$ 2d$_{5/2}$ &  &  & 0.05887 &  &  &  &  &  &  &  &  \\
 & 1g$_{7/2}$ 2d$_{3/2}$ &  &  & 0.01588 & 1 & 1 & 3 & 3 & 4 & 1 & -0.00343 &  &  & 1g$_{7/2}$ 2d$_{3/2}$ &  &  & -0.01905 & 1 & 1 & 3 & 3 & 4 & 1 & 0.00868 &  \\
 & 1g$_{7/2}$ 3s$_{1/2}$ &  &  & -0.01313 &  &  &  &  &  &  &  &  &  & 1g$_{7/2}$ 3s$_{1/2}$ &  &  & 0.03725 &  &  &  &  &  &  &  &  \\
 & (2d$_{5/2})^2$ &  &  & 0.02165 &  &  &  &  &  &  &  &  &  & (2d$_{5/2})^2$ &  &  & -0.09353 &  &  &  &  &  &  &  &  \\
 & 2d$_{5/2}$ 2d$_{3/2}$ &  &  & 0.02237 &  &  &  &  &  &  &  &  &  & 2d$_{5/2}$ 2d$_{3/2}$ &  &  & -0.05702 &  &  &  &  &  &  &  &  \\
 &  &  &  &  &  &  &  &  &  &  &  &  &  &  &  &  &  &  &  &  &  &  &  &  &  \\
 & (1g$_{7/2})^2$ &  &  & 0.07401 &  &  &  &  &  &  &  &  &  & (1g$_{7/2})^2$ &  &  & -0.06640 &  &  &  &  &  &  &  &  \\
 & 1g$_{7/2}$ 2d$_{5/2}$ &  &  & -0.02345 &  &  &  &  &  &  &  &  &  & 1g$_{7/2}$ 2d$_{5/2}$ &  &  & 0.05887 &  &  &  &  &  &  &  &  \\
 & 1g$_{7/2}$ 2d$_{3/2}$ &  &  & 0.01588 & 1 & 1 & 2 & 5 & 4 & 1 & -0.00694 &  &  & 1g$_{7/2}$ 2d$_{3/2}$ &  &  & -0.01905 & 1 & 1 & 2 & 5 & 4 & 1 & 0.01754 &  \\
 & 1g$_{7/2}$ 3s$_{1/2}$ &  &  & -0.01313 &  &  &  &  &  &  &  &  &  & 1g$_{7/2}$ 3s$_{1/2}$ &  &  & 0.03725 &  &  &  &  &  &  &  &  \\
 & (2d$_{5/2})^2$ &  &  & 0.02165 &  &  &  &  &  &  &  &  &  & (2d$_{5/2})^2$ &  &  & -0.09353 &  &  &  &  &  &  &  &  \\
 & 2d$_{5/2}$ 2d$_{3/2}$ &  &  & 0.02237 &  &  &  &  &  &  &  &  &  & 2d$_{5/2}$ 2d$_{3/2}$ &  &  & -0.05702 &  &  &  &  &  &  &  &  \\
 &  &  &  &  &  &  &  &  &  &  &  &  &  &  &  &  &  &  &  &  &  &  &  &  &  \\
 & (1g$_{7/2})^2$ &  &  & 0.07401 &  &  &  &  &  &  &  &  &  & (1g$_{7/2})^2$ &  &  & -0.06640 &  &  &  &  &  &  &  &  \\
 & 1g$_{7/2}$ 2d$_{5/2}$ &  &  & -0.02345 &  &  &  &  &  &  &  &  &  & 1g$_{7/2}$ 2d$_{5/2}$ &  &  & 0.05887 &  &  &  &  &  &  &  &  \\
 & 1g$_{7/2}$ 2d$_{3/2}$ &  &  & 0.01588 & 1 & 1 & 3 & 3 & 3 & 1 & -0.00151 &  &  & 1g$_{7/2}$ 2d$_{3/2}$ &  &  & -0.01905 & 1 & 1 & 3 & 3 & 3 & 1 & -0.01539 &  \\
 & (2d$_{5/2})^2$ &  &  & 0.02165 &  &  &  &  &  &  &  &  &  & (2d$_{5/2})^2$ &  &  & -0.09353 &  &  &  &  &  &  &  &  \\
 & 2d$_{5/2}$ 2d$_{3/2}$ &  &  & 0.02237 &  &  &  &  &  &  &  &  &  & 2d$_{5/2}$ 2d$_{3/2}$ &  &  & -0.05702 &  &  &  &  &  &  &  &  \\
 &  &  &  &  &  &  &  &  &  &  &  &  &  &  &  &  &  &  &  &  &  &  &  &  &  \\
 & (1g$_{7/2})^2$ &  &  & 0.07401 &  &  &  &  &  &  &  &  &  & (1g$_{7/2})^2$ &  &  & -0.06640 &  &  &  &  &  &  &  &  \\
 & 1g$_{7/2}$ 2d$_{5/2}$ &  &  & -0.02345 & 1 & 1 & 2 & 5 & 5 & 1 & 0.01706 &  &  & 1g$_{7/2}$ 2d$_{5/2}$ &  &  & 0.05887 & 1 & 1 & 2 & 5 & 5 & 1 & -0.01023 &  \\
 & 1g$_{7/2}$ 2d$_{3/2}$ &  &  & 0.01588 &  &  &  &  &  &  &  &  &  & 1g$_{7/2}$ 2d$_{3/2}$ &  &  & -0.01905 &  &  &  &  &  &  &  &  \\
 &  &  &  &  &  &  &  &  &  &  &  &  &  &  &  &  &  &  &  &  &  &  &  &  &  \\
 &  &  &  &  & 1 & 0 & 4 & 4 & 4 & 0 & 0 &  &  &  &  &  &  & 1 & 0 & 4 & 4 & 4 & 0 & 0 &  \\
\multirow{-102}{*}{$^{116}$Sn$_{1.290}$(2$^{+}$)} &  &  &  &  & 1 & 0 & 4 & 4 & 5 & 1 & 0 &  & \multirow{-102}{*}{$^{116}$Sn$_{1.290}$(2$^{+}$)} &  &  & \multirow{-103}{*}{$^{118}$Te$_{0.605}$(2$^{+}$)} &  & 1 & 0 & 4 & 4 & 5 & 1 & 0 &  \\
 & \multirow{-3}{*}{(1h$_{11/2})^2$} & \multirow{-42}{*}{4} & \multirow{-103}{*}{} &  & 1 & 1 & 3 & 5 & 5 & 1 & 0 &  &  & \multirow{-3}{*}{(1h$_{11/2})^2$} & \multirow{-42}{*}{4} &  &  & 1 & 1 & 3 & 5 & 5 & 1 & 0 &  \\
 &  &  &  & \multirow{-4}{*}{0} & 1 & 1 & 4 & 3 & 3 & 1 & 0 &  &  &  &  &  & \multirow{-4}{*}{0} & 1 & 1 & 4 & 3 & 3 & 1 & 0 &  \\ \cline{1-25}
\end{tabular}%
}
\end{table}

\pagebreak

\begin{table}[]
\centering
\setlength{\tabcolsep}{5pt}
\renewcommand{\arraystretch}{0.9}
\caption{Cluster spectroscopic factors ($\mathcal{S}$ (c.m.)) for
di-proton transfer involving the lighter nucleus $^{60}$Ni are
presented for the relevant overlaps. The calculations were performed
using \textsc{kshell} with the \texttt{KB3G} and the \texttt{fpd6npn}
effective interactions. Here, $j_1$ and $j_2$ stand for the total
angular momenta of nucleons 1 and 2, respectively. The quantum numbers
$n$, $\ell$ ($N$, $L$) represent the principal quantum numbers and
orbital angular momenta of the nucleons relative to each other
(and to the core), respectively. $J$ and $\Lambda$ denote the total
angular momentum and the orbital angular momentum of the cluster
with respect to the core, respectively, and $S$ is the intrinsic
spin of the two-nucleon cluster. The spins and parities of the
initial and final nuclei are given by $I^{\pi}_{\mathrm{i}}$
and $I^{\pi}_{\mathrm{f}}$, respectively.}

\label{tab:your-table}
\resizebox{\textwidth}{!}{%
\begin{tabular}{lllllllllllllllllllllllll}
\cmidrule{1-12} \cmidrule{14-25}
$I^\pi_\text{i}$ & $j_1$$j_2$ & $J$ & $I^\pi_\text{f}$ & $\mathcal{S}$ ($j-j$) \texttt{(KB3G)} & $n$ & $\ell$ & $N$ & $L$ & $\Lambda$ & $S$ & $\mathcal{S}$ (c.m.) \texttt{(KB3G)} & & $I^\pi_\text{i}$ & $j_1$$j_2$ & $J$ & $I^\pi_\text{f}$ & $\mathcal{S}$ ($j-j$)\texttt{(fpd6npn)} & $n$ & $\ell$ & $N$ & $L$ & $\Lambda$ & $S$ & $\mathcal{S}$ (c.m.) \texttt{(fpd6npn)}\\ \cmidrule{1-12} \cmidrule{14-25}

 & (1f$_{7/2})^2$ &  &  & -0.72087 &  &  &  &  &  &  &  &  &  & (1f$_{7/2})^2$ &  &  & -0.73762 &  &  &  &  &  &  &  \\
 & (2p$_{3/2})^2$ &  &  & -0.08238 &  &  &  &  &  &  &  &  &  & (2p$_{3/2})^2$ &  &  & -0.19678 &  &  &  &  &  &  &  \\
 & (1f$_{5/2})^2$ &  &  & -0.14864 &  &  &  &  &  &  &  &  &  & (1f$_{5/2})^2$ &  &  & -0.23955 &  &  &  &  &  &  &  \\
 & (2p$_{1/2})^2$ &  &  & -0.04649 & \multirow{-4}{*}{1} & \multirow{-4}{*}{0} & \multirow{-4}{*}{4} & \multirow{-4}{*}{0} & \multirow{-4}{*}{0} & \multirow{-4}{*}{0} & \multirow{-4}{*}{-0.19182} &  &  & (2p$_{1/2})^2$ &  &  & -0.11658 & \multirow{-4}{*}{1} & \multirow{-4}{*}{0} & \multirow{-4}{*}{4} & \multirow{-4}{*}{0} & \multirow{-4}{*}{0} & \multirow{-4}{*}{0} & \multirow{-4}{*}{-0.27654} \\
 &  &  &  &  &  &  &  &  &  &  &  &  &  &  &  &  &  &  &  &  &  &  &  &  \\
 & (1f$_{7/2})^2$ &  &  & -0.72087 &  &  &  &  &  &  &  &  &  & (1f$_{7/2})^2$ &  &  & -0.73762 &  &  &  &  &  &  &  \\
 & (2p$_{3/2})^2$ &  &  & -0.08238 &  &  &  &  &  &  &  &  &  & (2p$_{3/2})^2$ &  &  & -0.19678 &  &  &  &  &  &  &  \\
 & (1f$_{5/2})^2$ &  &  & -0.14864 &  &  &  &  &  &  &  &  &  & (1f$_{5/2})^2$ &  &  & -0.23955 &  &  &  &  &  &  &  \\
 & (2p$_{1/2})^2$ & \multirow{-9}{*}{0} & \multirow{-9}{*}{$^{58}$Fe$_\text{g.s.}$(0$^{+}$)} & -0.04649 & \multirow{-4}{*}{1} & \multirow{-4}{*}{1} & \multirow{-4}{*}{3} & \multirow{-4}{*}{1} & \multirow{-4}{*}{1} & \multirow{-4}{*}{1} & \multirow{-4}{*}{0.16482} &  &  & (2p$_{1/2})^2$ & \multirow{-9}{*}{0} & \multirow{-9}{*}{$^{58}$Fe$_\text{g.s.}$(0$^{+}$)} & -0.11658 & \multirow{-4}{*}{1} & \multirow{-4}{*}{1} & \multirow{-4}{*}{3} & \multirow{-4}{*}{1} & \multirow{-4}{*}{1} & \multirow{-4}{*}{1} & \multirow{-4}{*}{0.14268} \\ \cline{4-12} \cline{17-25} 
 &  &  &  &  &  &  &  &  &  &  &  &  &  &  &  &  &  &  &  &  &  &  &  &  \\
 & (1f$_{7/2})^2$ &  &  & 1.18879 &  &  &  &  &  &  &  &  &  & (1f$_{7/2})^2$ &  &  & -0.95227 &  &  &  &  &  &  &  \\
 & 1f$_{7/2}$ 2p$_{3/2}$ &  &  & 0.00031 &  &  &  &  &  &  &  &  &  & 1f$_{7/2}$ 2p$_{3/2}$ &  &  & -0.07134 &  &  &  &  &  &  &  \\
 & 1f$_{7/2}$ 1f$_{5/2}$ &  &  & 0.02185 &  &  &  &  &  &  &  &  &  & 1f$_{7/2}$ 1f$_{5/2}$ &  &  & -0.02189 &  &  &  &  &  &  &  \\
 & (2p$_{3/2})^2$ &  &  & 0.02428 & 1 & 0 & 3 & 2 & 2 & 0 & 0.13706 &  &  & (2p$_{3/2})^2$ &  &  & -0.02145 & 1 & 0 & 3 & 2 & 2 & 0 & -0.10704 \\
 & 2p$_{3/2}$ 1f$_{5/2}$ &  &  & 0.02815 &  &  &  &  &  &  &  &  &  & 2p$_{3/2}$ 1f$_{5/2}$ &  &  & -0.01031 &  &  &  &  &  &  &  \\
 & 2p$_{3/2}$ 2p$_{1/2}$ &  &  & 0.00788 &  &  &  &  &  &  &  &  &  & 2p$_{3/2}$ 2p$_{1/2}$ &  &  & 0.02524 &  &  &  &  &  &  &  \\
 & (1f$_{5/2})^2$ &  &  & 0.01875 &  &  &  &  &  &  &  &  &  & (1f$_{5/2})^2$ &  &  & 0.03553 &  &  &  &  &  &  &  \\
 & 1f$_{5/2}$ 2p$_{1/2}$ &  &  & 0.03276 &  &  &  &  &  &  &  &  &  & 1f$_{5/2}$ 2p$_{1/2}$ &  &  & 0.00562 &  &  &  &  &  &  &  \\
 &  &  &  &  &  &  &  &  &  &  &  &  &  &  &  &  &  &  &  &  &  &  &  &  \\
 & (1f$_{7/2})^2$ &  &  & 1.18879 &  &  &  &  &  &  &  &  &  & (1f$_{7/2})^2$ &  &  & -0.95227 &  &  &  &  &  &  &  \\
 & 1f$_{7/2}$ 2p$_{3/2}$ &  &  & 0.00031 &  &  &  &  &  &  &  &  &  & 1f$_{7/2}$ 2p$_{3/2}$ &  &  & -0.07134 &  &  &  &  &  &  &  \\
 & 1f$_{7/2}$ 1f$_{5/2}$ &  &  & 0.02185 &  &  &  &  &  &  &  &  &  & 1f$_{7/2}$ 1f$_{5/2}$ &  &  & -0.02189 &  &  &  &  &  &  &  \\
 & (2p$_{3/2})^2$ &  &  & 0.02428 & 1 & 0 & 3 & 2 & 2 & 1 & 0.00505 &  &  & (2p$_{3/2})^2$ &  &  & -0.02145 & 1 & 0 & 3 & 2 & 2 & 1 & -0.00893 \\
 & 2p$_{3/2}$ 1f$_{5/2}$ &  &  & 0.02815 &  &  &  &  &  &  &  &  &  & 2p$_{3/2}$ 1f$_{5/2}$ &  &  & -0.01031 &  &  &  &  &  &  &  \\
 & 2p$_{3/2}$ 2p$_{1/2}$ &  &  & 0.00788 &  &  &  &  &  &  &  &  &  & 2p$_{3/2}$ 2p$_{1/2}$ &  &  & 0.02524 &  &  &  &  &  &  &  \\
 & (1f$_{5/2})^2$ &  &  & 0.01875 &  &  &  &  &  &  &  &  &  & (1f$_{5/2})^2$ &  &  & 0.03553 &  &  &  &  &  &  &  \\
 & 1f$_{5/2}$ 2p$_{1/2}$ &  &  & 0.03276 &  &  &  &  &  &  &  &  &  & 1f$_{5/2}$ 2p$_{1/2}$ &  &  & 0.00562 &  &  &  &  &  &  &  \\
 &  &  &  &  &  &  &  &  &  &  &  &  &  &  &  &  &  &  &  &  &  &  &  &  \\
 & (1f$_{7/2})^2$ &  &  & 1.18879 &  &  &  &  &  &  &  &  &  & (1f$_{7/2})^2$ &  &  & -0.95227 &  &  &  &  &  &  &  \\
 & 1f$_{7/2}$ 2p$_{3/2}$ &  &  & 0.00031 &  &  &  &  &  &  & {\color[HTML]{000000} } &  &  & 1f$_{7/2}$ 2p$_{3/2}$ &  &  & -0.07134 &  &  &  &  &  &  &  \\
 & 1f$_{7/2}$ 1f$_{5/2}$ &  &  & 0.02185 &  &  &  &  &  &  &  &  &  & 1f$_{7/2}$ 1f$_{5/2}$ &  &  & -0.02189 &  &  &  &  &  &  &  \\
 & (2p$_{3/2})^2$ &  &  & 0.02428 & 1 & 1 & 2 & 3 & 2 & 1 & 0.00099 &  &  & (2p$_{3/2})^2$ &  &  & -0.02145 & 1 & 1 & 2 & 3 & 2 & 1 & -0.02111 \\
 & 2p$_{3/2}$ 1f$_{5/2}$ &  &  & 0.02815 &  &  &  &  &  &  &  &  &  & 2p$_{3/2}$ 1f$_{5/2}$ &  &  & -0.01031 &  &  &  &  &  &  &  \\
 & 2p$_{3/2}$ 2p$_{1/2}$ &  &  & 0.00788 &  &  &  &  &  &  &  &  &  & 2p$_{3/2}$ 2p$_{1/2}$ &  &  & 0.02524 &  &  &  &  &  &  &  \\
 & (1f$_{5/2})^2$ &  &  & 0.01875 &  &  &  &  &  &  &  &  &  & (1f$_{5/2})^2$ &  &  & 0.03553 &  &  &  &  &  &  &  \\
 & 1f$_{5/2}$ 2p$_{1/2}$ &  &  & 0.03276 &  &  &  &  &  &  &  &  &  & 1f$_{5/2}$ 2p$_{1/2}$ &  &  & 0.00562 &  &  &  &  &  &  &  \\
 &  &  &  &  &  &  &  &  &  &  &  &  &  &  &  &  &  &  &  &  &  &  &  &  \\
 & (1f$_{7/2})^2$ &  &  & 1.18879 &  &  &  &  &  &  &  &  &  & (1f$_{7/2})^2$ &  &  & -0.95227 &  &  &  &  &  &  &  \\
 & 1f$_{7/2}$ 2p$_{3/2}$ &  &  & 0.00031 &  &  &  &  &  &  &  &  &  & 1f$_{7/2}$ 2p$_{3/2}$ &  &  & -0.07134 &  &  &  &  &  &  &  \\
 & 1f$_{7/2}$ 1f$_{5/2}$ &  &  & 0.02185 & 1 & 1 & 2 & 3 & 3 & 1 & -0.15856 &  &  & 1f$_{7/2}$ 1f$_{5/2}$ &  &  & -0.02189 & 1 & 1 & 2 & 3 & 3 & 1 & 0.15163 \\
 & 2p$_{3/2}$ 1f$_{5/2}$ &  &  & 0.02815 &  &  &  &  &  &  &  &  &  & 2p$_{3/2}$ 1f$_{5/2}$ &  &  & -0.01031 &  &  &  &  &  &  &  \\
 & (1f$_{5/2})^2$ &  & \multirow{-32}{*}{$^{58}$Fe$_{0.839}$(2$^{+}$)} & 0.01875 &  &  &  &  &  &  &  &  &  & (1f$_{5/2})^2$ &  & \multirow{-32}{*}{$^{58}$Fe$_{0.839}$(2$^{+}$)} & 0.03553 &  &  &  &  &  &  &  \\
\multirow{-42}{*}{$^{60}$Ni$_\text{g.s.}$(0$^{+}$)} & 1f$_{5/2}$ 2p$_{1/2}$ & \multirow{-33}{*}{2} &  & 0.03276 &  &  &  &  &  &  &  &  & \multirow{-42}{*}{$^{60}$Ni$_\text{g.s.}$(0$^{+}$)} & 1f$_{5/2}$ 2p$_{1/2}$ & \multirow{-33}{*}{2} &  & 0.00562 &  &  &  &  &  &  &  \\ \cline{2-12} \cline{14-25} 
 &  &  &  &  &  &  &  &  &  &  &  &  &  &  &  &  &  &  &  &  &  &  &  &  \\
 & (1f$_{7/2})^2$ &  &  & 0.26283 &  &  &  &  &  &  &  &  &  & (1f$_{7/2})^2$ &  &  & 0.23206 &  &  &  &  &  &  &  \\
 & 1f$_{7/2}$ 2p$_{3/2}$ &  &  & -0.04255 &  &  &  &  &  &  &  &  &  & 1f$_{7/2}$ 2p$_{3/2}$ &  &  & -0.04563 &  &  &  &  &  &  &  \\
 & 1f$_{7/2}$ 1f$_{5/2}$ &  &  & -0.01594 &  &  &  &  &  &  &  &  &  & 1f$_{7/2}$ 1f$_{5/2}$ &  &  & -0.02217 &  &  &  &  &  &  &  \\
 & (2p$_{3/2})^2$ &  &  & -0.00588 & 1 & 0 & 3 & 2 & 2 & 0 & 0.01311 &  &  & (2p$_{3/2})^2$ &  &  & -0.03673 & 1 & 0 & 3 & 2 & 2 & 0 & -0.02341 \\
 & 2p$_{3/2}$ 1f$_{5/2}$ &  &  & 0.00037 &  &  &  &  &  &  &  &  &  & 2p$_{3/2}$ 1f$_{5/2}$ &  &  & -0.02866 &  &  &  &  &  &  &  \\
 & 2p$_{3/2}$ 2p$_{1/2}$ &  &  & -0.01087 &  &  &  &  &  &  &  &  &  & 2p$_{3/2}$ 2p$_{1/2}$ &  &  & -0.06303 &  &  &  &  &  &  &  \\
 & (1f$_{5/2})^2$ &  &  & -0.00731 &  &  &  &  &  &  &  &  &  & (1f$_{5/2})^2$ &  &  & -0.05022 &  &  &  &  &  &  &  \\
 & 1f$_{5/2}$ 2p$_{1/2}$ &  &  & -0.00060 &  &  &  &  &  &  &  &  &  & 1f$_{5/2}$ 2p$_{1/2}$ &  &  & -0.05101 &  &  &  &  &  &  &  \\
 &  &  &  &  &  &  &  &  &  &  &  &  &  &  &  &  &  &  &  &  &  &  &  &  \\
 & (1f$_{7/2})^2$ &  &  & 0.26283 &  &  &  &  &  &  &  &  &  & (1f$_{7/2})^2$ &  &  & 0.23206 &  &  &  &  &  &  &  \\
 & 1f$_{7/2}$ 2p$_{3/2}$ &  &  & -0.04255 &  &  &  &  &  &  &  &  &  & 1f$_{7/2}$ 2p$_{3/2}$ &  &  & -0.04563 &  &  &  &  &  &  &  \\
 & 1f$_{7/2}$ 1f$_{5/2}$ &  &  & -0.01594 &  &  &  &  &  &  &  &  &  & 1f$_{7/2}$ 1f$_{5/2}$ &  &  & -0.02217 &  &  &  &  &  &  &  \\
 & (2p$_{3/2})^2$ &  &  & -0.00588 & 1 & 0 & 3 & 2 & 2 & 1 & -0.01134 &  &  & (2p$_{3/2})^2$ &  &  & -0.03673 & 1 & 0 & 3 & 2 & 2 & 1 & -0.02567 \\
 & 2p$_{3/2}$ 1f$_{5/2}$ &  &  & 0.00037 &  &  &  &  &  &  &  &  &  & 2p$_{3/2}$ 1f$_{5/2}$ &  &  & -0.02866 &  &  &  &  &  &  &  \\
 & 2p$_{3/2}$ 2p$_{1/2}$ &  &  & -0.01087 &  &  &  &  &  &  &  &  &  & 2p$_{3/2}$ 2p$_{1/2}$ &  &  & -0.06303 &  &  &  &  &  &  &  \\
 & (1f$_{5/2})^2$ &  &  & -0.00731 &  &  &  &  &  &  &  &  &  & (1f$_{5/2})^2$ &  &  & -0.05022 &  &  &  &  &  &  &  \\
 & 1f$_{5/2}$ 2p$_{1/2}$ &  &  & -0.00060 &  &  &  &  &  &  &  &  &  & 1f$_{5/2}$ 2p$_{1/2}$ &  &  & -0.05101 &  &  &  &  &  &  &  \\
 &  &  &  &  &  &  &  &  &  &  &  &  &  &  &  &  &  &  &  &  &  &  &  &  \\
 & (1f$_{7/2})^2$ &  &  & 0.26283 &  &  &  &  &  &  &  &  &  & (1f$_{7/2})^2$ &  &  & 0.23206 &  &  &  &  &  &  &  \\
 & 1f$_{7/2}$ 2p$_{3/2}$ &  &  & -0.04255 &  &  &  &  &  &  &  &  &  & 1f$_{7/2}$ 2p$_{3/2}$ &  &  & -0.04563 &  &  &  &  &  &  &  \\
 & 1f$_{7/2}$ 1f$_{5/2}$ &  &  & -0.01594 &  &  &  &  &  &  &  &  &  & 1f$_{7/2}$ 1f$_{5/2}$ &  &  & -0.02217 &  &  &  &  &  &  &  \\
 & (2p$_{3/2})^2$ &  &  & -0.00588 & 1 & 1 & 2 & 3 & 2 & 1 & -0.01022 &  &  & (2p$_{3/2})^2$ &  &  & -0.03673 & 1 & 1 & 2 & 3 & 2 & 1 & -0.0088 \\
 & 2p$_{3/2}$ 1f$_{5/2}$ &  &  & 0.00037 &  &  &  &  &  &  &  &  &  & 2p$_{3/2}$ 1f$_{5/2}$ &  &  & -0.02866 &  &  &  &  &  &  &  \\
 & 2p$_{3/2}$ 2p$_{1/2}$ &  &  & -0.01087 &  &  &  &  &  &  &  &  &  & 2p$_{3/2}$ 2p$_{1/2}$ &  &  & -0.06303 &  &  &  &  &  &  &  \\
 & (1f$_{5/2})^2$ &  &  & -0.00731 &  &  &  &  &  &  &  &  &  & (1f$_{5/2})^2$ &  &  & -0.05022 &  &  &  &  &  &  &  \\
 & 1f$_{5/2}$ 2p$_{1/2}$ &  &  & -0.00060 &  &  &  &  &  &  &  &  &  & 1f$_{5/2}$ 2p$_{1/2}$ &  &  & -0.05101 &  &  &  &  &  &  &  \\
 &  &  &  &  &  &  &  &  &  &  &  &  &  &  &  &  &  &  &  &  &  &  &  &  \\
 & (1f$_{7/2})^2$ &  &  & 0.26283 &  &  &  &  &  &  &  &  &  & (1f$_{7/2})^2$ &  &  & 0.23206 &  &  &  &  &  &  &  \\
 & 1f$_{7/2}$ 2p$_{3/2}$ &  &  & -0.04255 &  &  &  &  &  &  &  &  &  & 1f$_{7/2}$ 2p$_{3/2}$ &  &  & -0.04563 &  &  &  &  &  &  &  \\
 & 1f$_{7/2}$ 1f$_{5/2}$ &  &  & -0.01594 & 1 & 1 & 2 & 3 & 3 & 1 & -0.03234 &  &  & 1f$_{7/2}$ 1f$_{5/2}$ &  &  & -0.02217 & 1 & 1 & 2 & 3 & 3 & 1 & -0.04364 \\
 & 2p$_{3/2}$ 1f$_{5/2}$ &  &  & 0.00037 &  &  &  &  &  &  &  &  &  & 2p$_{3/2}$ 1f$_{5/2}$ &  &  & -0.02866 &  &  &  &  &  &  &  \\
 & (1f$_{5/2})^2$ &  &  & -0.00731 &  &  &  &  &  &  &  &  &  & (1f$_{5/2})^2$ &  &  & -0.05022 &  &  &  &  &  &  &  \\
 & 1f$_{5/2}$ 2p$_{1/2}$ & \multirow{-33}{*}{2} & \multirow{-33}{*}{$^{58}$Fe$_\text{g.s.}$(0$^{+}$)} & -0.00060 &  &  &  &  &  &  &  &  &  & 1f$_{5/2}$ 2p$_{1/2}$ & \multirow{-33}{*}{2} & \multirow{-33}{*}{$^{58}$Fe$_\text{g.s.}$(0$^{+}$)} & -0.05101 &  &  &  &  &  &  &  \\ \cline{5-12} \cline{18-25} 
 &  &  &  &  &  &  &  &  &  &  &  &  &  &  &  &  &  &  &  &  &  &  &  &  \\
 & (1f$_{7/2})^2$ &  &  & -0.55772 &  &  &  &  &  &  &  &  &  & (1f$_{7/2})^2$ &  &  & 0.64768 &  &  &  &  &  &  &  \\
 & (2p$_{3/2})^2$ &  &  & -0.06315 &  &  &  &  &  &  &  &  &  & (2p$_{3/2})^2$ &  &  & 0.17939 &  &  &  &  &  &  &  \\
 & (1f$_{5/2})^2$ &  &  & -0.11752 &  &  &  &  &  &  &  &  &  & (1f$_{5/2})^2$ &  &  & 0.22818 &  &  &  &  &  &  &  \\
 & (2p$_{1/2})^2$ &  &  & -0.03473 & \multirow{-4}{*}{1} & \multirow{-4}{*}{0} & \multirow{-4}{*}{4} & \multirow{-4}{*}{0} & \multirow{-4}{*}{0} & \multirow{-4}{*}{0} & \multirow{-4}{*}{-0.14817} &  &  & (2p$_{1/2})^2$ &  &  & 0.11400 & \multirow{-4}{*}{1} & \multirow{-4}{*}{0} & \multirow{-4}{*}{4} & \multirow{-4}{*}{0} & \multirow{-4}{*}{0} & \multirow{-4}{*}{0} & \multirow{-4}{*}{0.25164} \\
 &  &  &  &  &  &  &  &  &  &  &  &  &  &  &  &  &  &  &  &  &  &  &  &  \\
 & (1f$_{7/2})^2$ &  &  & -0.55772 &  &  &  &  &  &  &  &  &  & (1f$_{7/2})^2$ &  &  & 0.64768 &  &  &  &  &  &  &  \\
 & (2p$_{3/2})^2$ &  &  & -0.06315 &  &  &  &  &  &  &  &  &  & (2p$_{3/2})^2$ &  &  & 0.17939 &  &  &  &  &  &  &  \\
 & (1f$_{5/2})^2$ &  &  & -0.11752 &  &  &  &  &  &  &  &  &  & (1f$_{5/2})^2$ &  &  & 0.22818 &  &  &  &  &  &  &  \\
 & (2p$_{1/2})^2$ & \multirow{-9}{*}{0} &  & -0.03473 & \multirow{-4}{*}{1} & \multirow{-4}{*}{1} & \multirow{-4}{*}{3} & \multirow{-4}{*}{1} & \multirow{-4}{*}{1} & \multirow{-4}{*}{1} & \multirow{-4}{*}{0.12695} &  &  & (2p$_{1/2})^2$ & \multirow{-9}{*}{0} &  & 0.11400 & \multirow{-4}{*}{1} & \multirow{-4}{*}{1} & \multirow{-4}{*}{3} & \multirow{-4}{*}{1} & \multirow{-4}{*}{1} & \multirow{-4}{*}{1} & \multirow{-4}{*}{-0.11687} \\
&  &  &  &  &  &  &  &  &  &  &  &  &  &  &  &  &  &  &  &  &  &  &  &  \\

\cline{5-12} \cline{18-25}
\end{tabular}%
}

\end{table}
\clearpage
\begin{table}[ht!]
\centering
\setlength{\tabcolsep}{4pt}
\renewcommand{\arraystretch}{0.9}
\resizebox{\linewidth}{!}{%
\begin{tabular}{llllllllllllllllllllllllll}
 & (1f$_{7/2})^2$ & \multirow{31}{*}{2} &  & 0.37195 &  &  &  &  &  &  &  &  &  & (1f$_{7/2})^2$ & \multirow{31}{*}{2} &  & -0.35064 &  &  &  &  &  &  &  &  \\
\multirow{49}{*}{$^{60}$Ni$_{1.322}$(2$^{+}$)} & 1f$_{7/2}$ 2p$_{3/2}$ &  & \multirow{49}{*}{$^{58}$Fe$_{0.839}$(2$^{+}$)} & 0.07272 &  &  &  &  &  &  &  &  & \multirow{49}{*}{$^{60}$Ni$_{1.322}$(2$^{+}$)} & 1f$_{7/2}$ 2p$_{3/2}$ &  & \multirow{50}{*}{$^{58}$Fe$_{0.839}$(2$^{+}$)} & -0.12193 &  &  &  &  &  &  &  &  \\
 & 1f$_{7/2}$ 1f$_{5/2}$ &  &  & 0.00559 &  &  &  &  &  &  &  &  &  & 1f$_{7/2}$ 1f$_{5/2}$ &  &  & -0.00166 &  &  &  &  &  &  &  &  \\
 & (2p$_{3/2})^2$ &  &  & 0.00588 & 1 & 0 & 3 & 2 & 2 & 0 & 0.01332 &  &  & (2p$_{3/2})^2$ &  &  & -0.00331 & 1 & 0 & 3 & 2 & 2 & 0 & -0.05266 &  \\
 & 2p$_{3/2}$ 1f$_{5/2}$ &  &  & 0.00914 &  &  &  &  &  &  &  &  &  & 2p$_{3/2}$ 1f$_{5/2}$ &  &  & -0.00194 &  &  &  &  &  &  &  &  \\
 & 2p$_{3/2}$ 2p$_{1/2}$ &  &  & 0.00564 &  &  &  &  &  &  &  &  &  & 2p$_{3/2}$ 2p$_{1/2}$ &  &  & 0.01794 &  &  &  &  &  &  &  &  \\
 & (1f$_{5/2})^2$ &  &  & 0.00848 &  &  &  &  &  &  &  &  &  & (1f$_{5/2})^2$ &  &  & 0.02030 &  &  &  &  &  &  &  &  \\
 & 1f$_{5/2}$ 2p$_{1/2}$ &  &  & 0.01620 &  &  &  &  &  &  &  &  &  & 1f$_{5/2}$ 2p$_{1/2}$ &  &  & 0.01219 &  &  &  &  &  &  &  &  \\
 &  &  &  &  &  &  &  &  &  &  &  &  &  &  &  &  &  &  &  &  &  &  &  &  &  \\
 & (1f$_{7/2})^2$ &  &  & 0.37195 &  &  &  &  &  &  &  &  &  & (1f$_{7/2})^2$ &  &  & -0.35064 &  &  &  &  &  &  &  &  \\
 & 1f$_{7/2}$ 2p$_{3/2}$ &  &  & 0.07272 &  &  &  &  &  &  &  &  &  & 1f$_{7/2}$ 2p$_{3/2}$ &  &  & -0.12193 &  &  &  &  &  &  &  &  \\
 & 1f$_{7/2}$ 1f$_{5/2}$ &  &  & 0.00559 &  &  &  &  &  &  &  &  &  & 1f$_{7/2}$ 1f$_{5/2}$ &  &  & -0.00166 &  &  &  &  &  &  &  &  \\
 & (2p$_{3/2})^2$ &  &  & 0.00588 & 1 & 0 & 3 & 2 & 2 & 1 & 0.01332 &  &  & (2p$_{3/2})^2$ &  &  & -0.00331 & 1 & 0 & 3 & 2 & 2 & 1 & -0.0165 &  \\
 & 2p$_{3/2}$ 1f$_{5/2}$ &  &  & 0.00914 &  &  &  &  &  &  &  &  &  & 2p$_{3/2}$ 1f$_{5/2}$ &  &  & -0.00194 &  &  &  &  &  &  &  &  \\
 & 2p$_{3/2}$ 2p$_{1/2}$ &  &  & 0.00564 &  &  &  &  &  &  &  &  &  & 2p$_{3/2}$ 2p$_{1/2}$ &  &  & 0.01794 &  &  &  &  &  &  &  &  \\
 & (1f$_{5/2})^2$ &  &  & 0.00848 &  &  &  &  &  &  &  &  &  & (1f$_{5/2})^2$ &  &  & 0.02030 &  &  &  &  &  &  &  &  \\
 & 1f$_{5/2}$ 2p$_{1/2}$ &  &  & 0.01620 &  &  &  &  &  &  &  &  &  & 1f$_{5/2}$ 2p$_{1/2}$ &  &  & 0.01219 &  &  &  &  &  &  &  &  \\
 &  &  &  &  &  &  &  &  &  &  &  &  &  &  &  &  &  &  &  &  &  &  &  &  &  \\
 & (1f$_{7/2})^2$ &  &  & 0.37195 &  &  &  &  &  &  &  &  &  & (1f$_{7/2})^2$ &  &  & -0.35064 &  &  &  &  &  &  &  &  \\
 & 1f$_{7/2}$ 2p$_{3/2}$ &  &  & 0.07272 &  &  &  &  &  &  &  &  &  & 1f$_{7/2}$ 2p$_{3/2}$ &  &  & -0.12193 &  &  &  &  &  &  &  &  \\
 & 1f$_{7/2}$ 1f$_{5/2}$ &  &  & 0.00559 &  &  &  &  &  &  &  &  &  & 1f$_{7/2}$ 1f$_{5/2}$ &  &  & -0.00166 &  &  &  &  &  &  &  &  \\
 & (2p$_{3/2})^2$ &  &  & 0.00588 & 1 & 1 & 2 & 3 & 2 & 1 & 0.01707 &  &  & (2p$_{3/2})^2$ &  &  & -0.00331 & 1 & 1 & 2 & 3 & 2 & 1 & -0.03264 &  \\
 & 2p$_{3/2}$ 1f$_{5/2}$ &  &  & 0.00914 &  &  &  &  &  &  &  &  &  & 2p$_{3/2}$ 1f$_{5/2}$ &  &  & -0.00194 &  &  &  &  &  &  &  &  \\
 & 2p$_{3/2}$ 2p$_{1/2}$ &  &  & 0.00564 &  &  &  &  &  &  &  &  &  & 2p$_{3/2}$ 2p$_{1/2}$ &  &  & 0.01794 &  &  &  &  &  &  &  &  \\
 & (1f$_{5/2})^2$ &  &  & 0.00848 &  &  &  &  &  &  &  &  &  & (1f$_{5/2})^2$ &  &  & 0.02030 &  &  &  &  &  &  &  &  \\
 & 1f$_{5/2}$ 2p$_{1/2}$ &  &  & 0.01620 &  &  &  &  &  &  &  &  &  & 1f$_{5/2}$ 2p$_{1/2}$ &  &  & 0.01219 &  &  &  &  &  &  &  &  \\
 &  &  &  &  &  &  &  &  &  &  &  &  &  &  &  &  &  &  &  &  &  &  &  &  &  \\
 & (1f$_{7/2})^2$ &  &  & 0.37195 &  &  &  &  &  &  &  &  &  & (1f$_{7/2})^2$ &  &  & -0.35064 &  &  &  &  &  &  &  &  \\
 & 1f$_{7/2}$ 2p$_{3/2}$ &  &  & 0.07272 &  &  &  &  &  &  &  &  &  & 1f$_{7/2}$ 2p$_{3/2}$ &  &  & -0.12193 &  &  &  &  &  &  &  &  \\
 & 1f$_{7/2}$ 1f$_{5/2}$ &  &  & 0.00559 & 1 & 1 & 2 & 3 & 3 & 1 & -0.05273 &  &  & 1f$_{7/2}$ 1f$_{5/2}$ &  &  & -0.00166 & 1 & 1 & 2 & 3 & 3 & 1 & 0.06573 &  \\
 & 2p$_{3/2}$ 1f$_{5/2}$ &  &  & 0.00914 &  &  &  &  &  &  &  &  &  & 2p$_{3/2}$ 1f$_{5/2}$ &  &  & -0.00194 &  &  &  &  &  &  &  &  \\
 & (1f$_{5/2})^2$ &  &  & 0.00848 &  &  &  &  &  &  &  &  &  & (1f$_{5/2})^2$ &  &  & 0.02030 &  &  &  &  &  &  &  &  \\
 & 1f$_{5/2}$ 2p$_{1/2}$ &  &  & 0.01620 &  &  &  &  &  &  &  &  &  & 1f$_{5/2}$ 2p$_{1/2}$ &  &  & 0.01219 &  &  &  &  &  &  &  &  \\ \cline{5-12} \cline{18-26} 
 &  &  &  &  &  &  &  &  &  &  &  &  &  &  &  &  &  &  &  &  &  &  &  &  &  \\
 & (1f$_{7/2})^2$ & 4 &  & -0.21898 &  &  &  &  &  &  &  &  &  & (1f$_{7/2})^2$ & 4 &  & 0.16457 &  &  &  &  &  &  &  &  \\
 & 1f$_{7/2}$ 2p$_{3/2}$ &  &  & 0.01556 &  &  &  &  &  &  &  &  &  & 1f$_{7/2}$ 2p$_{3/2}$ &  &  & -0.04199 &  &  &  &  &  &  &  &  \\
 & 1f$_{7/2}$ 1f$_{5/2}$ &  &  & 0.01400 & 1 & 0 & 2 & 4 & 4 & 0 & -0.02253 &  &  & 1f$_{7/2}$ 1f$_{5/2}$ &  &  & -0.03363 & 1 & 0 & 2 & 4 & 4 & 0 & 0.00464 &  \\
 & 1f$_{7/2}$ 2p$_{1/2}$ &  &  & -0.00195 &  &  &  &  &  &  &  &  &  & 1f$_{7/2}$ 2p$_{1/2}$ &  &  & -0.03874 &  &  &  &  &  &  &  &  \\
 & 2p$_{3/2}$ 1f$_{5/2}$ &  &  & 0.00046 &  &  &  &  &  &  &  &  &  & 2p$_{3/2}$ 1f$_{5/2}$ &  &  & 0.01092 &  &  &  &  &  &  &  &  \\
 & (1f$_{5/2})^2$ &  &  & -0.00619 &  &  &  &  &  &  &  &  &  & (1f$_{5/2})^2$ &  &  & 0.01642 &  &  &  &  &  &  &  &  \\
 &  &  &  &  &  &  &  &  &  &  &  &  &  &  &  &  &  &  &  &  &  &  &  &  &  \\
 & (1f$_{7/2})^2$ &  &  & -0.21898 &  &  &  &  &  &  &  &  &  & (1f$_{7/2})^2$ &  &  & 0.16457 &  &  &  &  &  &  &  &  \\
 & 1f$_{7/2}$ 2p$_{3/2}$ &  &  & 0.01556 &  &  &  &  &  &  &  &  &  & 1f$_{7/2}$ 2p$_{3/2}$ &  &  & -0.04199 &  &  &  &  &  &  &  &  \\
 & 1f$_{7/2}$ 1f$_{5/2}$ &  &  & 0.01400 & 1 & 0 & 2 & 4 & 4 & 1 & 0.00284 &  &  & 1f$_{7/2}$ 1f$_{5/2}$ &  &  & -0.03363 & 1 & 0 & 2 & 4 & 4 & 1 & -0.01432 &  \\
 & 1f$_{7/2}$ 2p$_{1/2}$ &  &  & -0.00195 &  &  &  &  &  &  &  &  &  & 1f$_{7/2}$ 2p$_{1/2}$ &  &  & -0.03874 &  &  &  &  &  &  &  &  \\
 & 2p$_{3/2}$ 1f$_{5/2}$ &  &  & 0.00046 &  &  &  &  &  &  &  &  &  & 2p$_{3/2}$ 1f$_{5/2}$ &  &  & 0.01092 &  &  &  &  &  &  &  &  \\
 & (1f$_{5/2})^2$ &  &  & -0.00619 &  &  &  &  &  &  &  &  &  & (1f$_{5/2})^2$ &  &  & 0.01642 &  &  &  &  &  &  &  &  \\
 &  &  &  &  &  &  &  &  &  &  &  &  &  &  &  &  &  &  &  &  &  &  &  &  &  \\
 & (1f$_{7/2})^2$ &  &  & -0.21898 &  &  &  &  &  &  &  &  &  & (1f$_{7/2})^2$ &  &  & 0.16457 &  &  &  &  &  &  &  &  \\
 & 1f$_{7/2}$ 2p$_{3/2}$ &  &  & 0.01556 &  &  &  &  &  &  &  &  &  & 1f$_{7/2}$ 2p$_{3/2}$ &  &  & -0.04199 &  &  &  &  &  &  &  &  \\
 & 1f$_{7/2}$ 1f$_{5/2}$ &  &  & 0.01400 & 1 & 1 & 1 & 5 & 4 & 1 & 0.00476 &  &  & 1f$_{7/2}$ 1f$_{5/2}$ &  &  & -0.03363 & 1 & 1 & 1 & 5 & 4 & 1 & -0.0256 &  \\
 & 1f$_{7/2}$ 2p$_{1/2}$ &  &  & -0.00195 &  &  &  &  &  &  &  &  &  & 1f$_{7/2}$ 2p$_{1/2}$ &  &  & -0.03874 &  &  &  &  &  &  &  &  \\
 & 2p$_{3/2}$ 1f$_{5/2}$ &  &  & 0.00046 &  &  &  &  &  &  &  &  &  & 2p$_{3/2}$ 1f$_{5/2}$ &  &  & 0.01092 &  &  &  &  &  &  &  &  \\
 & (1f$_{5/2})^2$ &  &  & -0.00619 &  &  &  &  &  &  &  &  &  & (1f$_{5/2})^2$ &  &  & 0.01642 &  &  &  &  &  &  &  & \\
 \hline \\
\end{tabular}%
}
\end{table}

\newpage
\begin{table}[ht]
\centering
\setlength{\tabcolsep}{4pt}
\renewcommand{\arraystretch}{0.8}
\caption{Cluster spectroscopic factors ($\mathcal{S}$ (c.m.)) for
di-neutron transfer involving the lighter nucleus $^{60}$Ni are presented
for the relevant overlaps. The calculations were performed using
\textsc{kshell} with the \texttt{KB3G} and the \texttt{fpd6npn}
effective interactions. Here, $j_1$ and $j_2$ stand for the total 
angular momenta of nucleons 1 and 2, respectively. The quantum numbers
$n$, $\ell$ ($N$, $L$) represent the principal quantum numbers and
orbital angular momenta of the nucleons relative to each other
(and to the core), respectively. $J$ and $\Lambda$ denote the total
angular momentum and the orbital angular momentum of the cluster
with respect to the core, respectively, and $S$ is the intrinsic
spin of the two-nucleon cluster. The spins and parities of the
initial and final nuclei are given by $I^{\pi}_{\mathrm{i}}$
and $I^{\pi}_{\mathrm{f}}$, respectively.}
\label{tab:60Ni_62Ni}
\resizebox{\textwidth}{!}{%
\begin{tabular}{lllllllllllllllllllllllll}
\cmidrule{1-12} \cmidrule{14-25}
$I^\pi_\text{i}$ & $j_1$$j_2$ & $J$ & $I^\pi_\text{f}$ & $\mathcal{S}$ ($j-j$) \texttt{(KB3G)} & $n$ & $\ell$ & $N$ & $L$ & $\Lambda$ & $S$ & $\mathcal{S}$ (c.m.) \texttt{(KB3G)} & & $I^\pi_\text{i}$ & $j_1$$j_2$ & $J$ & $I^\pi_\text{f}$ & $\mathcal{S}$ ($j-j$)\texttt{(fpd6npn)} & $n$ & $\ell$ & $N$ & $L$ & $\Lambda$ & $S$ & $\mathcal{S}$ (c.m.) \texttt{(fpd6npn)}\\ \cmidrule{1-12} \cmidrule{14-25}

\multirow{31}{*}{$^{60}$Ni$_\text{g.s.}$(0$^{+}$)} & (2p$_{3/2})^2$ &  &  & 0.81720 &  &  &  &  &  &  &  &  &  & (2p$_{3/2})^2$ &  &  & 0.64849 &  &  &  &  &  &  &  \\
 & (1f$_{5/2})^2$ &  & \multirow{6}{*}{$^{62}$Ni$_\text{g.s.}$(0$^{+}$)} & 0.81905 & 1 & 0 & 4 & 0 & 0 & 0 & 0.57613 &  & \multirow{31}{*}{$^{60}$Ni$_\text{g.s.}$(0$^{+}$)} & (1f$_{5/2})^2$ &  & \multirow{6}{*}{$^{62}$Ni$_\text{g.s.}$(0$^{+}$)} & 0.71899 & 1 & 0 & 4 & 0 & 0 & 0 & 0.50034 \\
 & (2p$_{1/2})^2$ &  &  & 0.38681 &  &  &  &  &  &  &  &  &  & (2p$_{1/2})^2$ &  &  & 0.41872 &  &  &  &  &  &  &  \\
 &  & 0 &  &  &  &  &  &  &  &  &  &  &  &  & 0 &  &  &  &  &  &  &  &  &  \\
 & (2p$_{3/2})^2$ &  &  & -0.08238 &  &  &  &  &  &  &  &  &  & (2p$_{3/2})^2$ &  &  & -0.19678 &  &  &  &  &  &  &  \\
 & (1f$_{5/2})^2$ &  &  & -0.14864 & 1 & 1 & 3 & 1 & 1 & 1 & 0.21165 &  &  & (1f$_{5/2})^2$ &  &  & -0.23955 & 1 & 1 & 3 & 1 & 1 & 1 & 0.22947 \\
 & (2p$_{1/2})^2$ &  &  & -0.04649 &  &  &  &  &  &  &  &  &  & (2p$_{1/2})^2$ &  &  & -0.11658 &  &  &  &  &  &  &  \\ \cline{2-12} \cline{15-25} 
 &  &  &  &  &  &  &  &  &  &  &  &  &  &  &  &  &  &  &  &  &  &  &  &  \\
 & (2p$_{3/2})^2$ &  &  & 0.04999 &  &  &  &  &  &  &  &  &  & (2p$_{3/2})^2$ &  &  & -0.08480 &  &  &  &  &  &  &  \\
 & 2p$_{3/2}$ 1f$_{5/2}$ &  &  & 0.07571 &  &  &  &  &  &  &  &  &  & 2p$_{3/2}$ 1f$_{5/2}$ &  &  & -0.08210 &  &  &  &  &  &  &  \\
 & 2p$_{3/2}$ 2p$_{1/2}$ &  &  & 0.24905 & 1 & 0 & 3 & 2 & 2 & 0 & 0.1392 &  &  & 2p$_{3/2}$ 2p$_{1/2}$ &  &  & -0.23097 & 1 & 0 & 3 & 2 & 2 & 0 & -0.12497 \\
 & (1f$_{5/2})^2$ &  &  & 0.40875 &  &  &  &  &  &  &  &  &  & (1f$_{5/2})^2$ &  &  & -0.16611 &  &  &  &  &  &  &  \\
 & 1f$_{5/2}$ 2p$_{1/2}$ &  &  & 0.19720 &  &  &  &  &  &  &  &  &  & 1f$_{5/2}$ 2p$_{1/2}$ &  &  & -0.21200 &  &  &  &  &  &  &  \\
 &  &  &  &  &  &  &  &  &  &  &  &  &  &  &  &  &  &  &  &  &  &  &  &  \\
 & (2p$_{3/2})^2$ &  &  & 0.04999 &  &  &  &  &  &  &  &  &  & (2p$_{3/2})^2$ &  &  & -0.08480 &  &  &  &  &  &  &  \\
 & 2p$_{3/2}$ 1f$_{5/2}$ &  &  & 0.07571 &  &  &  &  &  &  &  &  &  & 2p$_{3/2}$ 1f$_{5/2}$ &  &  & -0.08210 &  &  &  &  &  &  &  \\
 & 2p$_{3/2}$ 2p$_{1/2}$ & 2 & \multirow{6}{*}{$^{62}$Ni$_{1.172}$(2$^{+}$)} & 0.24905 & 1 & 0 & 3 & 2 & 2 & 1 & 0.05813 &  &  & 2p$_{3/2}$ 2p$_{1/2}$ & 2 & \multirow{6}{*}{$^{62}$Ni$_{1.172}$(2$^{+}$)} & -0.23097 & 1 & 0 & 3 & 2 & 2 & 1 & -0.05229 \\
 & (1f$_{5/2})^2$ &  &  & 0.40875 &  &  &  &  &  &  &  &  &  & (1f$_{5/2})^2$ &  &  & -0.16611 &  &  &  &  &  &  &  \\
 & 1f$_{5/2}$ 2p$_{1/2}$ &  &  & 0.19720 &  &  &  &  &  &  &  &  &  & 1f$_{5/2}$ 2p$_{1/2}$ &  &  & -0.21200 &  &  &  &  &  &  &  \\
 &  &  &  &  &  &  &  &  &  &  &  &  &  &  &  &  &  &  &  &  &  &  &  &  \\
 & (2p$_{3/2})^2$ &  &  & 0.04999 &  &  &  &  &  &  &  &  &  & (2p$_{3/2})^2$ &  &  & -0.08480 &  &  &  &  &  &  &  \\
 & 2p$_{3/2}$ 1f$_{5/2}$ &  &  & 0.07571 &  &  &  &  &  &  &  &  &  & 2p$_{3/2}$ 1f$_{5/2}$ &  &  & -0.08210 &  &  &  &  &  &  &  \\
 & 2p$_{3/2}$ 2p$_{1/2}$ &  &  & 0.24905 & 1 & 1 & 2 & 3 & 2 & 1 & -0.01797 &  &  & 2p$_{3/2}$ 2p$_{1/2}$ &  &  & -0.23097 & 1 & 1 & 2 & 3 & 2 & 1 & 0.01914 \\
 & (1f$_{5/2})^2$ &  &  & 0.40875 &  &  &  &  &  &  &  &  &  & (1f$_{5/2})^2$ &  &  & -0.16611 &  &  &  &  &  &  &  \\
 & 1f$_{5/2}$ 2p$_{1/2}$ &  &  & 0.19720 &  &  &  &  &  &  &  &  &  & 1f$_{5/2}$ 2p$_{1/2}$ &  &  & -0.21200 &  &  &  &  &  &  &  \\
 &  &  &  &  &  &  &  &  &  &  &  &  &  &  &  &  &  &  &  &  &  &  &  &  \\
 & 2p$_{3/2}$ 1f$_{5/2}$ &  &  & 0.07571 &  &  &  &  &  &  &  &  &  & 2p$_{3/2}$ 1f$_{5/2}$ &  &  & -0.08210 &  &  &  &  &  &  &  \\
 & (1f$_{5/2})^2$ &  &  & 0.40875 & 1 & 1 & 2 & 3 & 3 & 1 & 0.14351 &  &  & (1f$_{5/2})^2$ &  &  & -0.16611 & 1 & 1 & 2 & 3 & 3 & 1 & -0.07555 \\
 & 1f$_{5/2}$ 2p$_{1/2}$ &  &  & 0.19720 &  &  &  &  &  &  &  &  &  & 1f$_{5/2}$ 2p$_{1/2}$ &  &  & -0.21200 &  &  &  &  &  &  &  \\ \cline{2-12} \cline{15-25} 
 &  &  &  &  &  &  &  &  &  &  &  &  &  &  &  &  &  &  &  &  &  &  &  &  \\
 & 2p$_{3/2}$ 1f$_{5/2}$ & \multirow{2}{*}{4} & $^{62}$Ni$_{2.336}$(4$^{+}$) & -0.42637 & \multirow{2}{*}{1} & \multirow{2}{*}{0} & \multirow{2}{*}{2} & \multirow{2}{*}{4} & \multirow{2}{*}{4} & 0 & -0.12941 &  &  & 2p$_{3/2}$ 1f$_{5/2}$ & \multirow{2}{*}{4} &  & 0.25564 & \multirow{2}{*}{1} & \multirow{2}{*}{0} & \multirow{2}{*}{2} & \multirow{2}{*}{4} & \multirow{2}{*}{4} & 0 & 0.07257 \\
 & (1f$_{5/2})^2$ &  &  & -0.28237 &  &  &  &  &  & 1 & -0.11572 &  &  & (1f$_{5/2})^2$ &  &  & 0.11461 &  &  &  &  &  & 1 & 0.06939 \\ \cline{1-12} \cline{14-25} 
 &  &  &  &  &  &  &  &  &  &  &  &  &  &  &  &  &  &  &  &  &  &  &  &  \\
\multirow{75}{*}{$^{60}$Ni$_{1.332}$(2$^{+}$)} & (2p$_{3/2})^2$ &  &  & -0.72884 &  &  &  &  &  &  &  &  & \multirow{75}{*}{$^{60}$Ni$_{1.332}$(2$^{+}$)} & (2p$_{3/2})^2$ &  &  & -0.42001 &  &  &  &  &  &  &  \\
 & 2p$_{3/2}$ 1f$_{5/2}$ &  &  & -0.29791 &  &  &  &  &  &  &  &  &  & 2p$_{3/2}$ 1f$_{5/2}$ &  &  & -0.27034 &  &  &  &  &  &  &  \\
 & 2p$_{3/2}$ 2p$_{1/2}$ &  &  & -0.17587 & 1 & 0 & 3 & 2 & 2 & 0 & -0.2943 &  &  & 2p$_{3/2}$ 2p$_{1/2}$ &  &  & -0.23836 & 1 & 0 & 3 & 2 & 2 & 0 & -0.23935 \\
 & (1f$_{5/2})^2$ &  &  & -0.42145 &  &  &  &  &  &  &  &  &  & (1f$_{5/2})^2$ &  &  & -0.27031 &  &  &  &  &  &  &  \\
 & 1f$_{5/2}$ 2p$_{1/2}$ &  &  & -0.18002 &  &  &  &  &  &  &  &  &  & 1f$_{5/2}$ 2p$_{1/2}$ &  &  & -0.29105 &  &  &  &  &  &  &  \\
 &  &  &  &  &  &  &  &  &  &  &  &  &  &  &  &  &  &  &  &  &  &  &  &  \\
 & (2p$_{3/2})^2$ &  &  & -0.72884 &  &  &  &  &  &  &  &  &  & (2p$_{3/2})^2$ &  &  & -0.42001 &  &  &  &  &  &  &  \\
 & 2p$_{3/2}$ 1f$_{5/2}$ &  &  & -0.29791 &  &  &  &  &  &  &  &  &  & 2p$_{3/2}$ 1f$_{5/2}$ &  &  & -0.27034 &  &  &  &  &  &  &  \\
 & 2p$_{3/2}$ 2p$_{1/2}$ & 2 & $^{62}$Ni$_\text{g.s.}$(0$^{+}$) & -0.17587 & 1 & 0 & 3 & 2 & 2 & 1 & -0.07606 &  &  & 2p$_{3/2}$ 2p$_{1/2}$ & 2 & $^{62}$Ni$_\text{g.s.}$(0$^{+}$) & -0.23836 & 1 & 0 & 3 & 2 & 2 & 1 & -0.07549 \\
 & (1f$_{5/2})^2$ &  &  & -0.42145 &  &  &  &  &  &  &  &  &  & (1f$_{5/2})^2$ &  &  & -0.27031 &  &  &  &  &  &  &  \\
 & 1f$_{5/2}$ 2p$_{1/2}$ &  &  & -0.18002 &  &  &  &  &  &  &  &  &  & 1f$_{5/2}$ 2p$_{1/2}$ &  &  & -0.29105 &  &  &  &  &  &  &  \\
 &  &  &  &  &  &  &  &  &  &  &  &  &  &  &  &  &  &  &  &  &  &  &  &  \\
 & (2p$_{3/2})^2$ &  &  & -0.72884 &  &  &  &  &  &  &  &  &  & (2p$_{3/2})^2$ &  &  & -0.42001 &  &  &  &  &  &  &  \\
 & 2p$_{3/2}$ 1f$_{5/2}$ &  &  & -0.29791 &  &  &  &  &  &  &  &  &  & 2p$_{3/2}$ 1f$_{5/2}$ &  &  & -0.27034 &  &  &  &  &  &  &  \\
 & 2p$_{3/2}$ 2p$_{1/2}$ &  &  & -0.17587 & 1 & 1 & 2 & 3 & 2 & 1 & -0.0408 &  &  & 2p$_{3/2}$ 2p$_{1/2}$ &  &  & -0.23836 & 1 & 1 & 2 & 3 & 2 & 1 & -0.01314 \\
 & (1f$_{5/2})^2$ &  &  & -0.42145 &  &  &  &  &  &  &  &  &  & (1f$_{5/2})^2$ &  &  & -0.27031 &  &  &  &  &  &  &  \\
 & 1f$_{5/2}$ 2p$_{1/2}$ &  &  & -0.18002 &  &  &  &  &  &  &  &  &  & 1f$_{5/2}$ 2p$_{1/2}$ &  &  & -0.29105 &  &  &  &  &  &  &  \\
 &  &  &  &  &  &  &  &  &  &  &  &  &  &  &  &  &  &  &  &  &  &  &  &  \\
 & 2p$_{3/2}$ 1f$_{5/2}$ &  &  & -0.29791 &  &  &  &  &  &  &  &  &  & 2p$_{3/2}$ 1f$_{5/2}$ &  &  & -0.27034 &  &  &  &  &  &  &  \\
 & (1f$_{5/2})^2$ &  &  & -0.42145 & 1 & 1 & 2 & 3 & 3 & 1 & -0.09028 &  &  & (1f$_{5/2})^2$ &  &  & -0.27031 & 1 & 1 & 2 & 3 & 3 & 1 & -0.07825 \\
 & 1f$_{5/2}$ 2p$_{1/2}$ &  &  & -0.18002 &  &  &  &  &  &  &  &  &  & 1f$_{5/2}$ 2p$_{1/2}$ &  &  & -0.29105 &  &  &  &  &  &  &  \\ \cline{4-12} \cline{17-25} 
 &  &  &  &  &  &  &  &  &  &  &  &  &  &  &  &  &  &  &  &  &  &  &  &  \\
 & (2p$_{3/2})^2$ &  & \multirow{31}{*}{$^{62}$Ni$_{1.172}$(2$^{+}$)} & 0.67347 &  &  &  &  &  &  &  &  &  & (2p$_{3/2})^2$ &  & \multirow{31}{*}{$^{62}$Ni$_{1.172}$(2$^{+}$)} & -0.38715 &  &  &  &  &  &  &  \\
 & (1f$_{5/2})^2$ &  &  & 0.37536 & 1 & 0 & 4 & 0 & 0 & 0 & 0.4278 &  &  & (1f$_{5/2})^2$ &  &  & -0.40981 & 1 & 0 & 4 & 0 & 0 & 0 & -0.30682 \\
 & (2p$_{1/2})^2$ &  &  & 0.30817 &  &  &  &  &  &  &  &  &  & (2p$_{1/2})^2$ &  &  & -0.28704 &  &  &  &  &  &  &  \\
 &  & 0 &  &  &  &  &  &  &  &  &  &  &  &  & 0 &  &  &  &  &  &  &  &  &  \\
 & (2p$_{3/2})^2$ &  &  & 0.67347 &  &  &  &  &  &  &  &  &  & (2p$_{3/2})^2$ &  &  & -0.38715 &  &  &  &  &  &  &  \\
 & (1f$_{5/2})^2$ &  &  & 0.37536 & 1 & 1 & 3 & 1 & 1 & 1 & 0.06951 &  &  & (1f$_{5/2})^2$ &  &  & -0.40981 & 1 & 1 & 3 & 1 & 1 & 1 & -0.14309 \\
 & (2p$_{1/2})^2$ &  &  & 0.30817 &  &  &  &  &  &  &  &  &  & (2p$_{1/2})^2$ &  &  & -0.28704 &  &  &  &  &  &  &  \\ \cline{5-12} \cline{18-25} 
 &  &  &  &  &  &  &  &  &  &  &  &  &  &  &  &  &  &  &  &  &  &  &  &  \\
 & (2p$_{3/2})^2$ &  &  & -0.19376 &  &  &  &  &  &  &  &  &  & (2p$_{3/2})^2$ &  &  &  &  &  &  &  &  &  &  \\
 & 2p$_{3/2}$ 1f$_{5/2}$ &  &  & -0.11821 &  &  &  &  &  &  &  &  &  & 2p$_{3/2}$ 1f$_{5/2}$ &  &  &  &  &  &  &  &  &  &  \\
 & 2p$_{3/2}$ 2p$_{1/2}$ &  &  & -0.41062 & 1 & 0 & 3 & 2 & 2 & 0 & -0.19559 &  &  & 2p$_{3/2}$ 2p$_{1/2}$ &  &  &  & 1 & 0 & 3 & 2 & 2 & 0 & 0.25554 \\
 & (1f$_{5/2})^2$ &  &  & 0.02068 &  &  &  &  &  &  &  &  &  & (1f$_{5/2})^2$ &  &  &  &  &  &  &  &  &  &  \\
 & 1f$_{5/2}$ 2p$_{1/2}$ &  &  & -0.32941 &  &  &  &  &  &  &  &  &  & 1f$_{5/2}$ 2p$_{1/2}$ &  &  &  &  &  &  &  &  &  &  \\
 &  &  &  &  &  &  &  &  &  &  &  &  &  &  &  &  &  &  &  &  &  &  &  &  \\
 & (2p$_{3/2})^2$ &  &  & -0.19376 &  &  &  &  &  &  &  &  &  & (2p$_{3/2})^2$ &  &  &  &  &  &  &  &  &  &  \\
 & 2p$_{3/2}$ 1f$_{5/2}$ &  &  & -0.11821 &  &  &  &  &  &  &  &  &  & 2p$_{3/2}$ 1f$_{5/2}$ &  &  &  &  &  &  &  &  &  &  \\
 & 2p$_{3/2}$ 2p$_{1/2}$ &  &  & -0.41062 & 1 & 0 & 3 & 2 & 2 & 1 & -0.09424 &  &  & 2p$_{3/2}$ 2p$_{1/2}$ &  &  &  & 1 & 0 & 3 & 2 & 2 & 1 & 0.1004 \\
 & (1f$_{5/2})^2$ &  &  & 0.02068 &  &  &  &  &  &  &  &  &  & (1f$_{5/2})^2$ &  &  &  &  &  &  &  &  &  &  \\
 & 1f$_{5/2}$ 2p$_{1/2}$ &  &  & -0.32941 &  &  &  &  &  &  &  &  &  & 1f$_{5/2}$ 2p$_{1/2}$ &  &  &  &  &  &  &  &  &  &  \\
 &  & 2 &  &  &  &  &  &  &  &  &  &  &  &  & 2 &  &  &  &  &  &  &  &  &  \\
 & (2p$_{3/2})^2$ &  &  & -0.19376 &  &  &  &  &  &  &  &  &  & (2p$_{3/2})^2$ &  &  &  &  &  &  &  &  &  &  \\
 & 2p$_{3/2}$ 1f$_{5/2}$ &  &  & -0.11821 &  &  &  &  &  &  &  &  &  & 2p$_{3/2}$ 1f$_{5/2}$ &  &  &  &  &  &  &  &  &  &  \\
 & 2p$_{3/2}$ 2p$_{1/2}$ &  &  & -0.41062 & 1 & 1 & 2 & 3 & 2 & 1 & 0.03208 &  &  & 2p$_{3/2}$ 2p$_{1/2}$ &  &  &  & 1 & 1 & 2 & 3 & 2 & 1 & -0.03735 \\
 & (1f$_{5/2})^2$ &  &  & 0.02068 &  &  &  &  &  &  &  &  &  & (1f$_{5/2})^2$ &  &  &  &  &  &  &  &  &  &  \\
 & 1f$_{5/2}$ 2p$_{1/2}$ &  &  & -0.32941 &  &  &  &  &  &  &  &  &  & 1f$_{5/2}$ 2p$_{1/2}$ &  &  &  &  &  &  &  &  &  &  \\
 &  &  &  &  &  &  &  &  &  &  &  &  &  &  &  &  &  &  &  &  &  &  &  &  \\
 & 2p$_{3/2}$ 1f$_{5/2}$ &  &  & -0.11821 &  &  &  &  &  &  &  &  &  & 2p$_{3/2}$ 1f$_{5/2}$ &  &  &  &  &  &  &  &  &  &  \\
 & (1f$_{5/2})^2$ &  &  & 0.02068 & 1 & 1 & 2 & 3 & 3 & 1 & -0.0395 &  &  & (1f$_{5/2})^2$ &  &  &  & 1 & 1 & 2 & 3 & 3 & 1 & 0.12381 \\
 & 1f$_{5/2}$ 2p$_{1/2}$ &  &  & -0.32941 &  &  &  &  &  &  &  &  &  & 1f$_{5/2}$ 2p$_{1/2}$ &  &  &  &  &  &  &  &  &  &  \\ \cline{5-12} \cline{18-25} 
 &  &  &  &  &  &  &  &  &  &  &  &  &  &  &  &  &  &  &  &  &  &  &  &  \\
 & 2p$_{3/2}$ 1f$_{5/2}$ & \multirow{2}{*}{4} &  & -0.22751 & \multirow{2}{*}{1} & \multirow{2}{*}{0} & \multirow{2}{*}{2} & \multirow{2}{*}{4} & \multirow{2}{*}{4} & 0 & -0.07317 &  &  & 2p$_{3/2}$ 1f$_{5/2}$ & \multirow{2}{*}{4} &  & 0.25564 & \multirow{2}{*}{1} & \multirow{2}{*}{0} & \multirow{2}{*}{2} & \multirow{2}{*}{4} & \multirow{2}{*}{4} & 0 & 0.02457 \\
 & (1f$_{5/2})^2$ &  &  & -0.19550 &  &  &  &  &  & 1 & -0.06175 &  &  & (1f$_{5/2})^2$ &  &  &  &  &  &  &  &  & 1 & 0.01437 \\ \cline{2-12} \cline{17-25} 
 &  &  &  &  &  &  &  &  &  &  &  &  &  &  &  &  &  &  &  &  &  &  &  &  \\
 & (2p$_{3/2})^2$ &  & \multirow{23}{*}{$^{62}$Ni$_{2.336}$(4$^{+}$)} & 0.07826 &  &  &  &  &  &  &  &  &  & (2p$_{3/2})^2$ &  & \multirow{23}{*}{$^{62}$Ni$_{2.336}$(4$^{+}$)} & 0.00309 &  &  &  &  &  &  &  \\
 & 2p$_{3/2}$ 1f$_{5/2}$ &  &  & 0.12881 &  &  &  &  &  &  &  &  &  & 2p$_{3/2}$ 1f$_{5/2}$ &  &  & -0.00926 &  &  &  &  &  &  &  \\
 & 2p$_{3/2}$ 2p$_{1/2}$ &  &  & -0.07829 & 1 & 0 & 3 & 2 & 2 & 0 & 0.014 &  &  & 2p$_{3/2}$ 2p$_{1/2}$ &  &  & 0.12360 & 1 & 0 & 3 & 2 & 2 & 0 & 0.04851 \\
 & (1f$_{5/2})^2$ &  &  & 0.1041 &  &  &  &  &  &  &  &  &  & (1f$_{5/2})^2$ &  &  & 0.05038 &  &  &  &  &  &  &  \\
 & 1f$_{5/2}$ 2p$_{1/2}$ &  &  & -0.035 &  &  &  &  &  &  &  &  &  & 1f$_{5/2}$ 2p$_{1/2}$ &  &  & 0.10602 &  &  &  &  &  &  &  \\
 &  &  &  &  &  &  &  &  &  &  &  &  &  &  &  &  &  &  &  &  &  &  &  &  \\
 & (2p$_{3/2})^2$ &  &  & 0.07826 &  &  &  &  &  &  &  &  &  & (2p$_{3/2})^2$ &  &  & 0.00309 &  &  &  &  &  &  &  \\
 & 2p$_{3/2}$ 1f$_{5/2}$ &  &  & 0.12881 &  &  &  &  &  &  &  &  &  & 2p$_{3/2}$ 1f$_{5/2}$ &  &  & -0.00926 &  &  &  &  &  &  &  \\
 & 2p$_{3/2}$ 2p$_{1/2}$ & 2 &  & -0.07829 & 1 & 0 & 3 & 2 & 2 & 1 & 0.0034 &  &  & 2p$_{3/2}$ 2p$_{1/2}$ & 2 &  & 0.12360 & 1 & 0 & 3 & 2 & 2 & 1 & 0.02019 \\
 & (1f$_{5/2})^2$ &  &  & 0.1041 &  &  &  &  &  &  &  &  &  & (1f$_{5/2})^2$ &  &  & 0.05038 &  &  &  &  &  &  &  \\
 & 1f$_{5/2}$ 2p$_{1/2}$ &  &  & -0.035 &  &  &  &  &  &  &  &  &  & 1f$_{5/2}$ 2p$_{1/2}$ &  &  & 0.10602 &  &  &  &  &  &  &  \\
 &  &  &  &  &  &  &  &  &  &  &  &  &  &  &  &  &  &  &  &  &  &  &  &  \\
 & (2p$_{3/2})^2$ &  &  & 0.07826 &  &  &  &  &  &  &  &  &  & (2p$_{3/2})^2$ &  &  & 0.00309 &  &  &  &  &  &  &  \\
 & 2p$_{3/2}$ 1f$_{5/2}$ &  &  & 0.12881 &  &  &  &  &  &  &  &  &  & 2p$_{3/2}$ 1f$_{5/2}$ &  &  & -0.00926 &  &  &  &  &  &  &  \\
 & 2p$_{3/2}$ 2p$_{1/2}$ &  &  & -0.07829 & 1 & 1 & 2 & 3 & 2 & 1 & 0.03875 &  &  & 2p$_{3/2}$ 2p$_{1/2}$ &  &  & 0.12360 & 1 & 1 & 2 & 3 & 2 & 1 & -0.02215 \\
 & (1f$_{5/2})^2$ &  &  & 0.1041 &  &  &  &  &  &  &  &  &  & (1f$_{5/2})^2$ &  &  & 0.05038 &  &  &  &  &  &  &  \\
 & 1f$_{5/2}$ 2p$_{1/2}$ &  &  & -0.035 &  &  &  &  &  &  &  &  &  & 1f$_{5/2}$ 2p$_{1/2}$ &  &  & 0.10602 &  &  &  &  &  &  &  \\
 &  &  &  &  &  &  &  &  &  &  &  &  &  &  &  &  &  &  &  &  &  &  &  &  \\
 & 2p$_{3/2}$ 1f$_{5/2}$ &  &  & 0.12881 &  &  &  &  &  &  &  &  &  & 2p$_{3/2}$ 1f$_{5/2}$ &  &  & -0.00926 &  &  &  &  &  &  &  \\
 & (1f$_{5/2})^2$ &  &  & 0.1041 & 1 & 1 & 2 & 3 & 3 & 1 & -0.00866 &  &  & (1f$_{5/2})^2$ &  &  & 0.05038 & 1 & 1 & 2 & 3 & 3 & 1 & 0.0404 \\
 & 1f$_{5/2}$ 2p$_{1/2}$ &  &  & -0.035 &  &  &  &  &  &  &  &  &  & 1f$_{5/2}$ 2p$_{1/2}$ &  &  & 0.10602 &  &  &  &  &  &  &  \\
 &  &  &  &  &  &  &  &  &  &  &  &  &  &  &  &  &  &  &  &  &  &  &  &  \\ \cline{5-12} \cline{18-25} 
 & 2p$_{3/2}$ 1f$_{5/2}$ & \multirow{2}{*}{4} &  & 0.75045 & \multirow{2}{*}{1} & \multirow{2}{*}{0} & \multirow{2}{*}{2} & \multirow{2}{*}{4} & \multirow{2}{*}{4} & 0 & 0.2036 &  &  & 2p$_{3/2}$ 1f$_{5/2}$ & \multirow{2}{*}{4} &  & -0.49860 & \multirow{2}{*}{1} & \multirow{2}{*}{0} & \multirow{2}{*}{2} & \multirow{2}{*}{4} & \multirow{2}{*}{4} & 0 & -0.14011 \\
 & (1f$_{5/2})^2$ &  &  & 0.23368 &  &  &  &  &  & 1 & 0.20368 &  &  & (1f$_{5/2})^2$ &  &  & -0.20800 &  &  &  &  &  & 1 & -0.13533 \\ \cline{1-12} \cline{14-25} 
\end{tabular}%
}
\end{table}

\begin{table}[ht]
\centering
\setlength{\tabcolsep}{4pt}
\renewcommand{\arraystretch}{0.9}
\caption{Cluster spectroscopic factors ($\mathcal{S}$ (c.m.)) for
di-neutron transfer involving the lighter nucleus $^{60}$Ni are presented
for the relevant overlaps. The calculations were performed using
\textsc{kshell} with the \texttt{JUN45} and the \texttt{JJ44BPN}
effective interactions. Here, $j_1$ and $j_2$ stand for the total
angular momenta of nucleons 1 and 2, respectively. The quantum numbers
$n$, $\ell$ ($N$, $L$) represent the principal quantum numbers and
orbital angular momenta of the nucleons relative to each other
(and to the core), respectively. $J$ and $\Lambda$ denote the total
angular momentum and the orbital angular momentum of the cluster
with respect to the core, respectively, and $S$ is the intrinsic 
spin of the two-neutron cluster. The spins and parities of the
initial and final nuclei are given by $I^{\pi}_{\mathrm{i}}$ and
$I^{\pi}_{\mathrm{f}}$, respectively.}
\label{tab:my-table}
\resizebox{\columnwidth}{!}{%
\begin{tabular}{lllllllllllllllllllllllll}
\cmidrule{1-12} \cmidrule{14-25}
$I^\pi_\text{i}$ & $j_1$$j_2$ & $J$ & $I^\pi_\text{f}$ & $\mathcal{S}$ ($j-j$) \texttt{(JUN45)} & $n$ & $\ell$ & $N$ & $L$ & $\Lambda$ & $S$ & $\mathcal{S}$ (c.m.) \texttt{(JUN45)} & & $I^\pi_\text{i}$ & $j_1$$j_2$ & $J$ & $I^\pi_\text{f}$ & $\mathcal{S}$ ($j-j$)\texttt{(JJ44BPN)} & $n$ & $\ell$ & $N$ & $L$ & $\Lambda$ & $S$ & $\mathcal{S}$ (c.m.) \texttt{(JJ44BPN)}\\ \cmidrule{1-12} \cmidrule{14-25}

\multirow{38}{*}{$^{60}$Ni$_\text{g.s.}$(0$^{+}$)} & (2p$_{3/2})^2$ & \multirow{9}{*}{0} &  & -0.7391 &  &  &  &  &  &  &  &  & \multirow{38}{*}{$^{60}$Ni$_\text{g.s.}$(0$^{+}$)} & (2p$_{3/2})^2$ & \multirow{9}{*}{0} &  & 0.78559 &  &  &  &  &  &  &  \\
 & (1f$_{5/2})^2$ &  & \multirow{8}{*}{$^{62}$Ni$_\text{g.s.}$(0$^{+}$)} & -0.9443 & 1 & 0 & 4 & 0 & 0 & 0 & -0.56254 &  &  & (1f$_{5/2})^2$ &  & \multirow{8}{*}{$^{62}$Ni$_\text{g.s.}$(0$^{+}$)} & 0.95186 & 1 & 0 & 4 & 0 & 0 & 0 & 0.59742 \\
 & (2p$_{1/2})^2$ &  &  & -0.38933 &  &  &  &  &  &  &  &  &  & (2p$_{1/2})^2$ &  &  & 0.43779 &  &  &  &  &  &  &  \\
 &  &  &  &  &  &  &  &  &  &  &  &  &  &  &  &  &  &  &  &  &  &  &  &  \\
 & (2p$_{3/2})^2$ &  &  & -0.7391 &  &  &  &  &  &  &  &  &  & (2p$_{3/2})^2$ &  &  & 0.78559 &  &  &  &  &  &  &  \\
 & (1f$_{5/2})^2$ &  &  & -0.9443 & 1 & 1 & 3 & 1 & 1 & 1 & -0.27371 &  &  & (1f$_{5/2})^2$ &  &  & 0.95186 & 1 & 1 & 3 & 1 & 1 & 1 & 0.28159 \\
 & (2p$_{1/2})^2$ &  &  & -0.38933 &  &  &  &  &  &  &  &  &  & (2p$_{1/2})^2$ &  &  & 0.43779 &  &  &  &  &  &  &  \\
 &  &  &  &  &  &  &  &  &  &  &  &  &  &  &  &  &  &  &  &  &  &  &  &  \\
 & (1g$_{9/2})^2$ &  &  & \multirow{2}{*}{0.501} & 1 & 0 & 5 & 0 & 0 & 0 & 0.04463 &  &  & (1g$_{9/2})^2$ &  &  & \multirow{2}{*}{-0.46856} & 1 & 0 & 5 & 0 & 0 & 0 & -0.04174 \\
 &  &  &  &  & 1 & 1 & 4 & 1 & 1 & 1 & -0.08918 &  &  &  &  &  &  & 1 & 1 & 4 & 1 & 1 & 1 & 0.08341 \\ \cline{4-12} \cline{17-25} 
 & (2p$_{3/2})^2$ & \multirow{22}{*}{2} &  & -0.12107 &  &  &  &  &  &  &  &  &  & (2p$_{3/2})^2$ & \multirow{21}{*}{2} &  & 0.09295 &  &  &  &  &  &  &  \\
 & 2p$_{3/2}$ 1f$_{5/2}$ &  & \multirow{21}{*}{$^{62}$Ni$_{1.172}$(2$^{+}$)} & -0.20150 &  &  &  &  &  &  &  &  &  & 2p$_{3/2}$ 1f$_{5/2}$ &  & \multirow{21}{*}{$^{62}$Ni$_{1.172}$(2$^{+}$)} & -0.16413 &  &  &  &  &  &  &  \\
 & 2p$_{3/2}$ 2p$_{1/2}$ &  &  & -0.16659 & 1 & 0 & 3 & 2 & 2 & 0 & -0.18597 &  &  & 2p$_{3/2}$ 2p$_{1/2}$ &  &  & 0.24316 & 1 & 0 & 3 & 2 & 2 & 0 & 0.15638 \\
 & (1f$_{5/2})^2$ &  &  & -0.49352 &  &  &  &  &  &  &  &  &  & (1f$_{5/2})^2$ &  &  & 0.43511 &  &  &  &  &  &  &  \\
 & 1f$_{5/2}$ 2p$_{1/2}$ &  &  & -0.41070 &  &  &  &  &  &  &  &  &  & 1f$_{5/2}$ 2p$_{1/2}$ &  &  & 0.36822 &  &  &  &  &  &  &  \\
 &  &  &  &  &  &  &  &  &  &  &  &  &  &  &  &  &  &  &  &  &  &  &  &  \\
 & (2p$_{3/2})^2$ &  &  & -0.12107 &  &  &  &  &  &  &  &  &  & (2p$_{3/2})^2$ &  &  & 0.09295 &  &  &  &  &  &  &  \\
 & 2p$_{3/2}$ 1f$_{5/2}$ &  &  & -0.20150 &  &  &  &  &  &  &  &  &  & 2p$_{3/2}$ 1f$_{5/2}$ &  &  & -0.16413 &  &  &  &  &  &  &  \\
 & 2p$_{3/2}$ 2p$_{1/2}$ &  &  & -0.16659 & 1 & 0 & 3 & 2 & 2 & 1 & -0.02943 &  &  & 2p$_{3/2}$ 2p$_{1/2}$ &  &  & 0.24316 & 1 & 0 & 3 & 2 & 2 & 1 & -0.00373 \\
 & (1f$_{5/2})^2$ &  &  & -0.49352 &  &  &  &  &  &  &  &  &  & (1f$_{5/2})^2$ &  &  & 0.43511 &  &  &  &  &  &  &  \\
 & 1f$_{5/2}$ 2p$_{1/2}$ &  &  & -0.41070 &  &  &  &  &  &  &  &  &  & 1f$_{5/2}$ 2p$_{1/2}$ &  &  & 0.36822 &  &  &  &  &  &  &  \\
 &  &  &  &  &  &  &  &  &  &  &  &  &  &  &  &  &  &  &  &  &  &  &  &  \\
 & (2p$_{3/2})^2$ &  &  & -0.12107 &  &  &  &  &  &  &  &  &  & (2p$_{3/2})^2$ &  &  & 0.09295 &  &  &  &  &  &  &  \\
 & 2p$_{3/2}$ 1f$_{5/2}$ &  &  & -0.20150 &  &  &  &  &  &  &  &  &  & 2p$_{3/2}$ 1f$_{5/2}$ &  &  & -0.16413 &  &  &  &  &  &  &  \\
 & 2p$_{3/2}$ 2p$_{1/2}$ &  &  & -0.16659 & 1 & 1 & 2 & 3 & 2 & 1 & 0.02647 &  &  & 2p$_{3/2}$ 2p$_{1/2}$ &  &  & 0.24316 & 1 & 1 & 2 & 3 & 2 & 1 & -0.10992 \\
 & (1f$_{5/2})^2$ &  &  & -0.49352 &  &  &  &  &  &  &  &  &  & (1f$_{5/2})^2$ &  &  & 0.43511 &  &  &  &  &  &  &  \\
 & 1f$_{5/2}$ 2p$_{1/2}$ &  &  & -0.41070 &  &  &  &  &  &  &  &  &  & 1f$_{5/2}$ 2p$_{1/2}$ &  &  & 0.36822 &  &  &  &  &  &  &  \\
 &  &  &  &  &  &  &  &  &  &  &  &  &  &  &  &  &  &  &  &  &  &  &  &  \\
 & 2p$_{3/2}$ 1f$_{5/2}$ &  &  & -0.20150 &  &  &  &  &  &  &  &  &  & 2p$_{3/2}$ 1f$_{5/2}$ &  &  & -0.16413 &  &  &  &  &  &  &  \\
 & (1f$_{5/2})^2$ &  &  & -0.49352 & 1 & 1 & 2 & 3 & 3 & 1 & -0.18559 &  &  & (1f$_{5/2})^2$ &  &  & 0.43511 & 1 & 1 & 2 & 3 & 3 & 1 & 0.24658 \\
 & 1f$_{5/2}$ 2p$_{1/2}$ &  &  & -0.41070 &  &  &  &  &  &  &  &  &  & 1f$_{5/2}$ 2p$_{1/2}$ &  &  & 0.36822 &  &  &  &  &  &  &  \\
 &  &  &  & \multirow{2}{*}{0.20856} & 1 & 0 & 4 & 2 & 2 & 0 & 0.01073 &  &  &  &  &  & \multirow{2}{*}{-0.13477} & 1 & 0 & 4 & 2 & 2 & 0 & -0.00694 \\
 & (1g$_{9/2})^2$ &  &  &  & 1 & 1 & 4 & 1 & 1 & 1 & 0.07384 &  &  & (1g$_{9/2})^2$ &  &  &  & 1 & 1 & 4 & 1 & 1 & 1 & -0.04771 \\ \cline{4-12} \cline{17-25} 
 &  &  &  &  &  &  &  &  &  &  &  &  &  &  &  &  &  &  &  &  &  &  &  &  \\
 & 2p$_{3/2}$ 1f$_{5/2}$ & \multirow{4}{*}{4} & \multirow{4}{*}{$^{62}$Ni$_{2.336}$(4$^{+}$)} & -0.31461 & \multirow{2}{*}{1} & \multirow{2}{*}{0} & \multirow{2}{*}{2} & \multirow{2}{*}{4} & \multirow{2}{*}{4} & 0 & -0.12562 &  &  & 2p$_{3/2}$ 1f$_{5/2}$ & \multirow{4}{*}{4} & \multirow{4}{*}{$^{62}$Ni$_{2.336}$(4$^{+}$)} & -0.37124 & \multirow{2}{*}{1} & \multirow{2}{*}{0} & \multirow{2}{*}{2} & \multirow{2}{*}{4} & \multirow{2}{*}{4} & 0 & -0.04896 \\
 & (1f$_{5/2})^2$ &  &  & -0.53679 &  &  &  &  &  & 1 & -0.08539 &  &  & (1f$_{5/2})^2$ &  &  & 0.44852 &  &  &  &  &  & 1 & -0.10076 \\
 &  &  &  &  &  &  &  &  &  &  &  &  &  &  &  &  &  &  &  &  &  &  &  &  \\
 & (1g$_{9/2})^2$ &  &  & \multirow{2}{*}{0.06716} & 1 & 0 & 3 & 4 & 4 & 0 & 0.00087 &  &  & (1g$_{9/2})^2$ &  &  & \multirow{2}{*}{0.07409} & 1 & 0 & 3 & 4 & 4 & 0 & 0.00405 \\
 &  &  &  &  & 1 & 1 & 2 & 5 & 5 & 1 & -0.00119 &  &  &  &  &  &  & 1 & 1 & 2 & 5 & 5 & 1 & -0.00551 \\ \cline{1-12} \cline{14-25} 
 & (2p$_{3/2})^2$ &  &  & -0.69641 &  &  &  &  &  &  &  &  &  & (2p$_{3/2})^2$ &  &  & 0.36793 &  &  &  &  &  &  &  \\
\multirow{61}{*}{$^{60}$Ni$_{1.332}$(2$^{+}$)} & 2p$_{3/2}$ 1f$_{5/2}$ &  & \multirow{21}{*}{$^{62}$Ni$_\text{g.s.}$(0$^{+}$)} & -0.69318 &  &  &  &  &  &  &  &  & \multirow{61}{*}{$^{60}$Ni$_{1.332}$(2$^{+}$)} & 2p$_{3/2}$ 1f$_{5/2}$ &  & \multirow{21}{*}{$^{62}$Ni$_\text{g.s.}$(0$^{+}$)} & -0.32504 &  &  &  &  &  &  &  \\
 & 2p$_{3/2}$ 2p$_{1/2}$ &  &  & -0.12573 & 1 & 0 & 3 & 2 & 2 & 0 & -0.29872 &  &  & 2p$_{3/2}$ 2p$_{1/2}$ &  &  & 0.2633 & 1 & 0 & 3 & 2 & 2 & 0 & 0.20113 \\
 & (1f$_{5/2})^2$ &  &  & -0.39266 &  &  &  &  &  &  &  &  &  & (1f$_{5/2})^2$ &  &  & 0.46531 &  &  &  &  &  &  &  \\
 & 1f$_{5/2}$ 2p$_{1/2}$ &  &  & -0.14075 &  &  &  &  &  &  &  &  &  & 1f$_{5/2}$ 2p$_{1/2}$ &  &  & 0.28454 &  &  &  &  &  &  &  \\
 &  &  &  &  &  &  &  &  &  &  &  &  &  &  &  &  &  &  &  &  &  &  &  &  \\
 & (2p$_{3/2})^2$ &  &  & -0.69641 &  &  &  &  &  &  &  &  &  & (2p$_{3/2})^2$ &  &  & 0.36793 &  &  &  &  &  &  &  \\
 & 2p$_{3/2}$ 1f$_{5/2}$ &  &  & -0.69318 &  &  &  &  &  &  &  &  &  & 2p$_{3/2}$ 1f$_{5/2}$ &  &  & -0.32504 &  &  &  &  &  &  &  \\
 & 2p$_{3/2}$ 2p$_{1/2}$ & 2 &  & -0.12573 & 1 & 0 & 3 & 2 & 2 & 1 & -0.13149 &  &  & 2p$_{3/2}$ 2p$_{1/2}$ & 2 &  & 0.2633 & 1 & 0 & 3 & 2 & 2 & 1 & -0.01416 \\
 & (1f$_{5/2})^2$ &  &  & -0.39266 &  &  &  &  &  &  &  &  &  & (1f$_{5/2})^2$ &  &  & 0.46531 &  &  &  &  &  &  &  \\
 & 1f$_{5/2}$ 2p$_{1/2}$ &  &  & -0.14075 &  &  &  &  &  &  &  &  &  & 1f$_{5/2}$ 2p$_{1/2}$ &  &  & 0.28454 &  &  &  &  &  &  &  \\
 &  &  &  &  &  &  &  &  &  &  &  &  &  &  &  &  &  &  &  &  &  &  &  &  \\
 & (2p$_{3/2})^2$ &  &  & -0.69641 &  &  &  &  &  &  &  &  &  & (2p$_{3/2})^2$ &  &  & 0.36793 &  &  &  &  &  &  &  \\
 & 2p$_{3/2}$ 1f$_{5/2}$ &  &  & -0.69318 &  &  &  &  &  &  &  &  &  & 2p$_{3/2}$ 1f$_{5/2}$ &  &  & -0.32504 &  &  &  &  &  &  &  \\
 & 2p$_{3/2}$ 2p$_{1/2}$ &  &  & -0.12573 & 1 & 1 & 2 & 3 & 2 & 1 & -0.14696 &  &  & 2p$_{3/2}$ 2p$_{1/2}$ &  &  & 0.2633 & 1 & 1 & 2 & 3 & 2 & 1 & -0.13449 \\
 & (1f$_{5/2})^2$ &  &  & -0.39266 &  &  &  &  &  &  &  &  &  & (1f$_{5/2})^2$ &  &  & 0.46531 &  &  &  &  &  &  &  \\
 & 1f$_{5/2}$ 2p$_{1/2}$ &  &  & -0.14075 &  &  &  &  &  &  &  &  &  & 1f$_{5/2}$ 2p$_{1/2}$ &  &  & 0.28454 &  &  &  &  &  &  &  \\
 &  &  &  &  &  &  &  &  &  &  &  &  &  &  &  &  &  &  &  &  &  &  &  &  \\
 & 2p$_{3/2}$ 1f$_{5/2}$ &  &  & -0.69318 &  &  &  &  &  &  &  &  &  & 2p$_{3/2}$ 1f$_{5/2}$ &  &  & -0.32504 &  &  &  &  &  &  &  \\
 & (1f$_{5/2})^2$ &  &  & -0.39266 & 1 & 1 & 2 & 3 & 3 & 1 & 0.02114 &  &  & (1f$_{5/2})^2$ &  &  & 0.46531 & 1 & 1 & 2 & 3 & 3 & 1 & 0.27495 \\
 & 1f$_{5/2}$ 2p$_{1/2}$ &  &  & -0.14075 &  &  &  &  &  &  &  &  &  & 1f$_{5/2}$ 2p$_{1/2}$ &  &  & 0.28454 &  &  &  &  &  &  &  \\
 &  &  &  & \multirow{2}{*}{0.09680} & 1 & 0 & 4 & 2 & 2 & 0 & 0.00498 &  &  &  &  &  & \multirow{2}{*}{-0.03352} & 1 & 0 & 4 & 2 & 2 & 0 & -0.00173 \\
 & (1g$_{9/2})^2$ &  &  &  & 1 & 1 & 4 & 1 & 1 & 1 & 0.03427 &  &  & (1g$_{9/2})^2$ &  &  &  & 1 & 1 & 4 & 1 & 1 & 1 & -0.01187 \\ \cline{5-12} \cline{18-25} 
 &  &  & \multirow{39}{*}{$^{62}$Ni$_{1.172}$(2$^{+}$)} &  &  &  &  &  &  &  &  &  &  &  &  & \multirow{39}{*}{$^{62}$Ni$_{1.172}$(2$^{+}$)} &  &  &  &  &  &  &  &  \\
 & (2p$_{3/2})^2$ &  &  & 0.46309 &  &  &  &  &  &  &  &  &  & (2p$_{3/2})^2$ &  &  & -0.69122 &  &  &  &  &  &  &  \\
 & (1f$_{5/2})^2$ &  &  & 0.50860 & 1 & 0 & 4 & 0 & 0 & 0 & 0.34478 &  &  & (1f$_{5/2})^2$ &  &  & -0.55072 & 1 & 0 & 4 & 0 & 0 & 0 & -0.43404 \\
 & (2p$_{1/2})^2$ &  &  & 0.25906 &  &  &  &  &  &  &  &  &  & (2p$_{1/2})^2$ &  &  & -0.21736 &  &  &  &  &  &  &  \\
 &  & 0 &  &  &  &  &  &  &  &  &  &  &  &  & 0 &  &  &  &  &  &  &  &  &  \\
 & (2p$_{3/2})^2$ &  &  & 0.46309 &  &  &  &  &  &  &  &  &  & (2p$_{3/2})^2$ &  &  & -0.69122 &  &  &  &  &  &  &  \\
 & (1f$_{5/2})^2$ &  &  & 0.50860 & 1 & 1 & 3 & 1 & 1 & 1 & 0.14859 &  &  & (1f$_{5/2})^2$ &  &  & -0.55072 & 1 & 1 & 3 & 1 & 1 & 1 & -0.09349 \\
 & (2p$_{1/2})^2$ &  &  & 0.25906 &  &  &  &  &  &  &  &  &  & (2p$_{1/2})^2$ &  &  & -0.21736 &  &  &  &  &  &  &  \\
 &  &  &  &  &  &  &  &  &  &  &  &  &  &  &  &  &  &  &  &  &  &  &  &  \\
 & (1g$_{9/2})^2$ &  &  & \multirow{2}{*}{-0.32336} & 1 & 0 & 5 & 0 & 0 & 0 & -0.02881 &  &  & (1g$_{9/2})^2$ &  &  & \multirow{2}{*}{0.30981} & 1 & 0 & 5 & 0 & 0 & 0 & 0.0276 \\
 &  &  &  &  & 1 & 1 & 4 & 1 & 1 & 1 & 0.05756 &  &  &  &  &  &  & 1 & 1 & 4 & 1 & 1 & 1 & -0.05515 \\ \cline{5-12} \cline{18-25} 
 & (2p$_{3/2})^2$ &  &  & -0.14560 &  &  &  &  &  &  &  &  &  & (2p$_{3/2})^2$ &  &  & -0.03979 &  &  &  &  &  &  &  \\
 & 2p$_{3/2}$ 1f$_{5/2}$ &  &  & -0.06042 &  &  &  &  &  &  &  &  &  & 2p$_{3/2}$ 1f$_{5/2}$ &  &  & -0.10813 &  &  &  &  &  &  &  \\
 & 2p$_{3/2}$ 2p$_{1/2}$ &  &  & -0.32127 & 1 & 0 & 3 & 2 & 2 & 0 & -0.14769 &  &  & 2p$_{3/2}$ 2p$_{1/2}$ &  &  & 0.21426 & 1 & 0 & 3 & 2 & 2 & 0 & 0.0689 \\
 & (1f$_{5/2})^2$ &  &  & -0.09716 &  &  &  &  &  &  &  &  &  & (1f$_{5/2})^2$ &  &  & -0.10568 &  &  &  &  &  &  &  \\
 & 1f$_{5/2}$ 2p$_{1/2}$ &  &  & -0.18362 &  &  &  &  &  &  &  &  &  & 1f$_{5/2}$ 2p$_{1/2}$ &  &  & 0.31104 &  &  &  &  &  &  &  \\
 &  &  &  &  &  &  &  &  &  &  &  &  &  &  &  &  &  &  &  &  &  &  &  &  \\
 & (2p$_{3/2})^2$ &  &  & -0.14560 &  &  &  &  &  &  &  &  &  & (2p$_{3/2})^2$ &  &  & -0.03979 &  &  &  &  &  &  &  \\
 & 2p$_{3/2}$ 1f$_{5/2}$ &  &  & -0.06042 &  &  &  &  &  &  &  &  &  & 2p$_{3/2}$ 1f$_{5/2}$ &  &  & -0.10813 &  &  &  &  &  &  &  \\
 & 2p$_{3/2}$ 2p$_{1/2}$ & 2 &  & -0.32127 & 1 & 0 & 3 & 2 & 2 & 1 & -0.07756 &  &  & 2p$_{3/2}$ 2p$_{1/2}$ & 2 &  & 0.21426 & 1 & 0 & 3 & 2 & 2 & 1 & 0.00434 \\
 & (1f$_{5/2})^2$ &  &  & -0.09716 &  &  &  &  &  &  &  &  &  & (1f$_{5/2})^2$ &  &  & -0.10568 &  &  &  &  &  &  &  \\
 & 1f$_{5/2}$ 2p$_{1/2}$ &  &  & -0.18362 &  &  &  &  &  &  &  &  &  & 1f$_{5/2}$ 2p$_{1/2}$ &  &  & 0.31104 &  &  &  &  &  &  &  \\
 &  &  &  &  &  &  &  &  &  &  &  &  &  &  &  &  &  &  &  &  &  &  &  &  \\
 & (2p$_{3/2})^2$ &  &  & -0.14560 &  &  &  &  &  &  &  &  &  & (2p$_{3/2})^2$ &  &  & -0.03979 &  &  &  &  &  &  &  \\
 & 2p$_{3/2}$ 1f$_{5/2}$ &  &  & -0.06042 &  &  &  &  &  &  &  &  &  & 2p$_{3/2}$ 1f$_{5/2}$ &  &  & -0.10813 &  &  &  &  &  &  &  \\
 & 2p$_{3/2}$ 2p$_{1/2}$ &  &  & -0.32127 & 1 & 1 & 2 & 3 & 2 & 1 & 0.01925 &  &  & 2p$_{3/2}$ 2p$_{1/2}$ &  &  & 0.21426 & 1 & 1 & 2 & 3 & 2 & 1 & -0.08523 \\
 & (1f$_{5/2})^2$ &  &  & -0.09716 &  &  &  &  &  &  &  &  &  & (1f$_{5/2})^2$ &  &  & -0.10568 &  &  &  &  &  &  &  \\
 & 1f$_{5/2}$ 2p$_{1/2}$ &  &  & -0.18362 &  &  &  &  &  &  &  &  &  & 1f$_{5/2}$ 2p$_{1/2}$ &  &  & 0.31104 &  &  &  &  &  &  &  \\
 &  &  &  &  &  &  &  &  &  &  &  &  &  &  &  &  &  &  &  &  &  &  &  &  \\
 & 2p$_{3/2}$ 1f$_{5/2}$ &  &  & -0.06042 &  &  &  &  &  &  &  &  &  & 2p$_{3/2}$ 1f$_{5/2}$ &  &  & -0.10813 &  &  &  &  &  &  &  \\
 & (1f$_{5/2})^2$ &  &  & -0.09716 & 1 & 1 & 2 & 3 & 3 & 1 & -0.05457 &  &  & (1f$_{5/2})^2$ &  &  & -0.10568 & 1 & 1 & 2 & 3 & 3 & 1 & 0.06499 \\
 & 1f$_{5/2}$ 2p$_{1/2}$ &  &  & -0.18362 &  &  &  &  &  &  &  &  &  & 1f$_{5/2}$ 2p$_{1/2}$ &  &  & 0.31104 &  &  &  &  &  &  &  \\
 &  &  &  & \multirow{2}{*}{0.05993} & 1 & 0 & 4 & 2 & 2 & 0 & 0.00309 &  &  &  &  &  & \multirow{2}{*}{-0.10813} & 1 & 0 & 4 & 2 & 2 & 0 & -0.00189 \\
 & (1g$_{9/2})^2$ &  &  &  & 1 & 1 & 4 & 1 & 1 & 1 & 0.02122 &  &  & (1g$_{9/2})^2$ &  &  &  & 1 & 1 & 4 & 1 & 1 & 1 & -0.01299 \\ \cline{5-12} \cline{18-25} 
 &  &  &  &  &  &  &  &  &  &  &  &  &  &  &  &  &  &  &  &  &  &  &  &  \\
 & 2p$_{3/2}$ 1f$_{5/2}$ & 4 &  & -0.03860 & 1 & 0 & 2 & 4 & 4 & 0 & -0.03511 &  &  & 2p$_{3/2}$ 1f$_{5/2}$ & 4 &  & 0.28613 & 1 & 0 & 2 & 4 & 4 & 0 & 0.01527 \\
 & (1f$_{5/2})^2$ &  &  & -0.28046 &  &  &  &  &  & 1 & -0.01048 &  &  & (1f$_{5/2})^2$ &  &  & -0.59063 &  &  &  &  &  & 1 & 0.07766 \\
 &  &  &  &  &  &  &  &  &  &  &  &  &  &  &  &  &  &  &  &  &  &  &  &  \\
 & (1g$_{9/2})^2$ &  &  & \multirow{2}{*}{0.01593} & 1 & 0 & 3 & 4 & 4 & 0 & 0.00087 &  &  & (1g$_{9/2})^2$ &  &  & \multirow{2}{*}{-0.04033} & 1 & 0 & 3 & 4 & 4 & 0 & -0.00221 \\
 &  &  &  &  & 1 & 1 & 2 & 5 & 5 & 1 & -0.00119 &  &  &  &  &  &  & 1 & 1 & 2 & 5 & 5 & 1 & 0.003 \\ \cline{4-12} \cline{17-25} 
\end{tabular}%
}
\end{table}
\clearpage
\begin{table}[ht!]
\centering

\resizebox{\columnwidth}{!}{%
\begin{tabular}{@{}lllllllllllllllllllllllll@{}}
 & (2p$_{3/2})^2$ & \multirow{22}{*}{2} & \multirow{26}{*}{$^{62}$Ni$_{2.336}$(4$^{+}$)} & 0.00143 &  &  &  &  &  &  &  &  &  & (2p$_{3/2})^2$ & \multirow{22}{*}{2} & \multirow{26}{*}{$^{62}$Ni$_{2.336}$(4$^{+}$)} & -0.11791 &  &  &  &  &  &  &  \\
 & 2p$_{3/2}$ 1f$_{5/2}$ &  &  & 0.30766 &  &  &  &  &  &  &  &  &  & 2p$_{3/2}$ 1f$_{5/2}$ &  &  & 0.06488 &  &  &  &  &  &  &  \\
 & 2p$_{3/2}$ 2p$_{1/2}$ &  &  & 0.00549 & 1 & 0 & 3 & 2 & 2 & 0 & 0.03658 &  &  & 2p$_{3/2}$ 2p$_{1/2}$ &  &  & 0.05121 & 1 & 0 & 3 & 2 & 2 & 0 & -0.02648 \\
 & (1f$_{5/2})^2$ &  &  & 0.02409 &  &  &  &  &  &  &  &  &  & (1f$_{5/2})^2$ &  &  & -0.10166 &  &  &  &  &  &  &  \\
 & 1f$_{5/2}$ 2p$_{1/2}$ &  &  & 0.05507 &  &  &  &  &  &  &  &  &  & 1f$_{5/2}$ 2p$_{1/2}$ &  &  & -0.05121 &  &  &  &  &  &  &  \\
 &  &  &  &  &  &  &  &  &  &  &  &  &  &  &  &  &  &  &  &  &  &  &  &  \\
 & (2p$_{3/2})^2$ &  &  & 0.00143 &  &  &  &  &  &  &  &  &  & (2p$_{3/2})^2$ &  &  & -0.11791 &  &  &  &  &  &  &  \\
 & 2p$_{3/2}$ 1f$_{5/2}$ &  &  & 0.30766 &  &  &  &  &  &  &  &  &  & 2p$_{3/2}$ 1f$_{5/2}$ &  &  & 0.06488 &  &  &  &  &  &  &  \\
 & 2p$_{3/2}$ 2p$_{1/2}$ &  &  & 0.00549 & 1 & 0 & 3 & 2 & 2 & 1 & 0.04515 &  &  & 2p$_{3/2}$ 2p$_{1/2}$ &  &  & 0.05121 & 1 & 0 & 3 & 2 & 2 & 1 & 0.03126 \\
 & (1f$_{5/2})^2$ &  &  & 0.02409 &  &  &  &  &  &  &  &  &  & (1f$_{5/2})^2$ &  &  & -0.10166 &  &  &  &  &  &  &  \\
 & 1f$_{5/2}$ 2p$_{1/2}$ &  &  & 0.05507 &  &  &  &  &  &  &  &  &  & 1f$_{5/2}$ 2p$_{1/2}$ &  &  & -0.05121 &  &  &  &  &  &  &  \\
 &  &  &  &  &  &  &  &  &  &  &  &  &  &  &  &  &  &  &  &  &  &  &  &  \\
 & (2p$_{3/2})^2$ &  &  & 0.00143 &  &  &  &  &  &  &  &  &  & (2p$_{3/2})^2$ &  &  & -0.11791 &  &  &  &  &  &  &  \\
 & 2p$_{3/2}$ 1f$_{5/2}$ &  &  & 0.30766 &  &  &  &  &  &  &  &  &  & 2p$_{3/2}$ 1f$_{5/2}$ &  &  & 0.06488 &  &  &  &  &  &  &  \\
 & 2p$_{3/2}$ 2p$_{1/2}$ &  &  & 0.00549 & 1 & 1 & 2 & 3 & 2 & 1 & 0.06661 &  &  & 2p$_{3/2}$ 2p$_{1/2}$ &  &  & 0.05121 & 1 & 1 & 2 & 3 & 2 & 1 & 0.0258 \\
 & (1f$_{5/2})^2$ &  &  & 0.02409 &  &  &  &  &  &  &  &  &  & (1f$_{5/2})^2$ &  &  & -0.10166 &  &  &  &  &  &  &  \\
 & 1f$_{5/2}$ 2p$_{1/2}$ &  &  & 0.05507 &  &  &  &  &  &  &  &  &  & 1f$_{5/2}$ 2p$_{1/2}$ &  &  & -0.05121 &  &  &  &  &  &  &  \\
 &  &  &  &  &  &  &  &  &  &  &  &  &  &  &  &  &  &  &  &  &  &  &  &  \\
 & 2p$_{3/2}$ 1f$_{5/2}$ &  &  & 0.30766 &  &  &  &  &  &  &  &  &  & 2p$_{3/2}$ 1f$_{5/2}$ &  &  & 0.06488 &  &  &  &  &  &  &  \\
 & (1f$_{5/2})^2$ &  &  & 0.02409 & 1 & 1 & 2 & 3 & 3 & 1 & -0.05421 &  &  & (1f$_{5/2})^2$ &  &  & -0.10166 & 1 & 1 & 2 & 3 & 3 & 1 & -0.05616 \\
 & 1f$_{5/2}$ 2p$_{1/2}$ &  &  & 0.05507 &  &  &  &  &  &  &  &  &  & 1f$_{5/2}$ 2p$_{1/2}$ &  &  & -0.05121 &  &  &  &  &  &  &  \\
 &  &  &  & \multirow{2}{*}{-0.04854} & 1 & 0 & 4 & 2 & 2 & 0 & -0.0025 &  &  &  &  &  & \multirow{2}{*}{0.00203} & 1 & 0 & 4 & 2 & 2 & 0 & 0.00011 \\
 & (1g$_{9/2})^2$ &  &  &  & 1 & 1 & 4 & 1 & 1 & 1 & -0.01719 &  &  & (1g$_{9/2})^2$ &  &  &  & 1 & 1 & 4 & 1 & 1 & 1 & 0.00072 \\ \cmidrule(lr){5-12} \cmidrule(l){19-25} 
 &  &  &  &  &  &  &  &  &  &  &  &  &  &  &  &  &  &  &  &  &  &  &  &  \\
 & 2p$_{3/2}$ 1f$_{5/2}$ & \multirow{5}{*}{4} &  & -0.85859 & \multirow{2}{*}{1} & \multirow{2}{*}{0} & \multirow{2}{*}{2} & \multirow{2}{*}{4} & \multirow{2}{*}{4} & 0 & -0.24365 &  &  & 2p$_{3/2}$ 1f$_{5/2}$ & \multirow{5}{*}{4} &  & -0.37384 & \multirow{2}{*}{1} & \multirow{2}{*}{0} & \multirow{2}{*}{2} & \multirow{2}{*}{4} & \multirow{2}{*}{4} & 0 & -0.02885 \\
 & (1f$_{5/2})^2$ &  &  & -0.38406 &  &  &  &  &  & 1 & -0.23303 &  &  & (1f$_{5/2})^2$ &  &  & 0.6746 &  &  &  &  &  & 1 & -0.10147 \\
 %&  &  &  &  &  &  &  &  &  &  &  &  &  &  &  &  &  &  &  &  &  &  &  &  \\
 &  &  &  & \multirow{2}{*}{0.07073} & 1 & 0 & 3 & 4 & 4 & 0 & 0.00386 &  &  &  &  &  & \multirow{2}{*}{0.07235} & 1 & 0 & 3 & 4 & 4 & 0 & 0.00395 \\
 & (1g$_{9/2})^2$ &  &  &  & 1 & 1 & 2 & 5 & 5 & 1 & -0.00526 &  &  & (1g$_{9/2})^2$ &  &  &  & 1 & 1 & 2 & 5 & 5 & 1 & -0.00538 \\
 &  &  &  &  &  &  &  &  &  &  &  &  &  &  &  &  &  &  &  &  &  &  &  &  \\ \bottomrule
\end{tabular}%
}
\end{table}
\end{document}